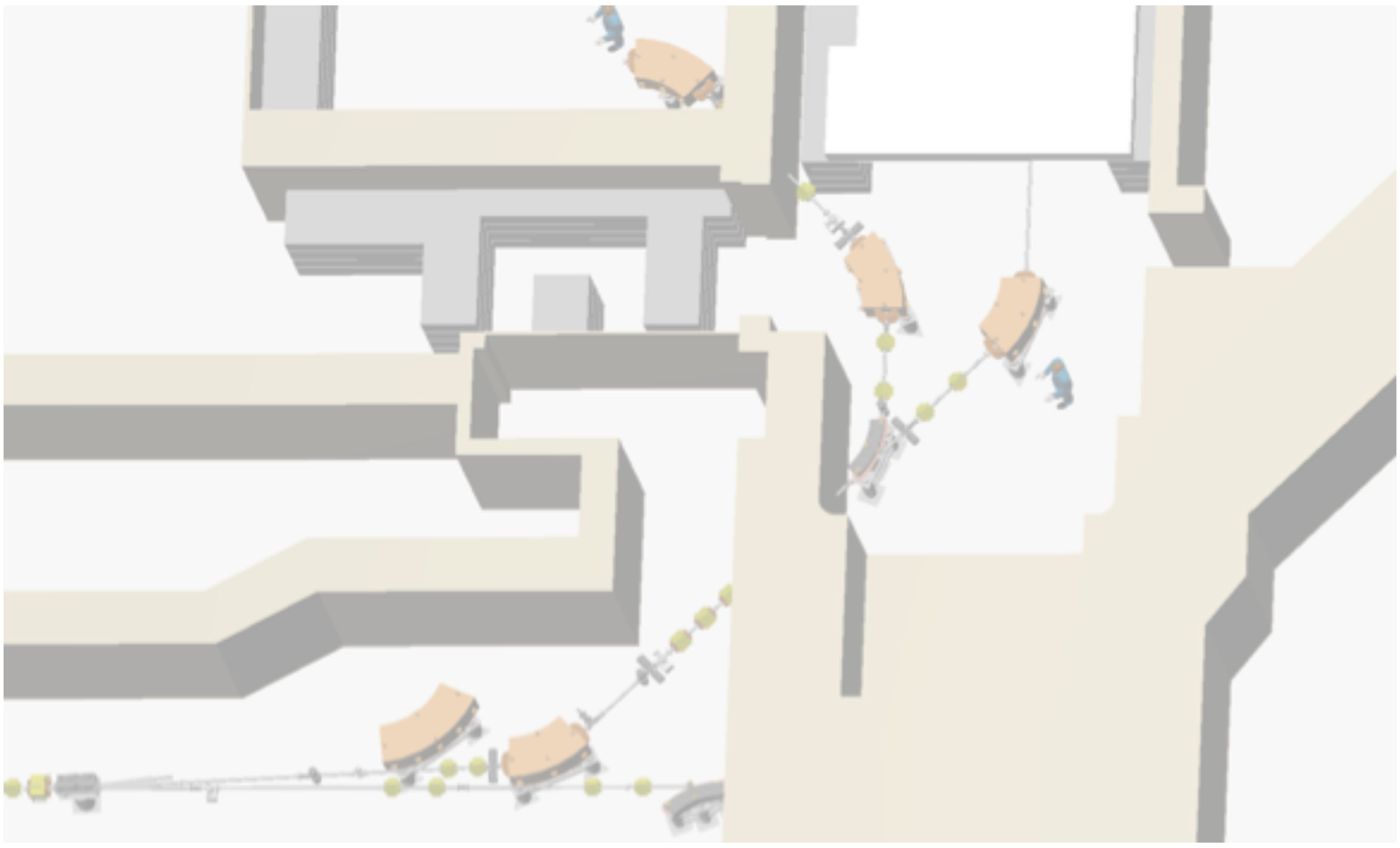

**Linear Accelerator Test Facility**

# Conceptual Design Report

Paolo Valente
*Istituto Nazionale di Fisica Nucleare, Sezione di Roma*

Maurizio Belli, Bruno Bolli, Bruno Buonomo, Sergio Cantarella, Riccardo Ceccarelli, Alberto Cecchinelli, Oreste Cerafogli, Renato Clementi, Claudio Di Giulio, Adolfo Esposito, Oscar Frasciello, Luca Foggetta, Andrea Ghigo, Simona Incremona, Franco Iungo, Roberto Mascio, Stefano Martelli, Graziano Piermarini, Lucia Sabbatini, Franco Sardone, Giancarlo Sensolini, Ruggero Ricci, Luis Antonio Rossi, Ugo Rotundo, Angelo Stella, Serena Strabioli, Raffaele Zarlenga

*Istituto Nazionale di Fisica Nucleare, Laboratori Nazionali di Frascati*

*Supported by the H2020 project AIDA-2020, GA no. 654168.*

*15/3/2016*

# 0 INTRODUCTION

Test beam and irradiation facilities are the key enabling infrastructures for research in high energy physics (HEP) and astro-particles.

In the last 11 years the Beam-Test Facility (BTF) of the DAΦNE accelerator complex in the Frascati laboratory has gained an important role in the European infrastructures devoted to the development and testing of particle detectors. At the same time the BTF operation has been largely shadowed, in terms of resources, by the running of the DAΦNE electron-positron collider.

The present proposal is aimed at improving the present performance of the facility from two different points of view:

- Extending the range of application for the LINAC beam extracted to the BTF lines, in particular in the (in some sense opposite) directions of hosting fundamental physics and providing electron irradiation also for industrial users;
- Extending the life of the LINAC beyond or independently from its use as injector of the DAΦNE collider, as it is also a key element of the electron/positron beam facility.

The main lines of these two developments can be identified as:

- Consolidation of the LINAC infrastructure, in order to guarantee a stable operation in the longer term;
- Upgrade of the LINAC energy, in order to increase the facility capability (especially for the almost unique extracted positron beam);
- Doubling of the BTF beam-lines, in order to cope with the signicant increase of users due to the much wider range of applications.

Even though such a project stems from a facility already existing and operational since more than a decade, based on an accelerator complex designed more than 20 years ago, it is probably useful considering the resulting infrastructure as a new facility, more than an improvement of the existing DAΦNE BTF: **BTF2** or **TALYA**, Linear Accelerator Test fAcilitY are the suggested names (Fig. 0.1).

**Figure 0.1:** Proposed logo for BTF2 or TALYA, a Linear Accelerator Test fAcilitY.

# 1 PRESENT SITUATION

## 1.1 THE FRASCATI ACCELERATOR COMPLEX

The DAΦNE complex of the INFN Frascati laboratory (LNF), has been designed at the beginning of the '90s, with the main purpose of carrying on high statistics experiments with Kaons, abundantly produced by the decay of the Φ meson [1]. The $e^+ e^-$ annihilation cross-section has a narrow peak at the meson mass of 1020 MeV/$c^2$, so that the collider is often called a "Φ-factory". The complex has been completed in 1996 and has been running with stable collisions for experiments since 1999, mainly with KLOE (and its upgraded version KLOE2) [2].

The high current LINAC accelerates electrons and positrons up to 510 MeV energy, that are then stacked and accumulated in a damping ring, for emittance reduction, prior to be injected into the two separated main rings. One out of the two collision points is generally used for experiments (since 2007, the KLOE one), while the beams get separated in the opposite one. The general layout of the complex is schematically shown in Fig. 1.1.

Three synchrotron radiation lines are also operational (DAΦNE-Light lab) [3].

The electron or positron beam can be extracted before the injection into the damping ring to a dedicated transfer line, where a system composed by a target plus a dipole and collimating slits, can attenuate and select the momentum of secondary particles in narrow (<1%) band. The secondary beam is thus driven to a dedicated experimental hall for beam-test activities (BTF, beam-test facility) [4].

The BTF line is fully operational since 2002, and since 2004 can operate in "parasitic" mode during the running of the collider. The full LINAC beam can also be extracted to the BTF line without being intercepted by the target (within the $3 \cdot 10^{10}$ particles/s limit established by radio-protection rules for the current shielding configuration).

In the following sections of this chapter a detailed description of the LINAC and the transfer lines is given, as well as the operation, beam parameters and users experience of the BTF.

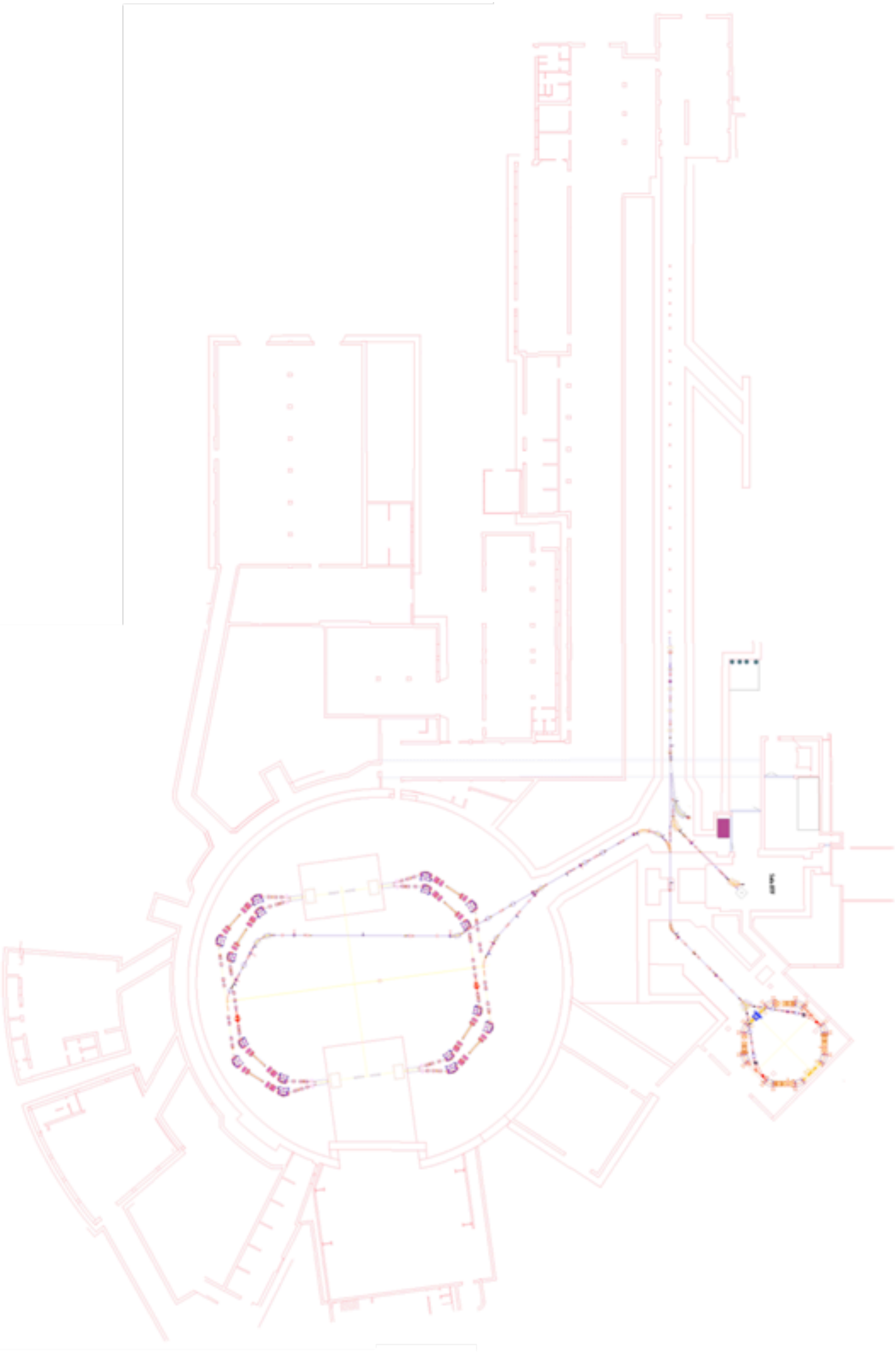

**Figure 1.1:** The DAΦNE accelerator complex in the Frascati laboratory.

## 1.2 LINAC

The LINAC of the DAΦNE complex has been designed, built, and installed by TITAN BETA (USA) and it has been in operation since 1996, as injector of the collider, and, since 2003, for the extraction of electron beam to the Beam Test Facility [5].

It is a ~60 m long, S-band (2856 MHz) linear accelerator, made up by a thermionic gun, four 45 MW klystrons (Thales TH-2128C) with SLED (SLAC Energy Doubler) compression, and by 15 travelling-wave, 2/3π, SLAC-type, 3 m long accelerating sections.

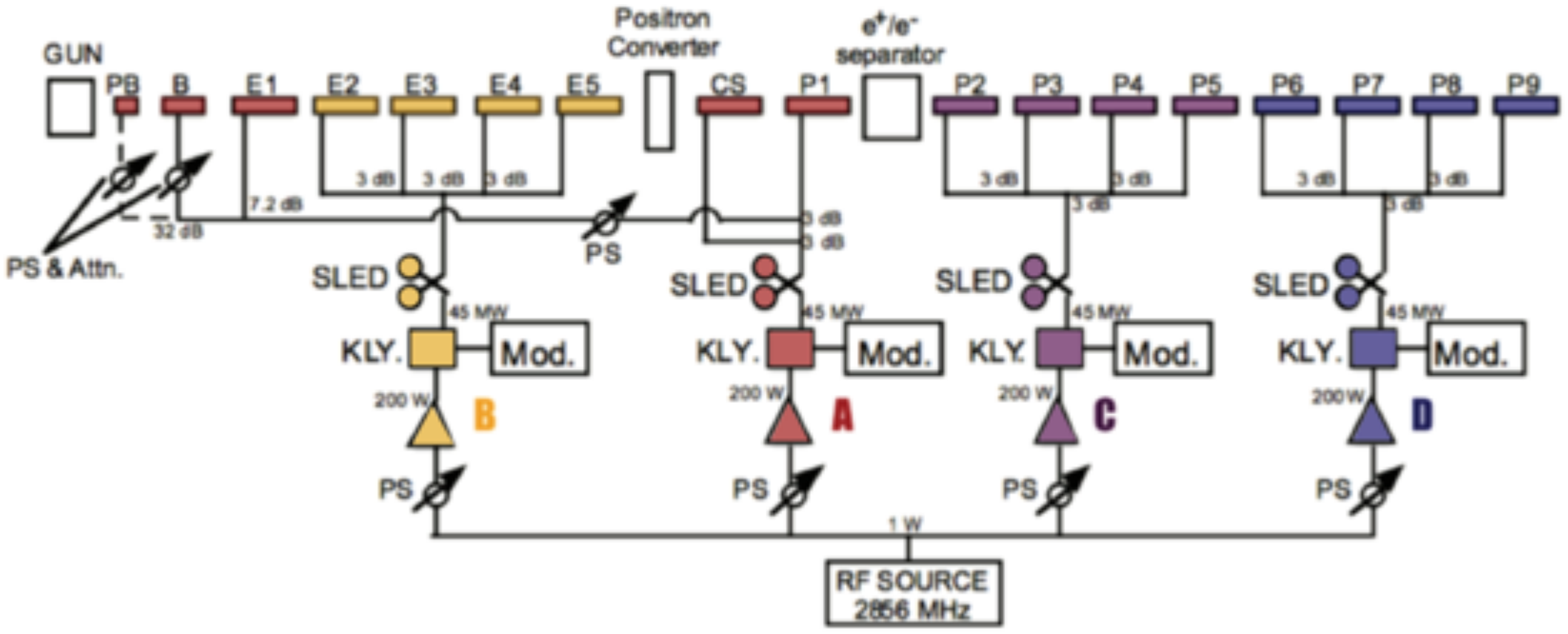

**Figure 1.2:** RF distribution scheme of the LINAC.

The RF distribution scheme is shown in Fig. 1.2: three of the klystrons have exactly the same configuration, consisting of an evacuated rectangular waveguide network with three 3 dB splitters arranged in order to divide the power into four equal parts, feeding each one an accelerating section. The configuration of the fourth klystron is different: half the power is sent to the capture section (CS), the first section downstream the positron converter (PC), while the second half is equally divided between two branches, the first one feeding the accelerating section P1, the second one feeding the pre-buncher, the buncher and the accelerating section E1.

All the 15 accelerating sections (E1-E5, CS, P1-P9) are of the same type: the well known 3 m long, 2/3 π travelling wave, constant gradient, SLAC design structures. In our configuration, with the above illustrated splitting and 45 MW coming out the klystron, the nominal accelerating component of the electric field is 24 MV/m in the CS and 18 MV/m in the remaining accelerating sections.

The phase between the sections can be adjusted by means of low power 360° phase shifters upstream the RF amplifiers of each klystron, and by a high power 360° phase shifter uncoupling the CS from E1. The relative phasing between accelerating sections belonging to the same klystron network were originally tuned by adjusting the rectangular waveguides upstream the section inputs.

The four modulators produce a pulse of 4.5 μs flat top with a repetition rate at 50 Hz: a HV power supply with resonant circuit charges the pulse forming network (PFN), composed by 9 LC cells up to 50 kV, and a switching thyratron.

Electrons are produced by a gridded electron gun with replaceable cathode, high voltage deck with up to 150 KV power supply, isolation transformer and corona shielding, allowing to reach full voltage in air. The deck contains all the necessary electronics to operate the electron gun, and includes fiber optical links to the low level control chassis.

Typical operation values are 6 A, 120 kV in positron mode, and 0.5 A, 120 kV in electron mode. In normal operation (DAΦNE injection at 510 MeV) the gun is pulsed at 50 Hz with a rectangular waveform of 10 ns.

A complete upgrade of the electronic control chassis of the electron gun has been performed in 2014 [6]. A new socket cathode has been installed to obtain the best electrical connection between the high voltage deck and cathode; in addition, inside the isolated high voltage station the original, custom TITAN BETA electronics has been replaced, and the new gun electronics (grid pulser and multiple outputs DC power supply system) has been installed (see Fig. 1.3). The pulse can now be precisely adjusted both in duration, between 1.4 and 40 ns, and amplitude, between 300 and 750 V (as shown in Fig. 1.4).

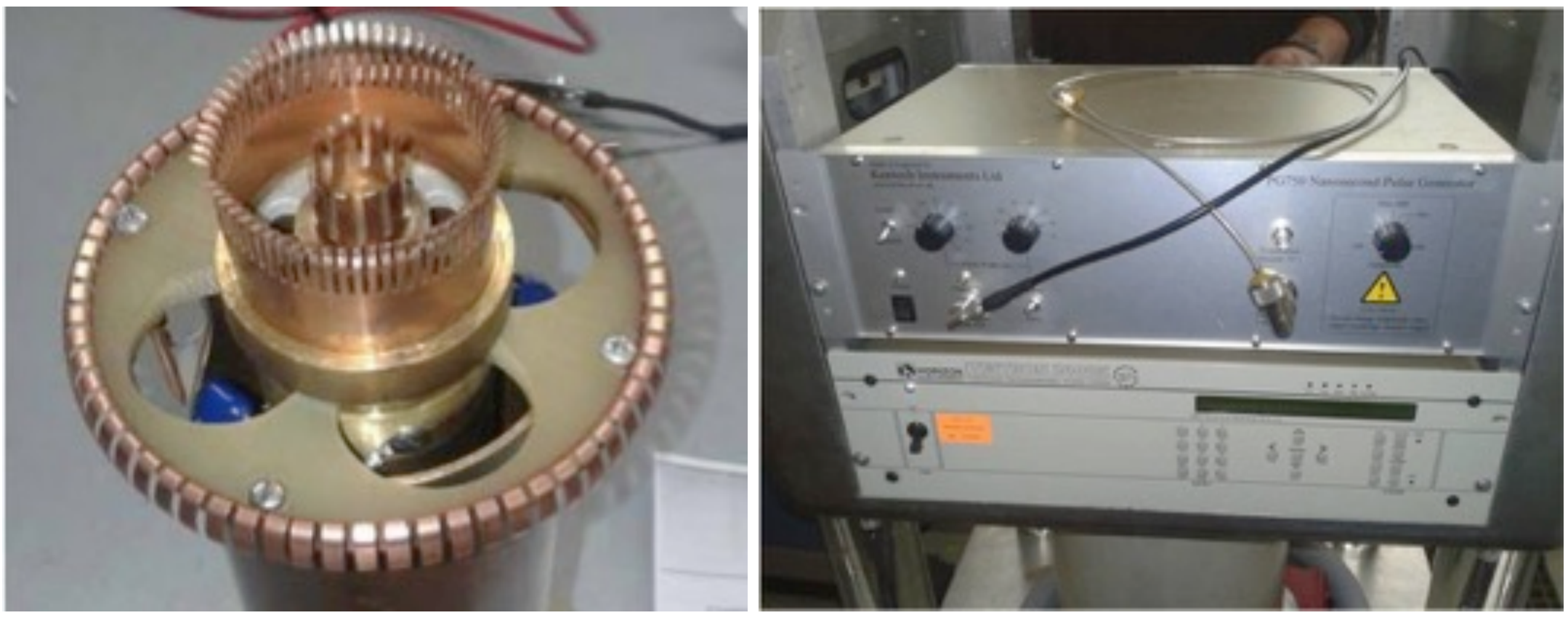

**Figure 1.3:** Electron gun cathode socket (left), and isolated high voltage station with new pulsing system (right).

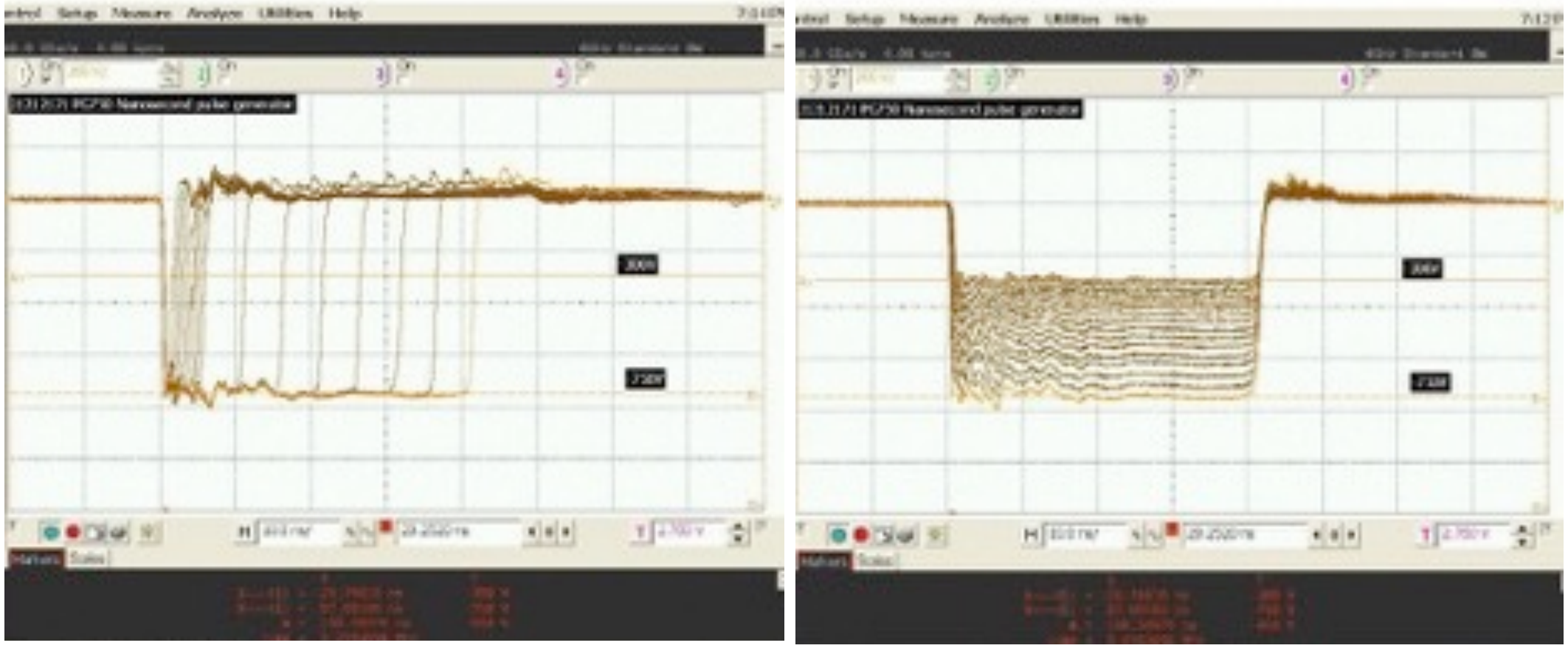

**Figure 1.4:** Rectangular waveform pulsing the electron gun, with variable length between 1.4 and 40 ns (left) and amplitude between 300 and 750 V (right).

Positrons are produced by a system based on the SLAC scheme: a Tungsten-Rhenium, adjustable thickness (around 2 $X_0$) target intercepts the electron beam after the first five accelerating sections (when they have reached an energy of about 220

MeV) and the produced positrons are collected by a flux concentrator jointly with DC solenoid magnets, generating a 5 T peak magnetic field. A remotely controlled actuator allows extracting the target from the beam path during the electron mode of operation.

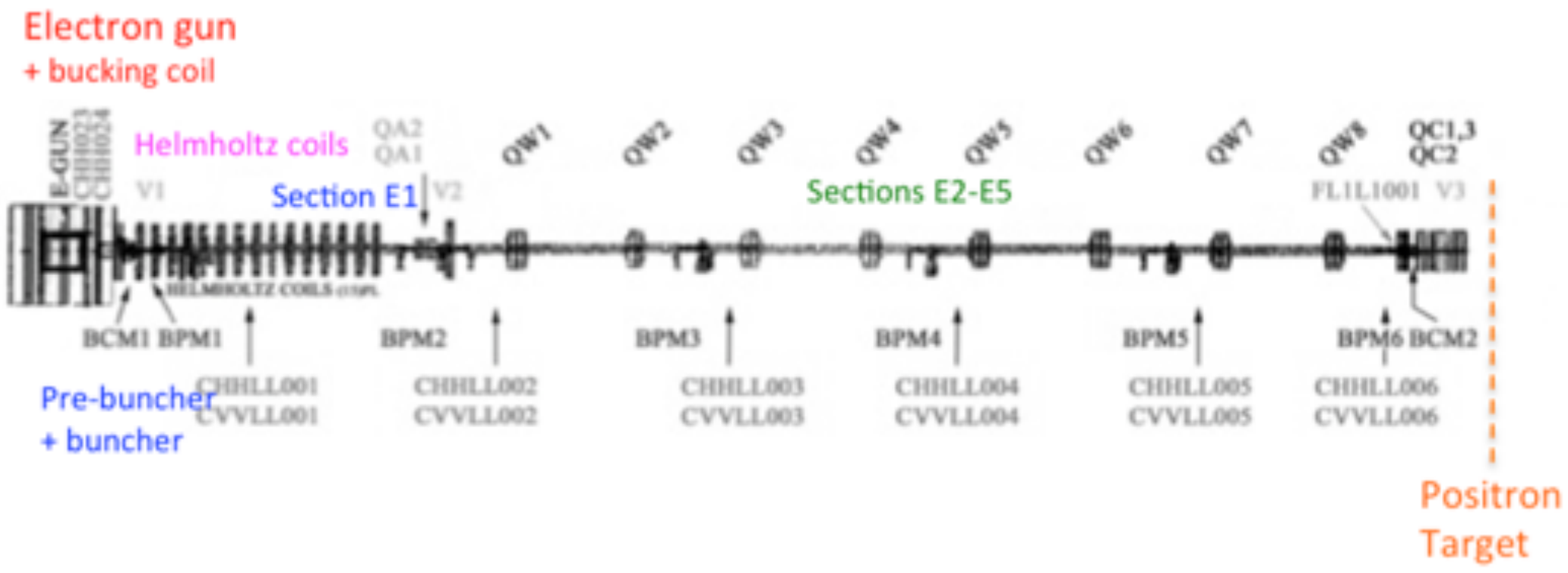

**Figure 1.5:** Scheme of the first part of the LINAC, from the gun to the positron converter.

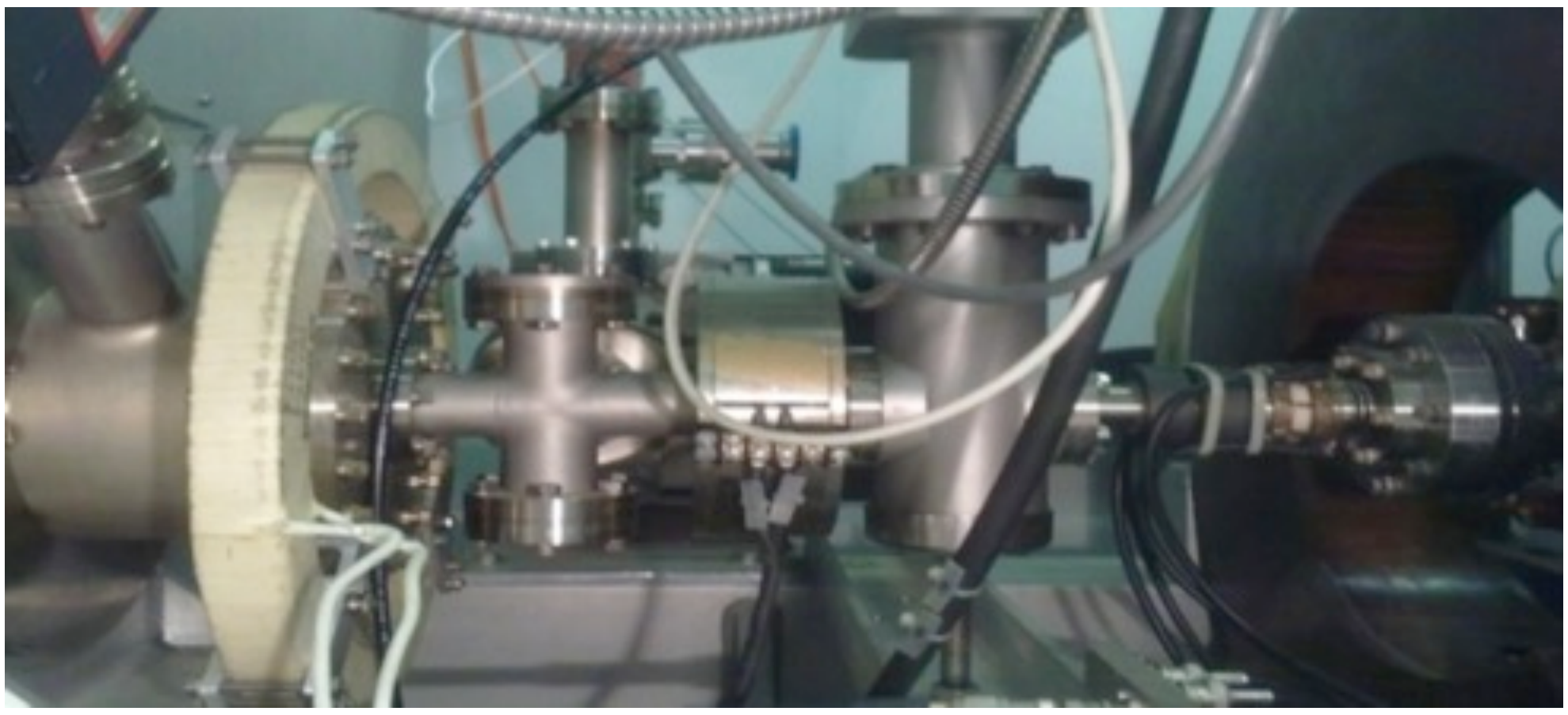

**Figure 1.6:** The effect of first focussing coil (in the middle of the picture) has to be compensated near the cathode by the bucking coil (on the left). On the right, the first of the 14 Helmholtz coils is visible.

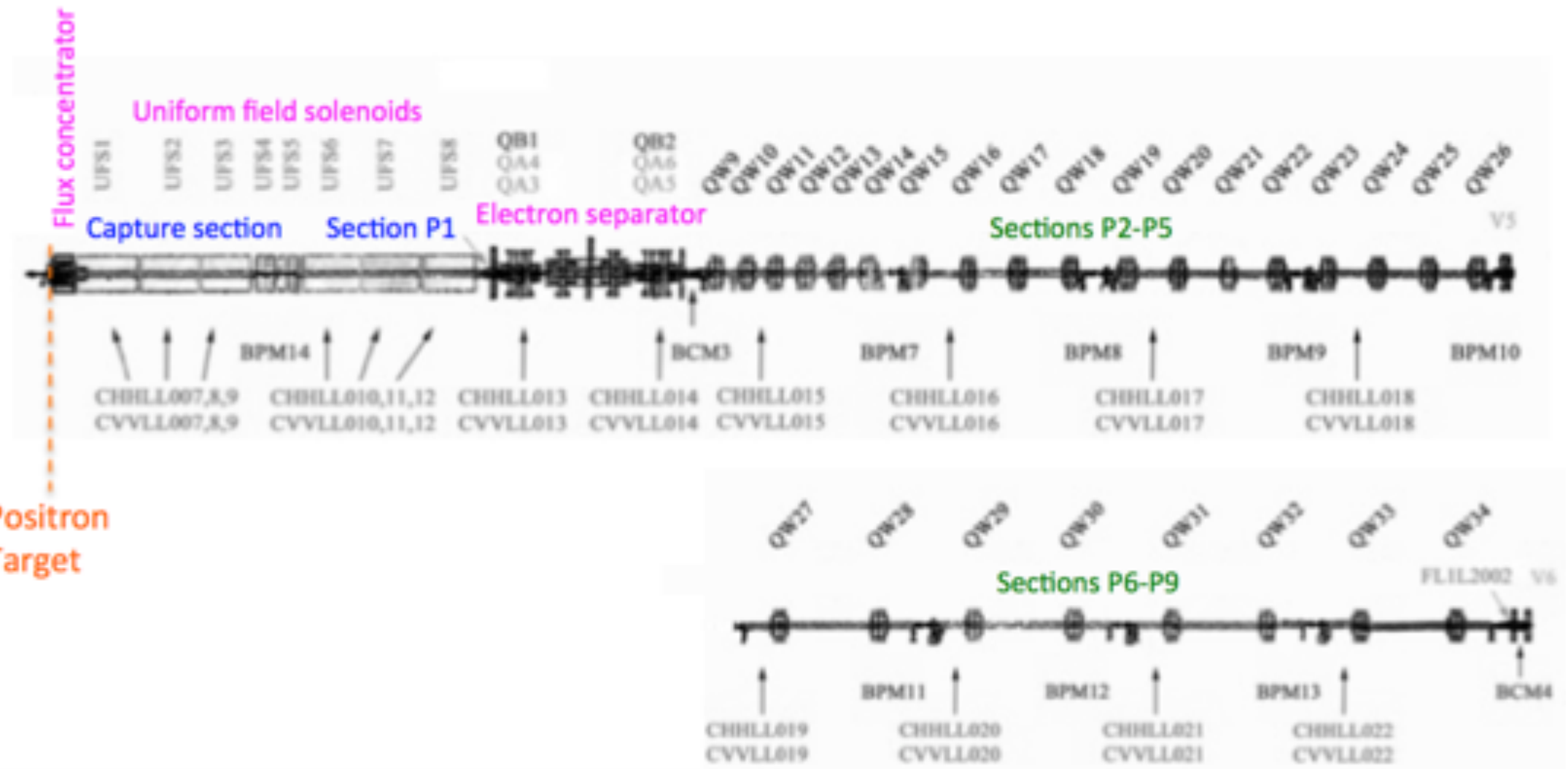

**Figure 1.7:** Scheme of the second part of the LINAC, downstream the positron converter.

Different focussing systems are used in the different portions of the LINAC: a bucking coil, reducing the presence of magnetic field in the gun cathode region, is followed by a solenoid driving the beam to the pre-buncher. The pre-buncher, buncher and the accelerating section E1 are immersed in a solenoidal field produced by 14 Helmholtz coils. A quadrupole doublet between E1 and E2 allows matching with the FODO that transports the beam to the positron converter, composed by two quadrupoles per each of the four accelerating sections (E1-E4). A scheme of the first part of the LINAC focussing is shown in Fig. 1.5, while a picture of the portion close to the gun is shown in Fig. 1.6

Downstream the converter and the first two sections (the capture section and P1), a magnetic chicane, composed by four dipoles is present which, when the target is inserted, separates secondary positrons and electrons in two different paths and drives the electrons onto a beam stopper.

In the remaining part of the LINAC, from section P2 to P9, a FODO, composed by 26 quadrupoles with steps tapered according to the beam energy, completes the focusing scheme, as shown in Fig. 1.7. A network of vertical and horizontal correctors, generally a couple per section, is used for the correction of the beam trajectory.

The beam diagnostics includes 14 beam position monitors, one at the end of each of the accelerating sections, four fluorescent screens (at the end of E5 section, on the converter target, at the separator output and at the LINAC end), and four wall current monitors of the resistive type (at the gun output, at the positron converter, at the separator output and at the LINAC end).

The transport efficiency from the capture section to the end of the LINAC is at the level of 90%, corresponding to a maximum current of 180 mA/85 mA in electron/positron mode, in excess of the design values. The overall momentum spread measured at the spectrometer, is at the design level of about 0.5%/1% for the electron/positron beam. In Tab. 1.1 the nominal and measured main parameters of the LINAC are listed.

| **Parameter** | **Design** | **Operational** |
|---|---|---|
| Maximum $e^-$ energy | 800 MeV | 750 MeV |
| Maximum $e^+$ energy | 550 MeV | 530 MeV |
| RF frequency | 2856 MHz | |
| $e^+$ conversion energy | 250 MeV | 220 MeV |
| Beam pulse length | 10 ns | 1.4 to 40 ns |
| Gun current | 8.0 A | 8.0A |
| RMS energy spread $e^-$ | 0.5 % | 0.56% |
| RMS energy spread $e^+$ | 1.0% | 0.95% |
| $e^-$ current on $e^+$ converter | 5.0 A | 5.2 A |
| Maximum $e^-$ current (10 ns) | >150 mA | 180 mA |
| Maximum $e^+$ current (10 ns) | 36 mA | 85 mA |
| Transport efficiency from capture section to end of LINAC | 90% | 90% |

**Table 1.1**: LINAC parameters summary.

## 1.3 BEAM TEST FACILITY, BTF

### 1.3.1 SINGLE PARTICLE PRODUCTION AND HIGH INTENSITY

The BTF has been designed as a part of the DAΦNE complex [7]: it is composed of a transfer line driven by a dipole magnet allowing the diversion of electrons or positrons, usually injected into to the damping ring, from the high intensity LINAC towards a dedicated experimental hall. The facility can provide runtime tuneable electron and positron beams in a defined range of different parameters, depending on the choice of one of the following two main operation modes:

- "**Single particle**" regime: in this operation mode, a step Copper target, allowing the selection of three different radiation lengths (1.7, 2 or 2.3 $X_0$), is inserted in the initial portion of the BTF line for intercepting the beam. This produces a secondary beam with a continuous full-span energy (from LINAC energy down to few MeV) and intensity, down to a regime in which the particle multiplicity per bunch follows a Poisson distribution.
- **High-intensity** beam extraction: the LINAC beam is directly steered in the BTF hall with a fixed energy (i.e. the final LINAC one) and with a reduced capability in multiplicity selection (typically from $10^{10}$ down to $10^4$ particles/bunch) by means of collimating Tungsten slits.

The dipole magnet steering the beam away from the main transfer line from the LINAC end to the accumulator ring has also the task, in single particle mode, of reselecting secondary particles emerging from the BTF target, with the momentum band defined by a downstream horizontal collimator (made of a pair of Tungsten slits).

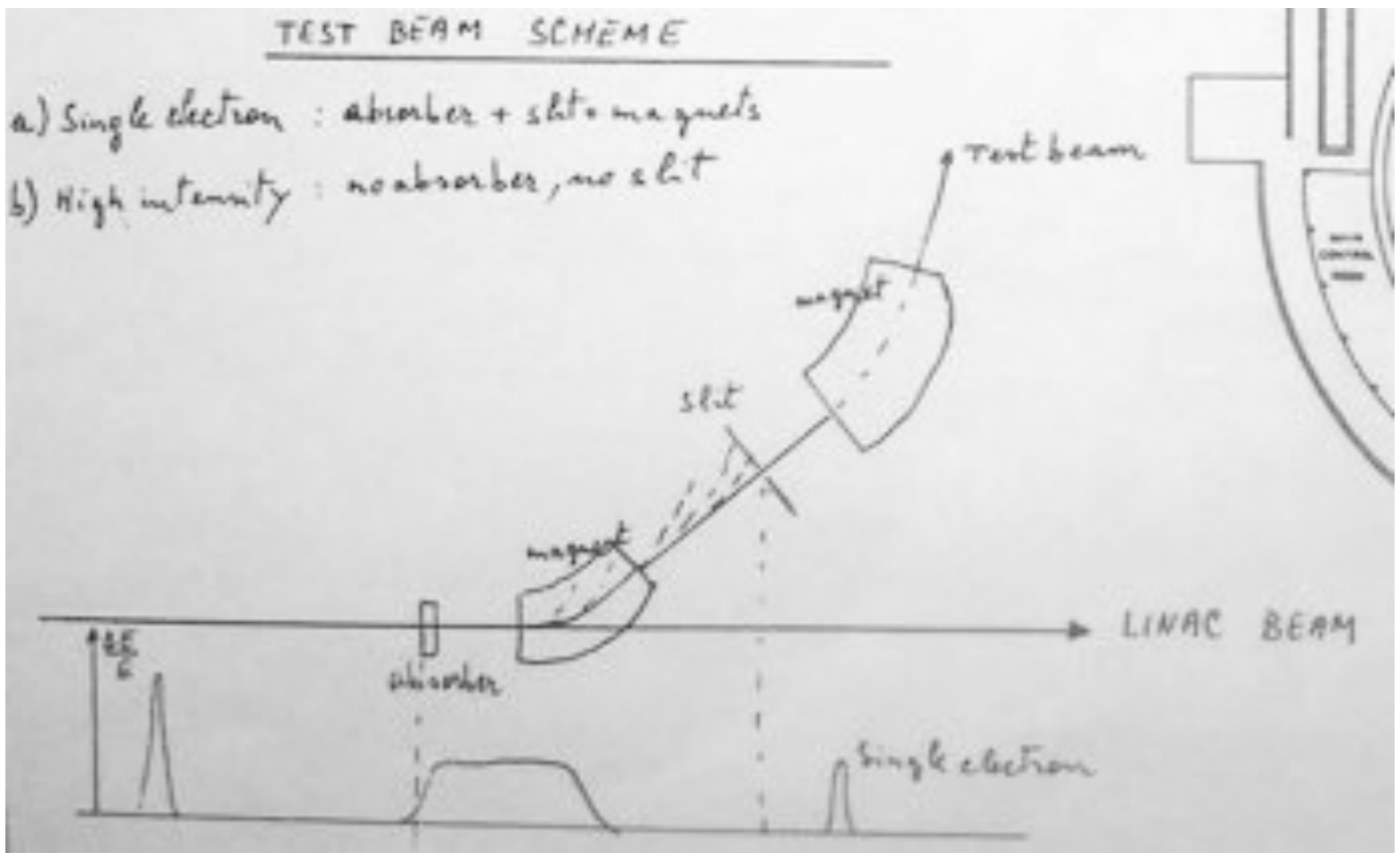

**Figure 1.8:** Original sketch of the beam-test facility line.

In the original design (the very first hand-written sketch shown in Fig. 1.8), this static bending magnet (DHSTB01) was placed on the main transfer line between the end of the LINAC and the accumulator transfer line, with a vacuum chamber with a twofold exit: one, straight, with magnet switched off, for allowing the injection in the damping ring, and one at 45°, driving electrons or positrons of the chosen energy (depending on the current setting) to the BTF line.

The collimator before the beam attenuating target, the target itself and the second collimator, limiting the angular acceptance at the entrance of the bending magnet, were sitting on the same transfer line used for the injections for DAΦNE, so that this configuration had an impact on both the DAΦNE injection efficiency and BTF duty factor, due to at least two important limiting factors:

- The time needed for switching on and off, at the beginning and at the end of each injection, the large, static dipole DHSTB01;
- The time needed for the insertion and removal from the beam line of the attenuating Copper target.

In addition to the not negligible dead-time introduced by those operations at each injection, another very important point was that with this operation sequence **none** among the target, collimators and magnetic configurations was kept from one BTF beam period (of a few minutes) to the following one, thus making the reproducibility very poor.

Fig. 1.9 shows the original layout of the transfer lines, with the two 45° bending magnets: the first, static, towards the BTF, the second, pulsed, towards the damping ring. In Fig. 1.10 a zoom of the same layout shows the main elements of the BTF beam attenuation and driving (collimators, target in addition to the DHSTB01 dipole). Another 45° dipole in the BTF experimental hall allows turning the beam along the main axis of the room. The BTF line is described with some more details in Sec. 1.3.3.

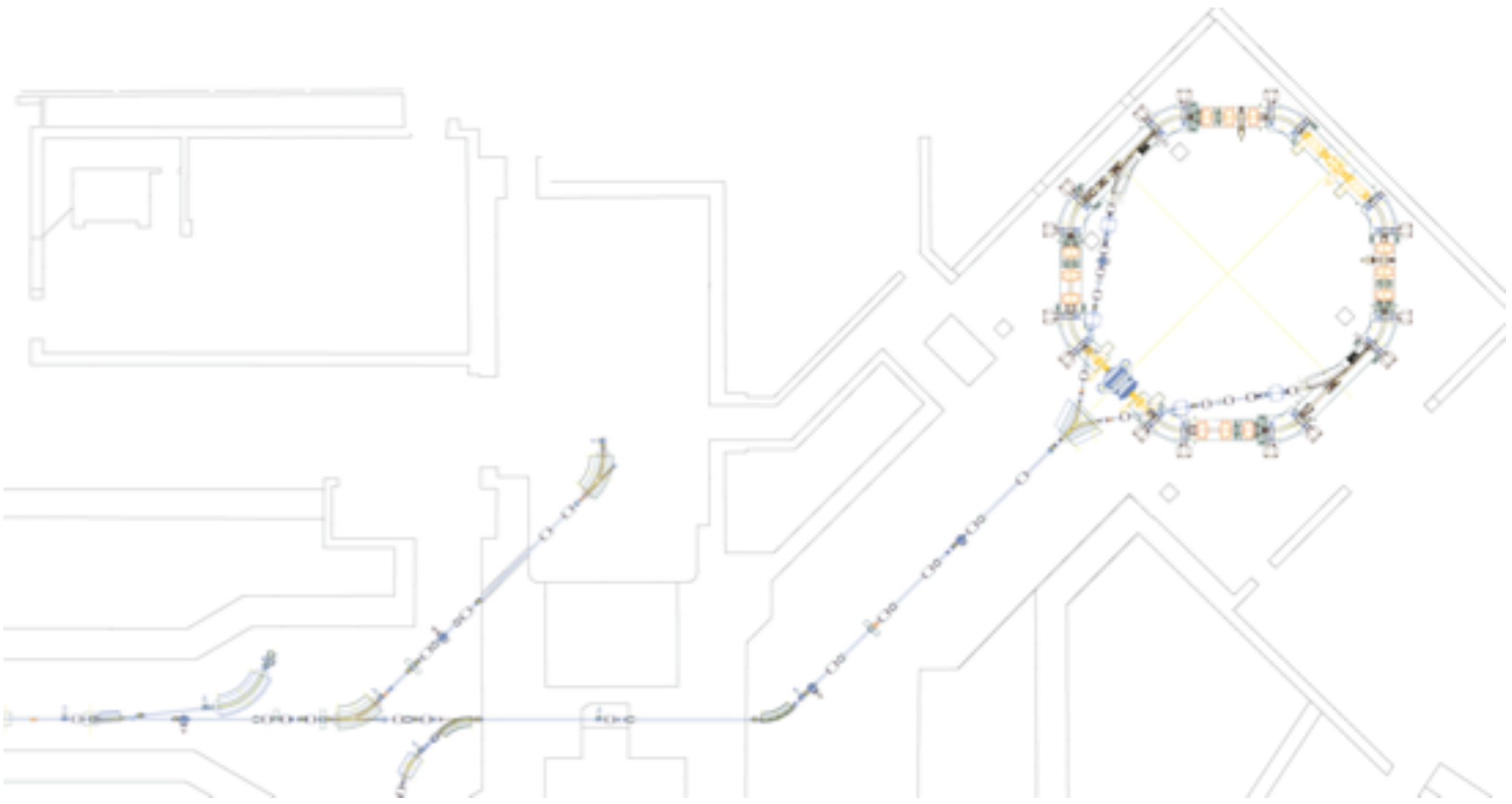

**Figure 1.9:** Layout of the transfer-lines from the end of the LINAC to the damping ring, with the original configuration of the line towards the Beam-Test Facility.

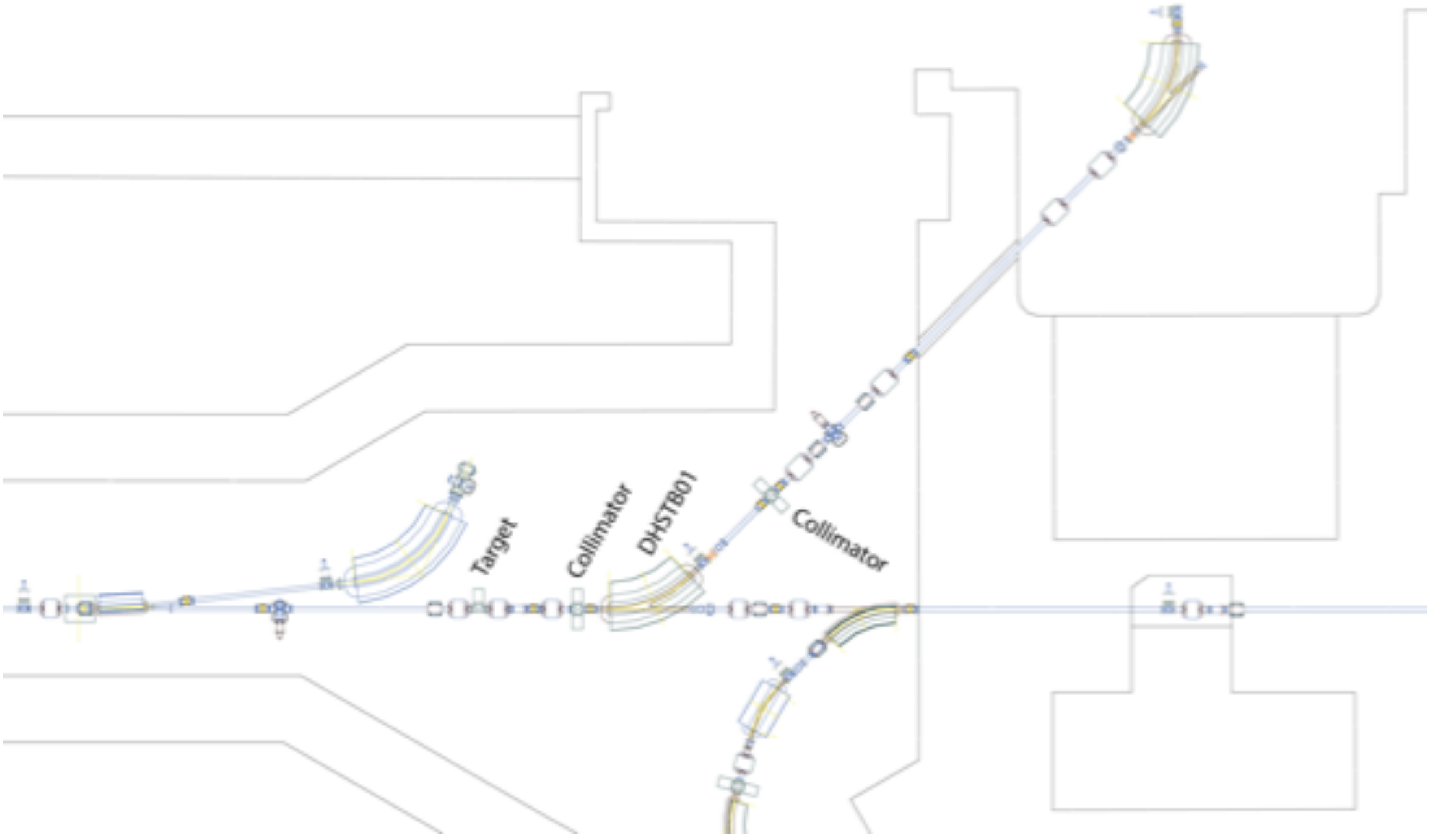

**Figure 1.10:** Layout of the original BTF beam attenuation and extraction system (until 2004) with the main elements (target, dipole, collimator) on the main injection line.

### 1.3.2 BTF OPERATION

In 2004 the BTF, LINAC spectrometer and damping ring transfer lines have been completely redesigned, dismantled and reinstalled in a new configuration, allowing a complete separation among three different lines [8]:

- The straight line, connecting the end of the LINAC with the 45° pulsed dipole towards the damping ring;
- The BTF line, with the beam attenuation and selection system, bent by 3° by means of a small dipole, fed by a DC power supply until 2006 (dubbed DHRTB101 until then), then replaced by a pulsed supply (with the new label DHPTB101);
- The spectrometer at 6°, for the precise measurement of the beam momentum at the end of the LINAC, by means of a 60° static dipole coupled to a metallic strip segmented detector.

The present layout of the transfer lines at the end of the LINAC is shown in Fig. 1.11.

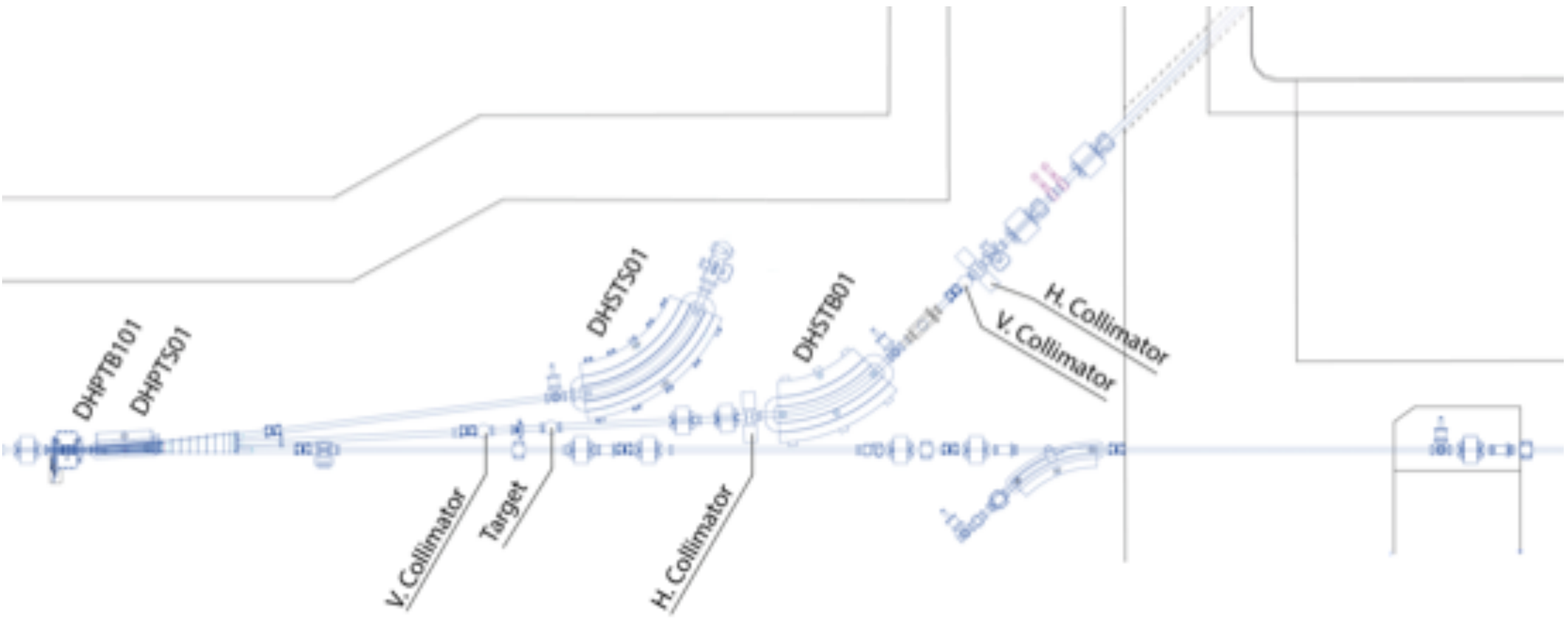

**Figure 1.11:** Present layout of the transfer lines at the end of the LINAC.

During DAΦNE collider operations, one of the 50 bunches accelerated by the LINAC in one second is driven to the 6° spectrometer line (by the DHPTS01 pulsed 45° bending), conventionally the last of a 1 second long sequence.

Out of injection, both in electron and positron mode, all the remaining 49 bunches are available and are driven to the BTF line at 3° by the DHPTB101 dipole.

During injections, after that the required number *n* of electron or positron packets have been driven on the straight line towards the damping ring (both DHPTB101 and DHPTS01 off), and subtracted the two buckets reserved for the energy measurement (since the extraction from the damping ring to the main rings runs at 2 Hz, also the number of pulses to the spectrometer in 1 second is doubled), the remaining 48-2*n* bunches can be driven to the BTF line (DHPTS01 off and DHPTB101 on).

During the switching of the LINAC and of all the transfer lines between electrons and positrons, no beam is available instead (gun and modulators off).

### 1.3.3 BTF LINE AND EXPERIMENTAL HALL

The beam attenuation and energy selection mechanism is exactly the one described in Sec. 1.3.1, with the only difference that the 45° sector magnet DHSTB01 is slightly displaced and rotated in order to compensate the 3° bending introduced by the pulsed magnet, and of course is also operated at a lower current in order to produce (for a given energy setting) only the additional 42° needed to drive the beam in the beam-pipe entering the LINAC tunnel walls at 45°, and exiting inside the BTF experimental hall.

The energy distribution of secondary particles emerging from the target is very wide, and can be estimated using the electromagnetic shower model by Rossi, or with a Monte Carlo simulation [9]. In Fig. 1.12 the spectrum of electrons emerging from the target is shown, as obtained by a GEANT3 simulation: by selecting secondaries in the forward direction (relatively loose cuts of 40 mrad and 10 mrad in the polar angle are shown, since divergent particles can be focussed by the following quadrupoles) the momentum distribution gets practically flat, apart from the very low (corresponding to the critical energy) and high end of the spectrum (see Fig. 1.13).

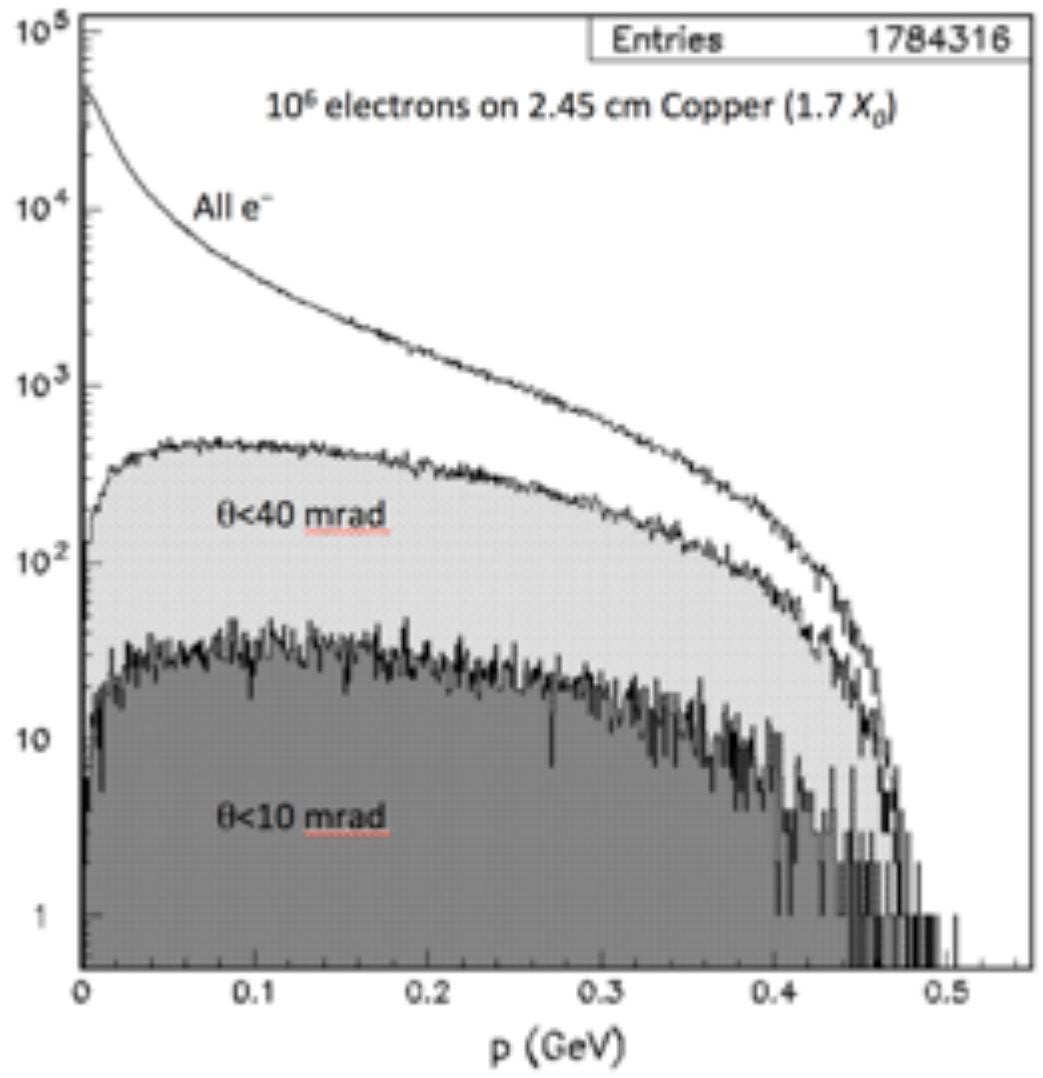

**Figure 1.12:** Momentum distribution of electrons emerging from the BTF Copper target, for a 1.7 $X_0$ depth (2.45 cm), from a GEANT3 simulation.

Indeed the polar angle of secondary particles from the target is correlated with the energy, as shown in Fig. 1.11 (left), but the distribution for forward electrons is much more spread (right).

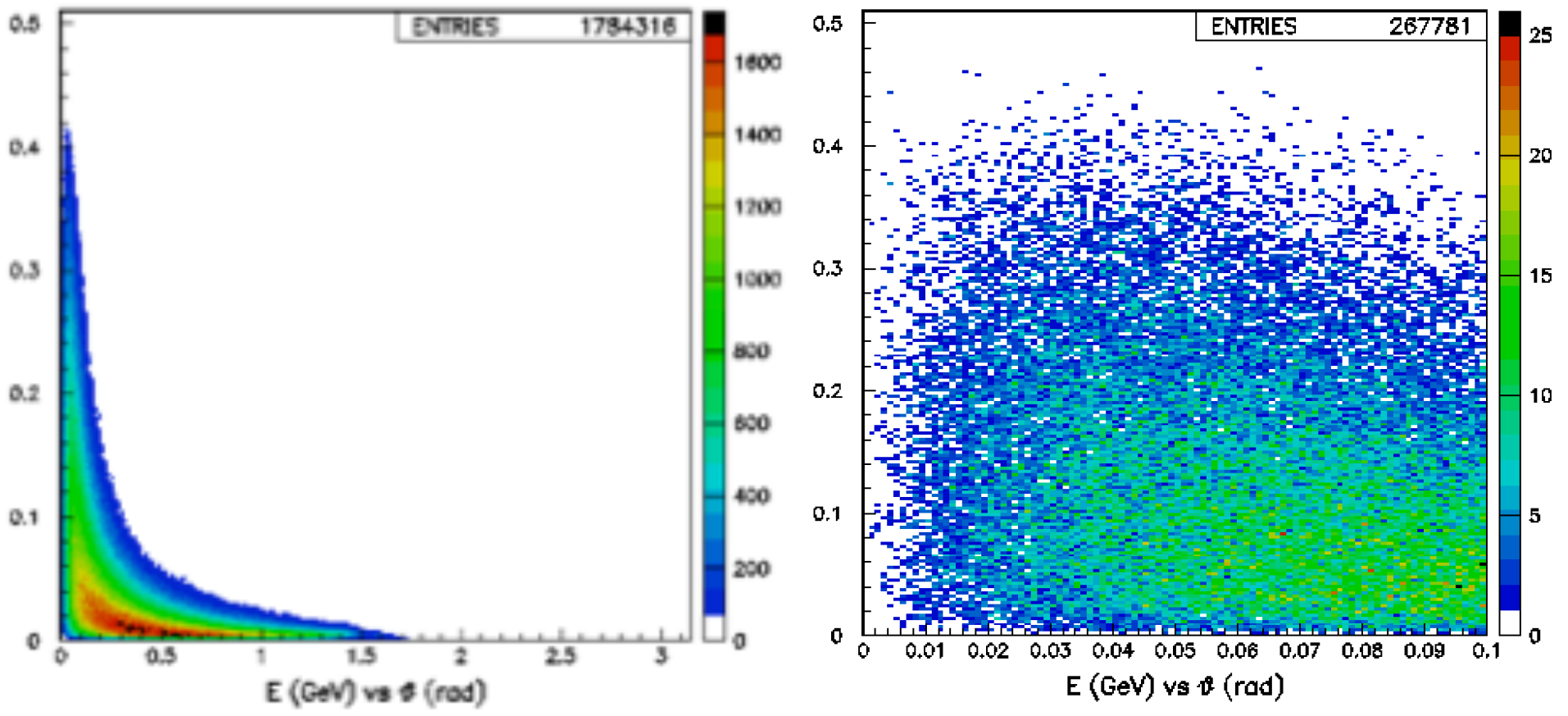

**Figure 1.13:** Energy vs. polar angle distribution of electrons emerging from the BTF Copper target, for a 1.7 $X_0$ depth (2.45 cm), from a GEANT3 simulation.

The yield of secondary electrons (or positrons) can be optimized by the FODO quadrupole doublet at 1 m from the BTF target (QUATB101 and QUATB102, as shown in the general layout in Fig. 1.14). Only electrons (or positrons) with energy around the selected value $E_{sel}$ will be actually transported to the downstream portion of the BTF line, depending on the current set on the 42° dipole DHSTB01.

The energy spread around $E_{sel}$ depends on the opening of the collimators in the horizontal plane, both the one downstream of the selecting dipole (SLTTB04), cleaning particles far from the central trajectory, and the one immediately upstream of the DHSTB01 (SLTTB02), limiting the spread of the entrance angle (or the divergence $x'$) inside the magnetic field. Being $h$ the total opening of the downstream collimator, and $L$ the path length, the energy resolution is given by the expression:

$$\Delta E/E = h/2L + \sqrt{2}|x'|_{max}.$$

Using the typical values <1 mm of collimator opening, a band well below 1% is reached.

The BTF beam optics can be optimized by the elements downstream the selecting dipole DHSTB01: two FODO quadrupole doublets (QUATB01-02 and QUATB03-04, as shown in Fig. 1.14) are placed along the ~10 m long line, one in the LINAC tunnel, and one inside the BTF experimental hall, where the beam pipe enters at 45° with respect to the room walls. In order to drive the beam along the main axis of the room, another 45° static dipole (DHSTB02) is used, identical to the one used for energy selection, after the last quadrupole doublet inside the BTF hall.

This last dipole (DHSTB02) has a two-fold exit, so that the beam can also be available on a straight section of the vacuum pipe (dipole off), essentially for diagnostics purposes, due to the limited space in front of this beam exit. In the years, this line has been used either as photon beam-line, when using the “Tagged photon source” (see Sec. 2.2), or alternatively to drive the high intensity electron beam onto an optimized, shielded Tungsten target for neutron production. In the latter case, since the neutron extraction line is at 90° with respect to the incoming electron beam, the problem due to the very limited space in the forward direction can be overcome.

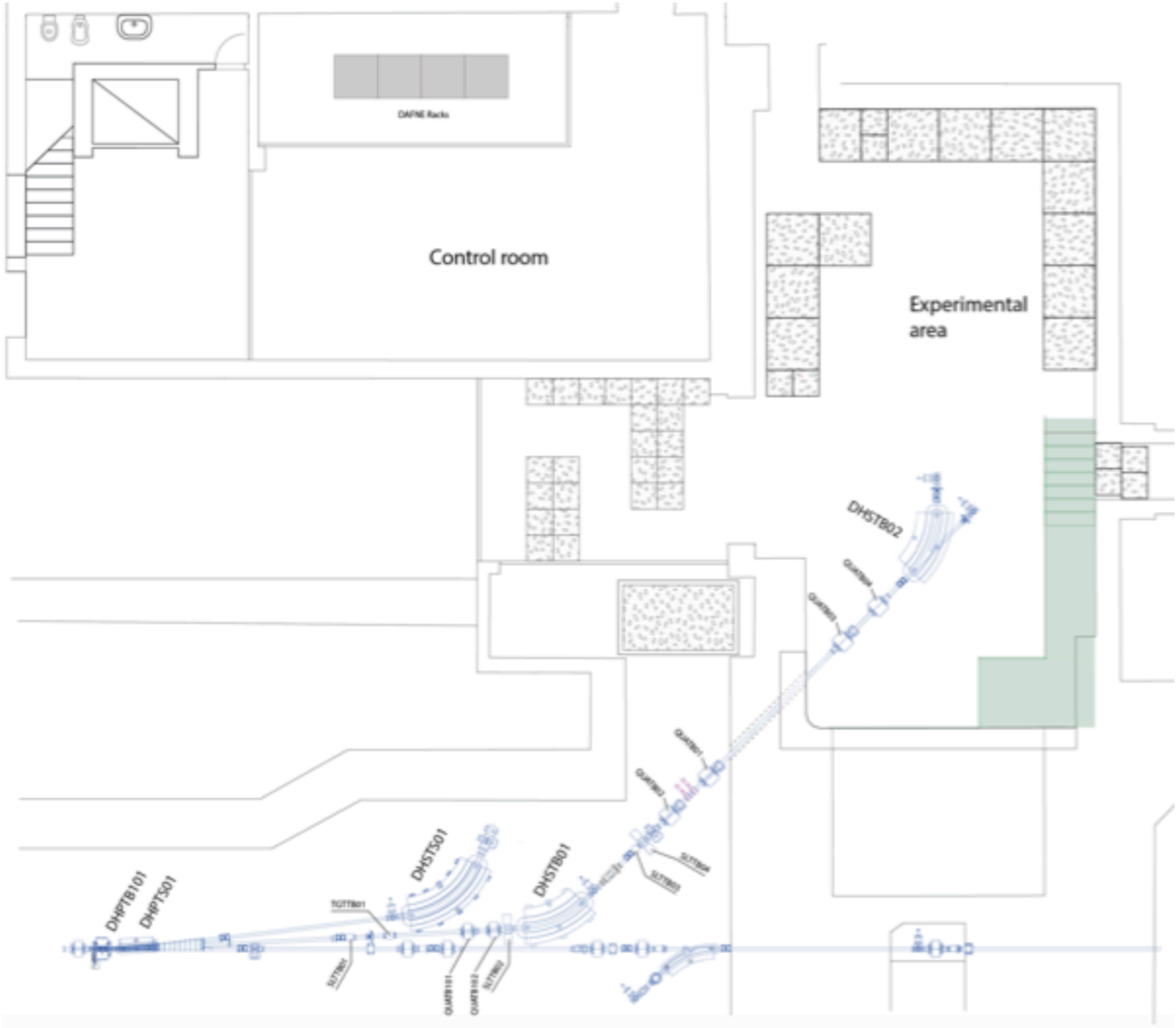

**Figure 1.14:** General layout of the present BTF line.

### 1.3.4 BTF DIAGNOSTICS AND BEAM PARAMETERS

In order to characterize the main beam parameters, a number of diagnostics tools are available, both for the high intensity and the "single particle" mode [10].

When transporting the primary beam from the DAΦNE LINAC without intercepting it with the BTF target, the number of electrons (or positrons, if the LINAC is operated in positron mode) can be modulated both acting on the transport (adjusting the optics of the LINAC itself or of the transport line) and by scrapering the beam with collimators, acting on the horizontal plane (SLTTB02 and SLTTB04) and on the vertical plane (SLTTB01 and SLTTB03).

Considering a typical beam charge of 2 nC/1 nC for the full electron/positron beam, optimized for the injection into the DAΦNE damping ring, driving the full LINAC beam to the BTF line corresponds to $6 \cdot 10^9$ particles/pulse, while keeping all the collimators at the minimum aperture (determined by the mechanical accuracy of the slits and of the stepping motors) of ≈0.2 mm, a few $10^3$ particles/pulse range can be reached.

The sensitivity limit for standard beam current diagnostics, like integrating current toroid (ICT) is of the order of $10^7$ particles range, for 10 ns long pulses. Three dedicated ICT monitors are available and integrated in the BTF diagnostics: WCMTB01 immediately upstream of the BTF target, already on the the 3° BTF line, WCMTB02 close to the main BTF exit (after the 45° bending of the DHSTB02 dipole), integrated in the vacuum segment together with the pre-vacuum port and the flange with the thin Beryllium window (see Fig. 1.15), and WCMTB03 on the straight line, used when DHSTB02 is kept off, with the main purpose of monitoring the electron beam intensity on the neutron production target.

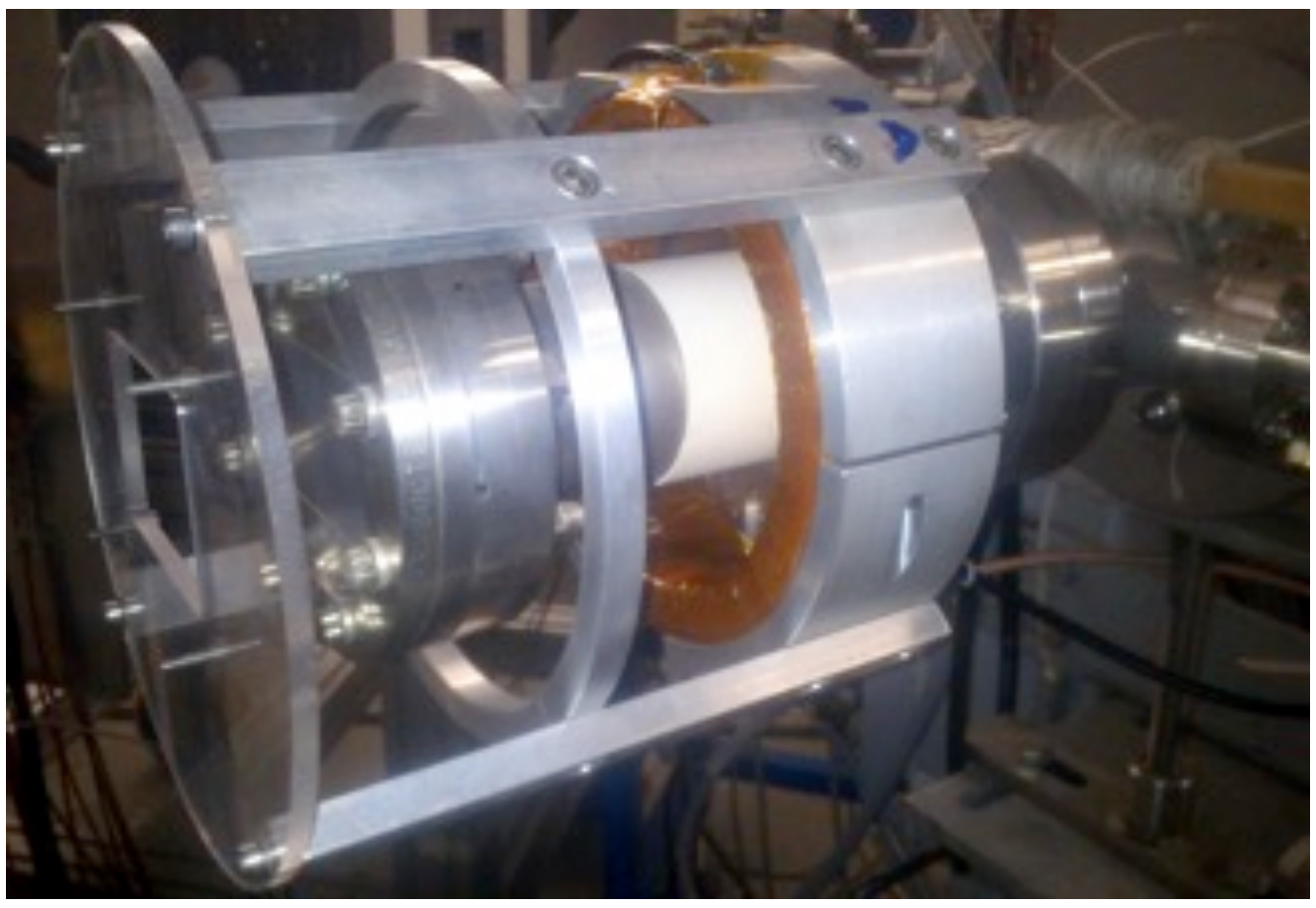

**Figure 1.15:** Main BTF exit (after the 45° bending of the DHSTB02 dipole) showing (from right to left): the pre-vacuum port and valve, the integrating current toroid around the ceramic gap in the beam pipe, the flange with the thin (0.5 mm Beryllium) window and relative Plexiglas protection.

The ICT detectors are readout and acquired in the control system by means of digital oscilloscopes. During operation, the intensity on the BTF target is continuously monitored (as shown in Fig. 1.16) by comparing the intensity detected by the WCMTB01 toroid (red trace) with respect to the one measured by WCMTM01 (yellow trace), positioned at the very end of the LINAC, before the three-way switch-yard and the two pulsed magnets, driving the beam alternatively to the spectrometer or to the BTF line. Very high transport efficiency is generally obtained (green trace in Fig. 1.16) with the standard LINAC setting for injection into DAΦNE.

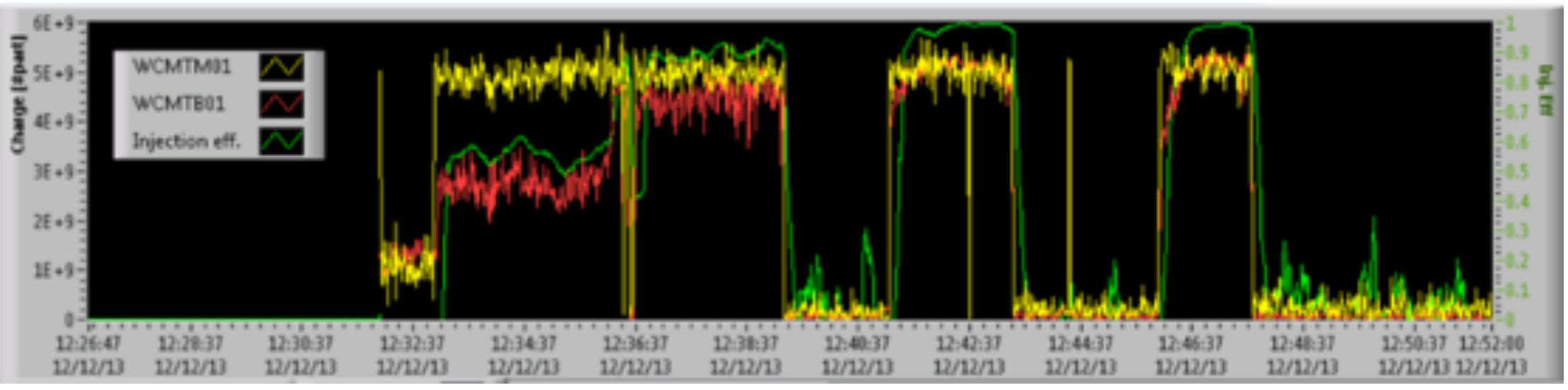

**Figure 1.16:** Beam intensity measured by the WCMTB01 on the BTF line (red trace), compared with the intensity at the end of the LINAC measured by WCMTM01 (yellow trace), and their ratio (green trace), during a typical operation cycle (no beam periods correspond to switches among electron and positron mode of the LINAC).

For intermediate intensities – below $10^7$ particles/bunch, but well above the operation of standard HEP detectors – a variety of detectors have been used, developed by the BTF staff or provided by the users groups during their running: diamond detectors, Cherenkov light detectors, fluorescence chambers, ionization chambers, etc. [11].

For lower intensities, of few hundreds particles and below, standard particle detectors are integrated in the BTF diagnostics and routinely used, in particular Silicon pixel imaging detectors (of the Medipix family) and calorimeters (usually a 11×11 cm$^2$ cross section, 21 $X_0$ deep lead glass block, from the former OPAL experiment electromagnetic calorimeter).

Using the average value of the calorimeter signal for a single electron (in ADC counts), of a given energy $E_{sel}$ fixed by the setting of the DHSTB01 dipole, we are able to count the number of particles by dividing the total deposited by the single electron value: $n$=ADC$_{tot}$/ADC$_1$ on a bunch by bunch basis, as shown in the screen-shot of the BTF online monitoring system in Fig. 1.17.

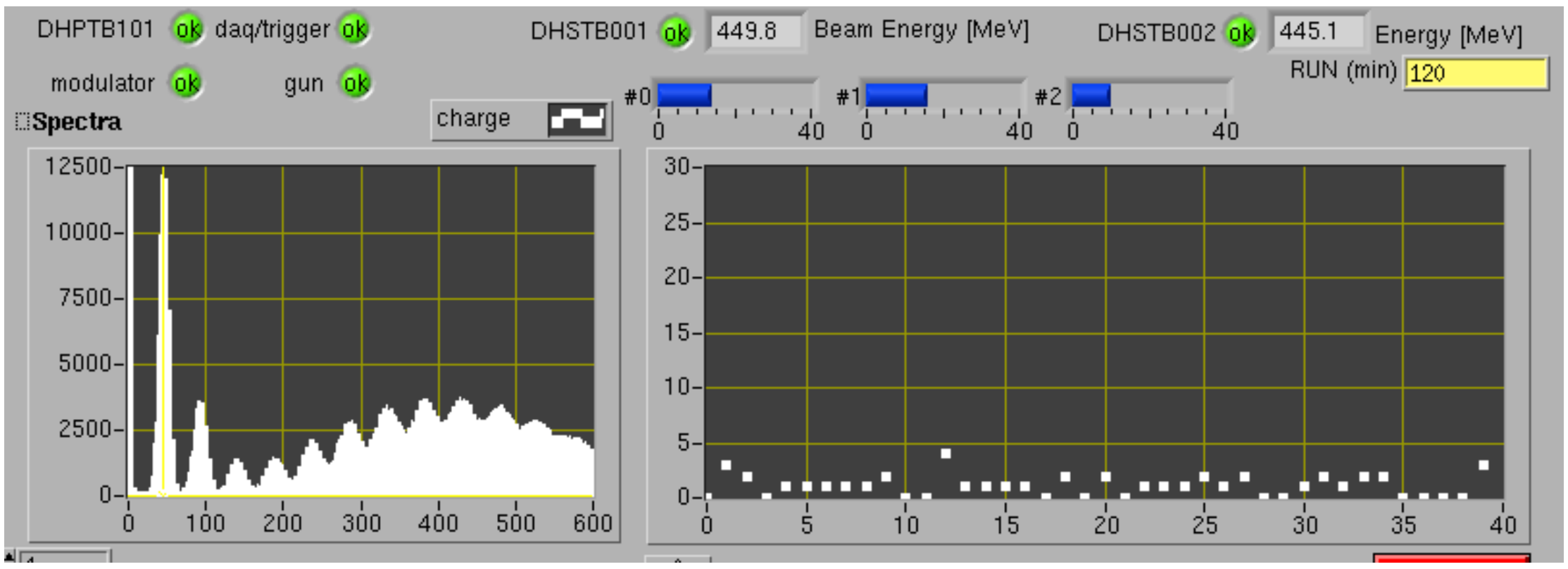

**Figure 1.17:** BTF online monitoring window, showing the calorimeter energy spectrum (left) and the measured number of electrons for each pulse (right), obtained by dividing the total ADC counts by the value of the first electron peak ADC$_1$ (see text). In this case, the selected energy by DHSTB01 was 450 MeV (dipole DHSTB02 can show a slightly different energy value due to the horizontal tilt angle).

A tracking device, a compact time projection chamber realized by modifying a triple GEM detector (or TPG) [12], with pad readout and drift time measurement, has been realized in the framework of the AIDA project (7$^{th}$ European Program) and is also available. Combining the transverse (with respect to the drift field) coordinates information with millimetric precision from the hit pads ($x$-$y$), with the longitudinal coordinate (along the drift) reconstructed from the measured drift time ($z=v_{drift}\times t_{meas}$), with ≈100 μm resolution, a 3D reconstruction of the electron tracks can be obtained, as shown in Fig. 1.18, where the incoming BTF positron beam is split in two population by the effect of multiple channeling in thin straight crystals [13,14].

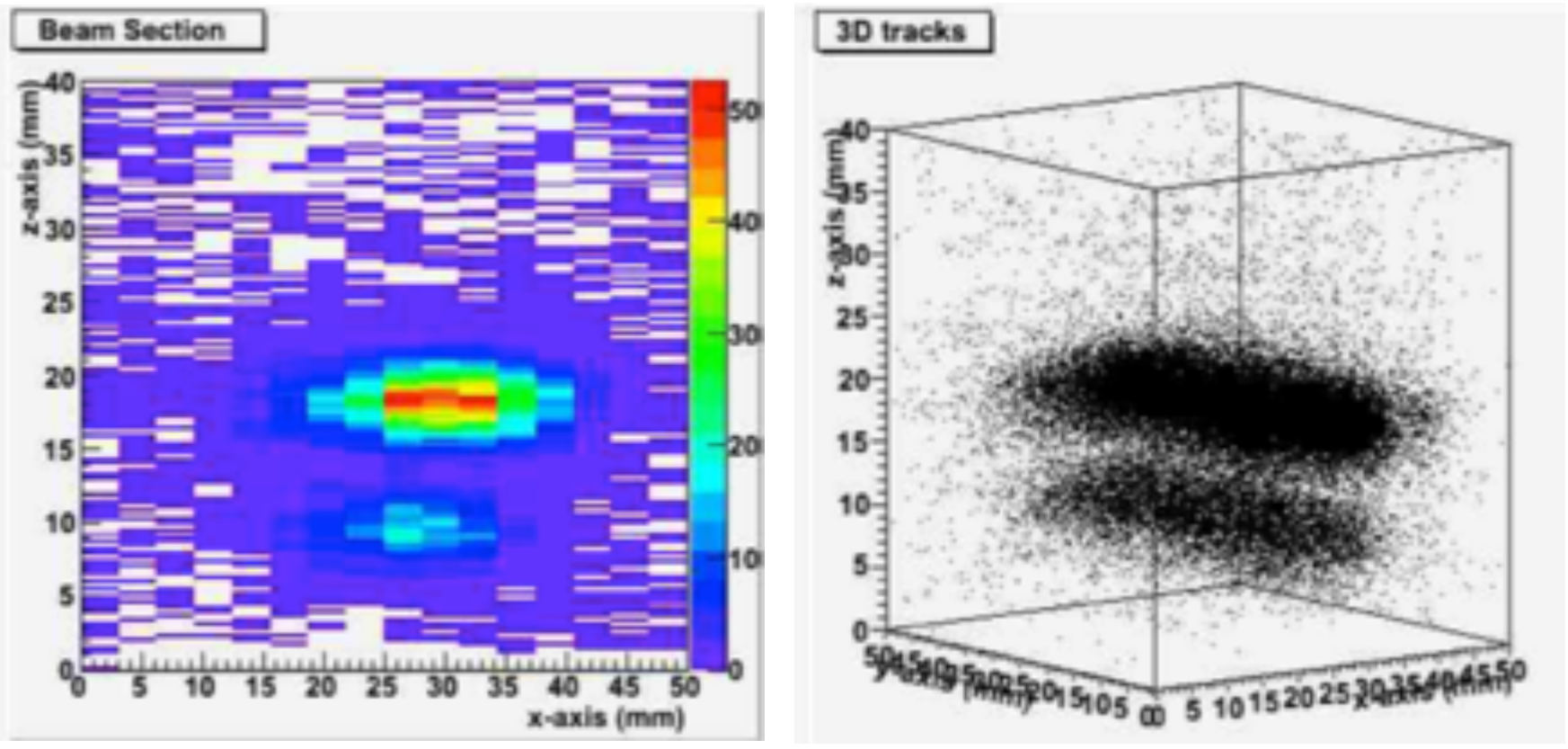

**Figure 1.18:** 477 MeV positrons split in two beam spots by channeling in a five-fold fan of planar crystals ($xz$ and 3D view).

Better resolution energy measurements have been obtained using users devices, in particular BGO or LYSO crystal matrix calorimeters. By properly setting the SLTTB02 and SLTTB04 aperture, <1% contribution from the energy spread of the selecting system can be obtained in the range 100-500 MeV. In Figg. 1.19 and 1.20 examples of energy measurements using a BGO crystal matrix of the BGO-OD experiment are shown [15].

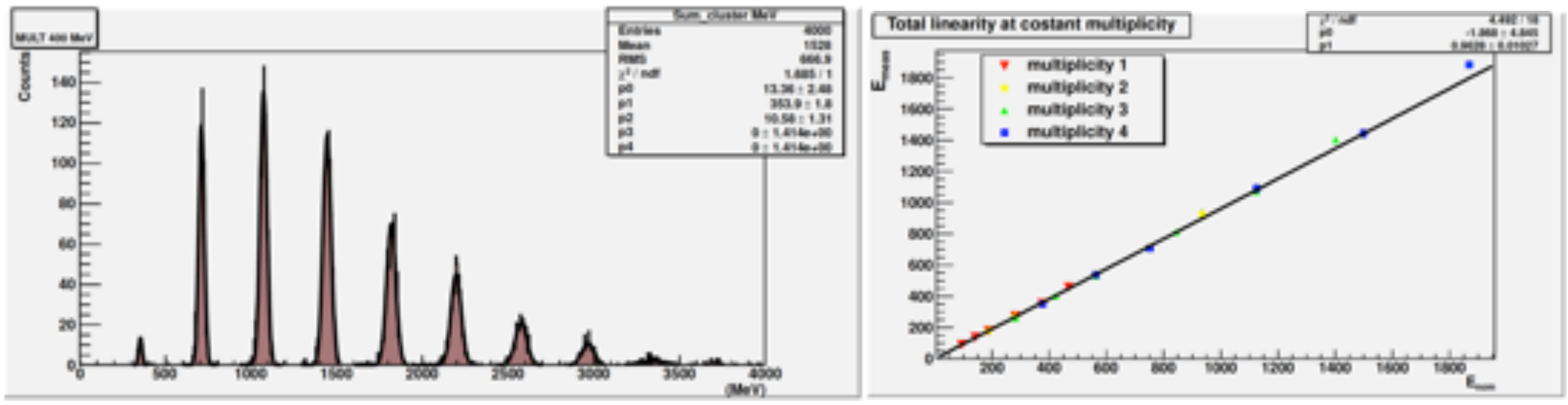

**Figure 1.19:** BGO-OD calorimeter beam-test at BTF. Left: energy spectrum at $E_{sel}$=400 MeV, up to $n$=10 multiplicity; right: linearity of response in the correlation of the measured energy as a function of the nominal total energy $n \times E_{sel}$.

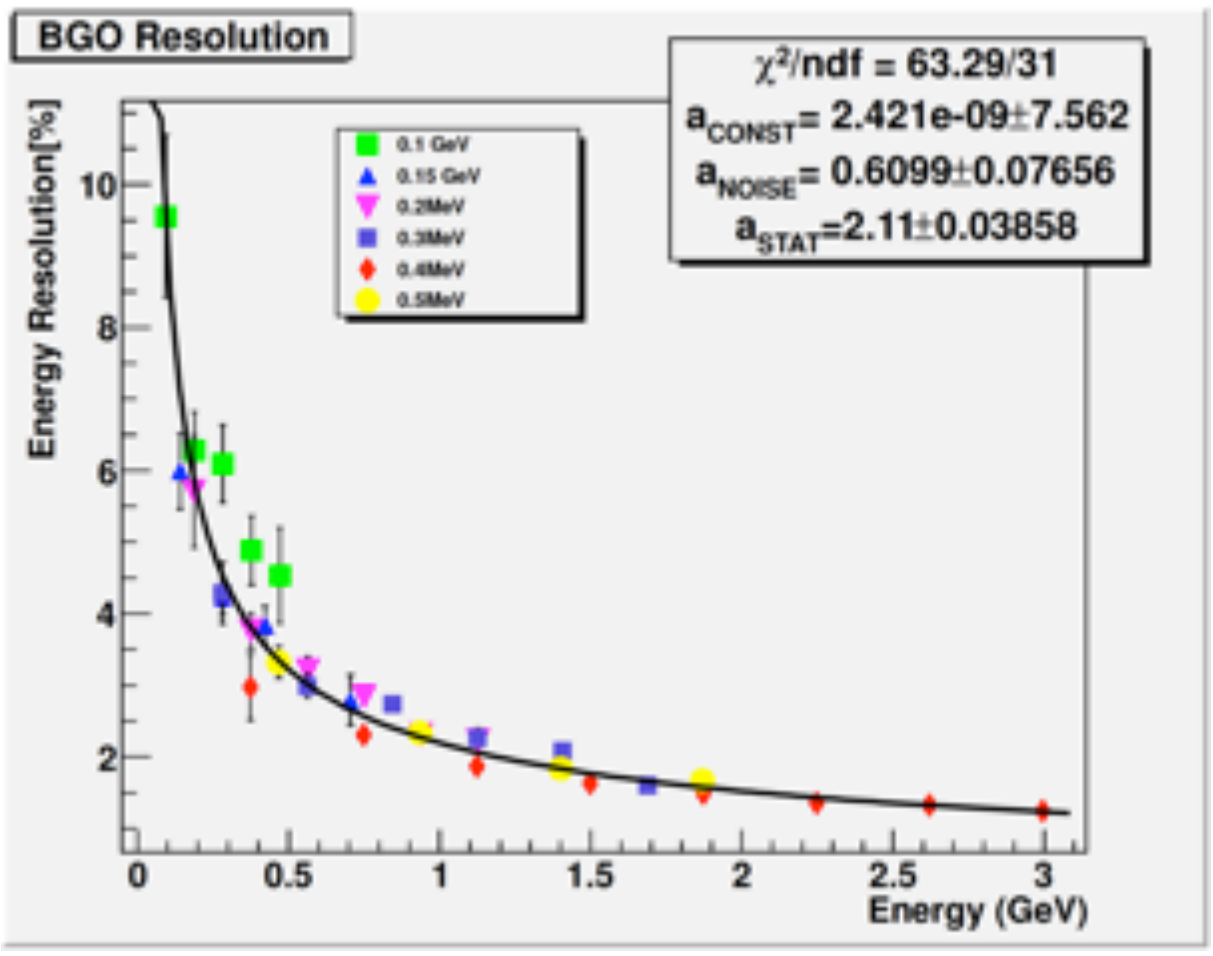

**Figure 1.20:** BGO-OD calorimeter beam-test at BTF. Energy resolution as a function of the nominal total energy $n \times E_{sel}$.

In order to summarize the overall beam parameters obtainable in the BTF line, one has to take into account the two main operation modes of the facility:

- “parasitic” mode, in parallel to the operation of the DAFNE collider, profiting of the LINAC pulses not injected into the damping ring;
- dedicated mode, running the LINAC alone, without injecting for DAFNE.

Not only in the first case the primary beam parameters will be fixed at the collider nominal values (510 MeV energy, pulse duration 10 ns), but also the LINAC operations will be driven by the injection sequence (electron injection, beam off for switching the LINAC and transfer lines, positron injection, switching, and so on), changing the available number of bunches and duty cycle of the BTF.

In the latter case, instead, the LINAC can be set specifically for the optimization of the BTF beam, both with the target (at low intensity) and when driving the primary beam to the BTF line. Of course there are limitations on the range of beam parameters from the LINAC. Concerning, for instance, the primary beam energy, the minimum energy that can be stably obtained using a reduced RF power in the four klystrons, or using only a portion of the accelerating sections, or using some of the four RF stations with a large phase shift, is in the range of 200-250 MeV. Lower energy can be obtained, at the price of a reduced beam intensity and requiring a careful optimization of the optics of the LINAC (not only of the transfer lines); this requires a time-consuming setting of the beam with very reduced diagnostics capabilities, due to the lower beam charge. The maximum energy to which electrons (positrons) can be accelerated, adjusting the power and phases of all the four RF stations has been measured to be 730 MeV (530) MeV, without accurate optimization of all the other LINAC parameters, including the initial focussing, transport, etc.

Another important degree of freedom that is gained running the LINAC without the constraints of the DAΦNE injection is the beam pulse duration. Thanks to the improved LINAC gun pulsing system, the pulse can be reduced down to 1.5 ns (corresponding to a few micro-bunches at 2856 MHz) or increased, with respect to the standard DAΦNE 10 ns value, up to 40 ns.

In Tab. 1.2 a summary of the beam parameters of the BTF is reported: the two main operation modes are further distinguished in two sub-cases, depending on the presence of the attenuating Copper target. Not only with the target inserted in the BTF line the downstream energy selection system (DHSTB01 dipole + collimators) can reselect secondary particles in a wide range, approximately among the Copper critical energy and the incoming primary LINAC energy, but also any of the two polarities can be selected in the secondary particles of the shower.

Of course the particle species (electrons or positrons) is instead fixed by the LINAC mode, depending on the presence of the upstream positron converter and setting of the LINAC optics, when running without the BTF target at higher intensity.

| **Parameter** | **"Parasitic" operation** | | **Dedicated operation** | |
|---|---|---|---|---|
| | With target | Without target | With target | Without target |
| **Particle species** | $e^+$ or $e^-$ Selectable | $e^+$ or $e^-$ Same as DAΦNE | $e^+$ or $e^-$ Selectable | |
| **Energy (MeV)** | 25–500 | 510 | 25–700 ($e^-/e^+$) | 250–730 ($e^-$) 250–530 ($e^+$) |
| **Energy spread** | 1% at 500 MeV | 0.5% | 0.5% | |
| **Rep. rate (Hz)** | Variable between 10 and 49, Depending on DAΦNE status | | 1–49 Selectable | |
| **Pulse duration (ns)** | 10 | | 1.5–40 (In 0.5 ns steps) | |
| **Intensity (particles/bunch)** | $1–10^5$ Depending on energy | $10^3–1.5\ 10^{10}$ | $1–10^5$ Depending on energy | $10^3–3\ 10^{10}$ |
| **Max. average flux** | $3.125\ 10^{10}$ particles/s | | | |
| **Spot size (mm)** | $\sigma(y)\times\sigma(x)$=0.5–25×0.5–55 | | | |
| **Divergence (mrad)** | 1.5-2 | | | |

**Table 1.2**: BTF beam parameters summary.

The last two rows of Tab. 1.2 refer to the parameters of the BTF beam spot size and divergence. Those parameters are measured routinely and with excellent resolution during the "single particle" operation, i.e. running with the BTF target, using imaging or even tracking particle detectors (Silicon pixel, gas detectors, scintillating fiber hodoscope, etc.). High intensity imaging of the beam is generally performed using fluorescence metallic plates with video-camera readout (flags). A user group has also performed a measurement of the emittance of the full electron primary beam [16].

The transverse shape and size of the beam depend on the BTF line optics, including the FODO quadrupoles, the two 45° bendings, and is of course strongly influenced by the setting of the four collimators (the two vertical, SLTTB01-03, and the two horizontal ones, SLTTB02-04). For instance, when running with the BTF target, if one keeps the SLTTB02 slits open, accepting particles entering the DHSTB01 magnet with a large angular spread (in the horizontal, bending plane), will result in a large momentum dispersion; this will enlarge the horizontal beam size due to the effect of the last (DHSTB02) dipole.

For a given energy and beam intensity, fixed by the DHSTB01 and DHSTB02 dipole settings and collimators configuration, the vertical and horizontal dimensions of the beam are of course controlled by the two FODO doublets. A number of datasets are available to the users with different focus positions.

In Fig. 1.21 an optimized beam spot in single particle mode at 450 MeV energy, imaged by a Medipix detector (55×55 μm$^2$ Silicon pixels), is shown with SLTTB02 slits kept with a small (0.5 mm) or large (2 mm) aperture. The effect of the dispersion due to the DHSTB02 last bending is apparent. The shape of the optimized beam (when removing the cited dispersion effect) in the transverse plane is fairly Gaussian, with standard deviation ≈ 0.5 mm in both coordinates; since the Medipix imaging detector is placed in air, close to the beam exit window, the dominant contribution to the beam size and divergence is the Coulomb multiple scattering on the 0.5 mm Beryllium window.

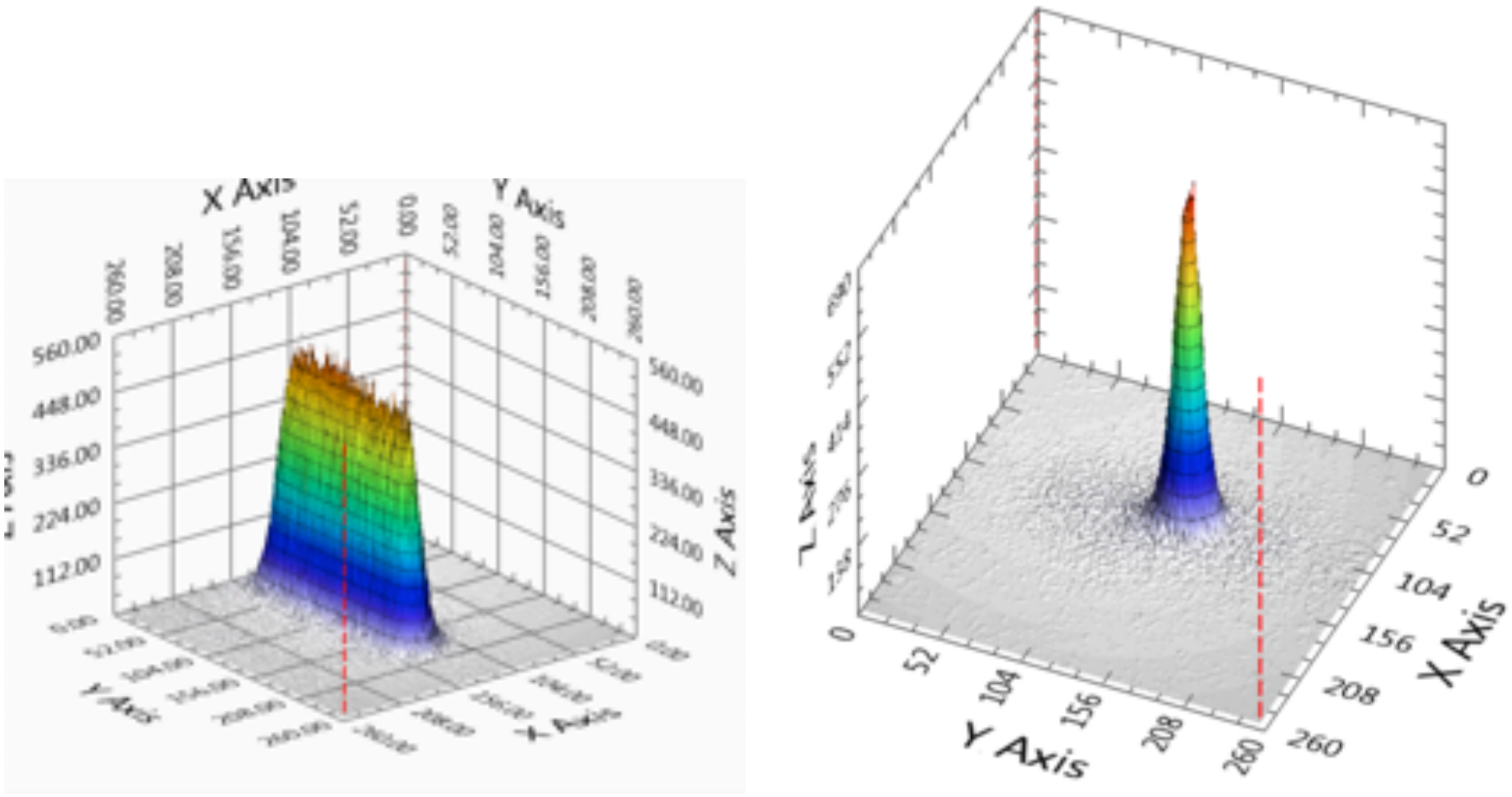

**Figure 1.21:** 450 MeV electrons imaged by Medipix Silicon pixel detector with SLTTB02 slits open at 2 mm (left) and 0.5 mm (right). The optimized beam on the right pane has a fairly Gaussian shape with $\sigma_x \times \sigma_y$=0.5×0.5 mm$^2$.

Dealing with electrons, a halo or penumbra, due to the edges of the collimators, is also present. Due to the straight shape of our collimators, this can give a rectangular shape penumbra, as shown in Fig. 1.22, in a high statistics image of the optimized beam.

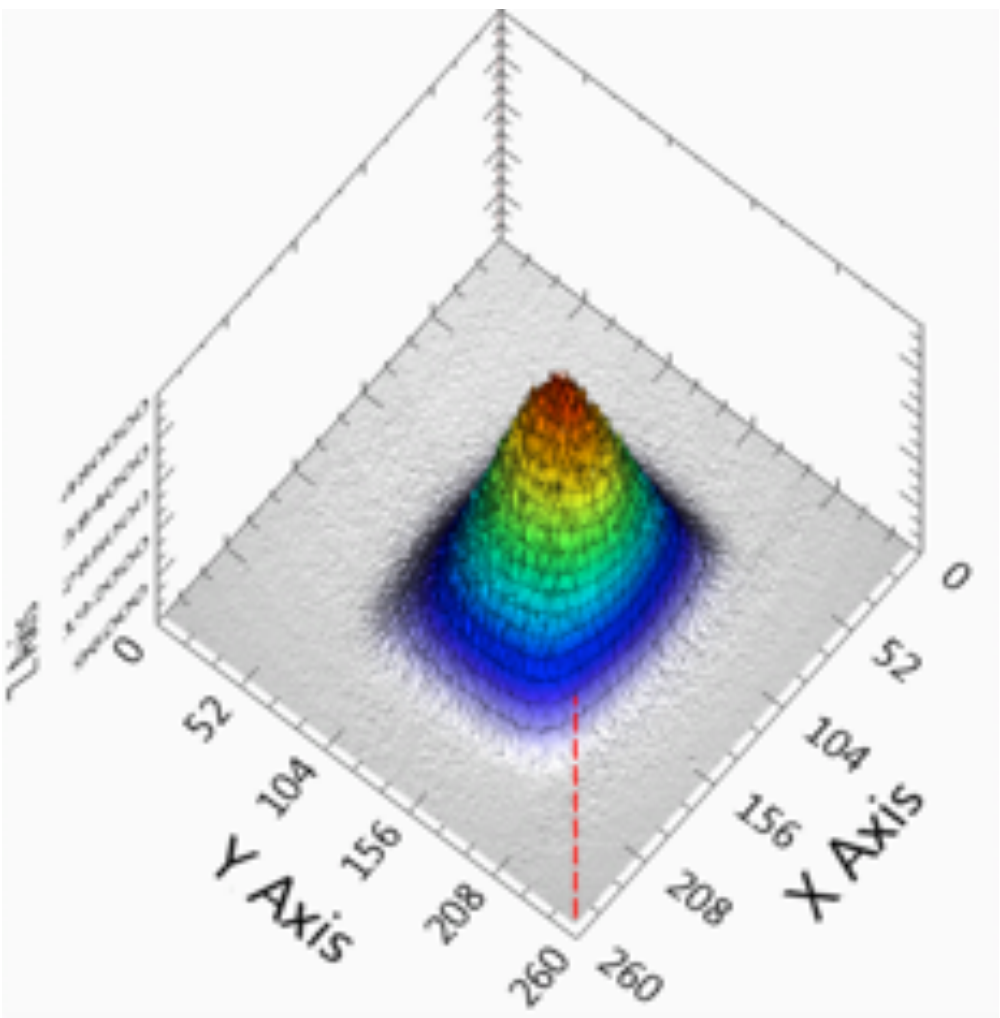

**Figure 1.22:** 450 MeV electrons imaged by Medipix Silicon pixel detector: with an high statistics run a rectangular shape halo due to the collimators is visible.

In order to further reduce the beam dimensions one possibility is to run the user setup in vacuum, thus avoiding the crossing of the 0.5 mm Beryllium window. A possible alternative is to use more transparent exit windows, like Kapton; in this case the full LINAC vacuum ($10^{-9}$ bar) cannot be sustained (also for safety reasons), and secondary (poorer) vacuum should be created in the BTF vacuum pipe last section, downstream the valve close to the entrance of the BTF experimental hall. A ionic pump is installed and connected to the BTF beam-pipe final section for this purpose.

A simple and quick solution for improving the beam size (typically in only one coordinate) is to place an additional (static) high-$Z$ collimator in air, downstream the exit window. However, in this case the detector will see a large fraction of the halo and the high photon flux due to the edges of the collimator. An example of the resulting beam with this kind of setup, realized with 70 mm thick Tungsten blocks placed with a ≈ 200 μm gap, is shown in Fig. 1.23, yielding a beam size of the 330 μm standard deviation (with a "standard" 450 MeV configuration), and clearly showing the (low statistics but significant) background from the collimator edges.

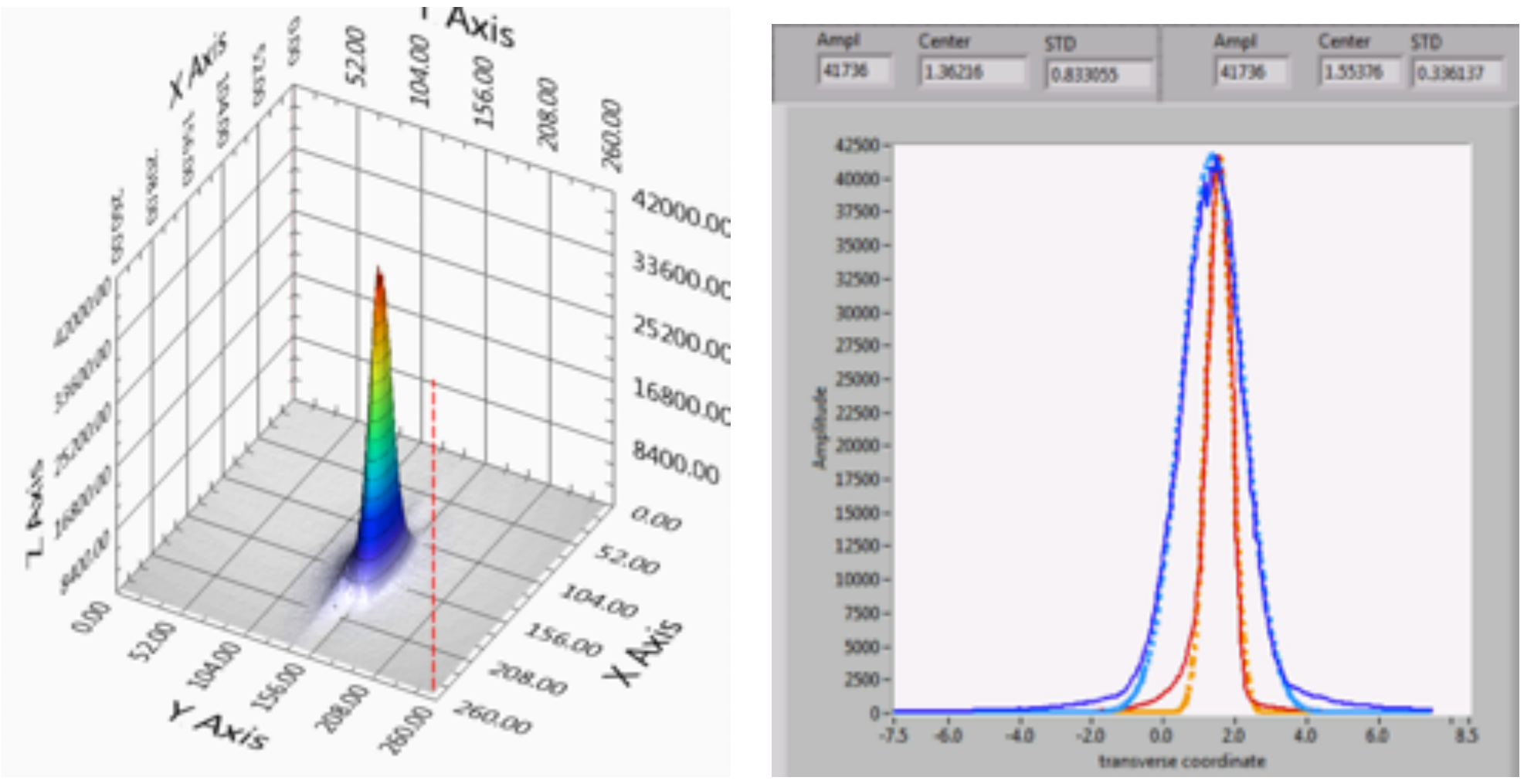

**Figure 1.23:** 450 MeV electrons imaged by Medipix Silicon pixel detector: an additional Tungsten collimator is placed downstream of the exit window, reducing the beam size in the horizontal coordinate (red histogram in right plot) to $\sigma_x$=340 μm, leaving the vertical one (blue histogram in right plot) unchanged.

A dependence of the beam spot from the selected energy is expected, as soon as the beam is observed in air, downstream of the Beryllium window. In Fig. 1.24 an example for a beam energy $E_{sel}$=80 MeV of the $x$ and $y$ projections of the spot measured by the Medipix detector is shown, together with the corresponding Gaussian fits, yielding for the standard deviation 2.7 and 2.4 mm respectively, well consistent with the expectation from the multiple scattering scaling with respect to the 450 MeV values.

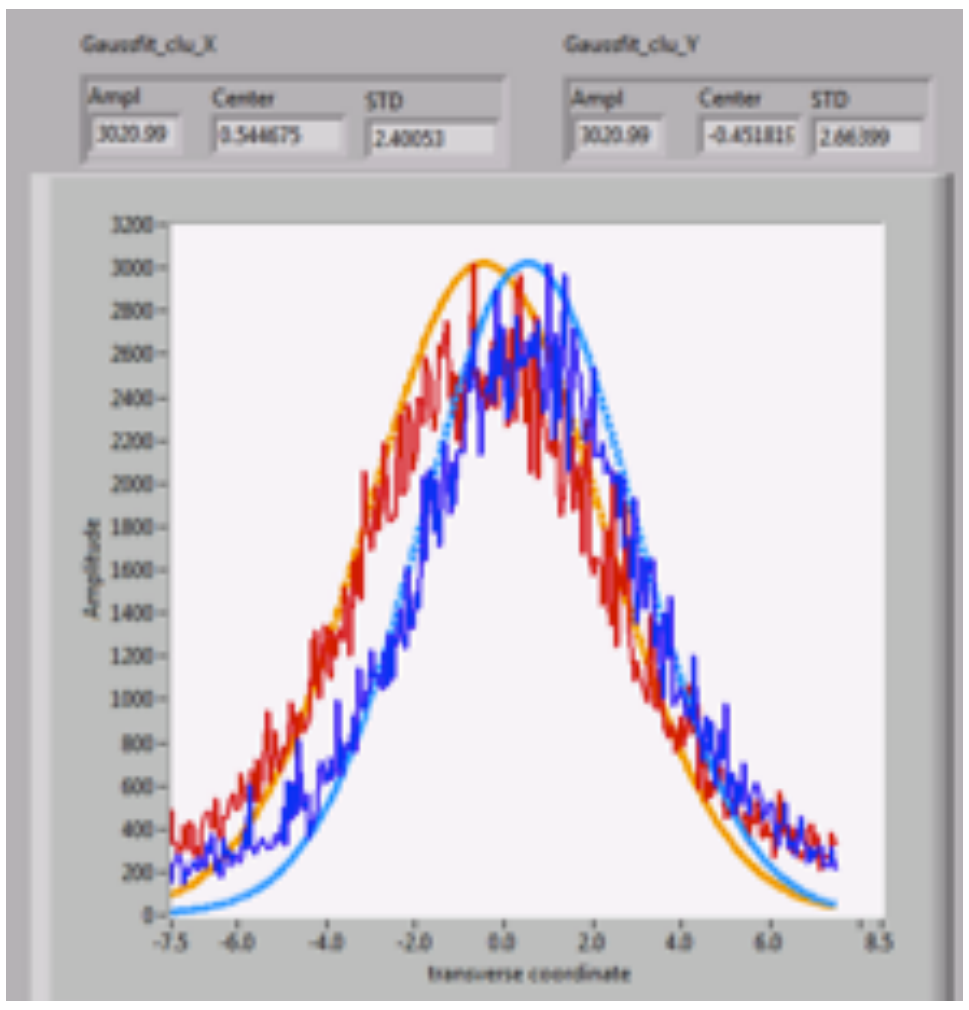

**Figure 1.24:** 80 MeV electrons imaged by Medipix Silicon pixel detector: the $x$ (red histogram) and $y$ (blue histogram) projections of the beam spot are shown together with their Gaussian fits.

An estimate of the beam divergence can be obtained by repeating the beam spot size measurement as a function of the distance of the Medipix detector, or by using more than one detector in order to the determine the track angle. Depending on the distance between the detectors, the contribution of the multiple scattering of the interposed air, as well as of the upstream detector on the downstream one, have to be taken into account.

The divergence is for any configuration influenced by the setting of the collimators: typical values range from a few mrad down to 1.5(2) mrad, in the vertical (horizontal) coordinate (as shown in Fig. 1.25).

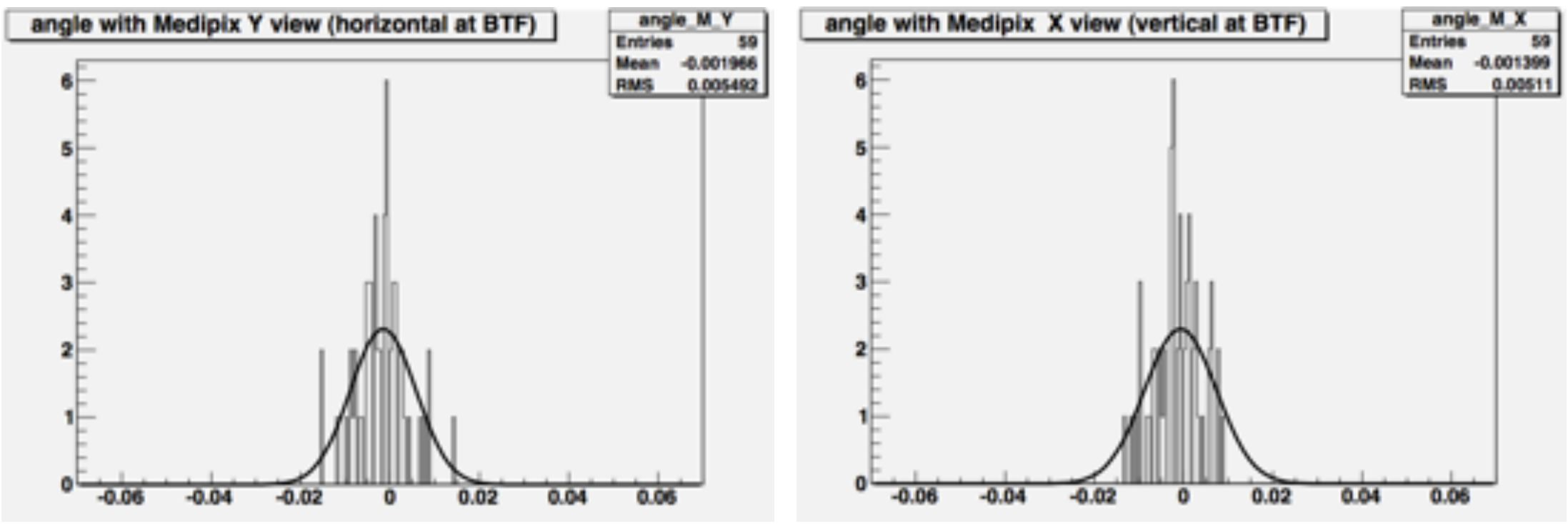

**Figure 1.25:** Angular divergence of 450 MeV electrons reconstructed by a couple of Medipix Silicon pixel detectors in the horizontal (left) and vertical (right) coordinate. In this plots the usual convention $x$ for horizontal, $y$ for vertical was exchanged.

### 1.3.5 BTF USERS

The commissioning of the Beam-Test line [9] has been carried on in 2002, and in 2003 a first running with experimental groups was already possible, although with the target and selecting dipole DHSTB01 still on the main injection line (as discussed in the previous sections) and thus with a heavily reduced duty-cycle.

After the separation of the BTF transfer line in 2004, the facility has been providing access continuously to a large number of experimental teams, with an average number of beam-days close to 200/year. The full statistics for the last five years of BTF operation of the total number of beam-days and working hours (both scheduled and realized) is reported in Fig. 1.26.

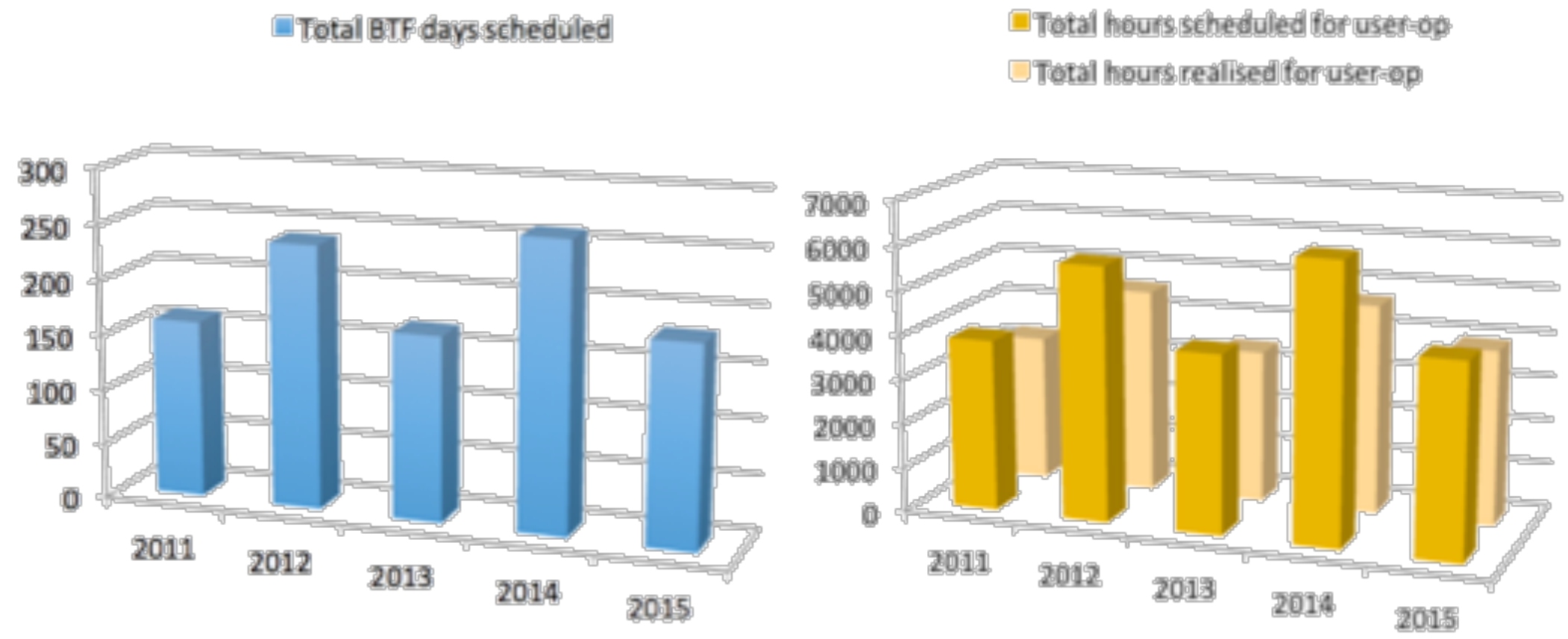

**Figure 1.26:** Total number of beam-days assigned (left) and beam-hours, scheduled and realized (right), for the last five years of BTF operation.

The facility usually allocates slots of 1 week, Monday to Monday, and operates 24/7. More complex experimental setups of course required much longer beam-periods, up to the approximately five months of total allocation for the AGILE satellite pay-load calibration in 2005 (beam was not delivered 100% of this period, however).

The total number of groups for each of the last five years is shown in Fig. 1.27. Approximately 1/3 of the 150 average users/year comes from a not Italian institution.

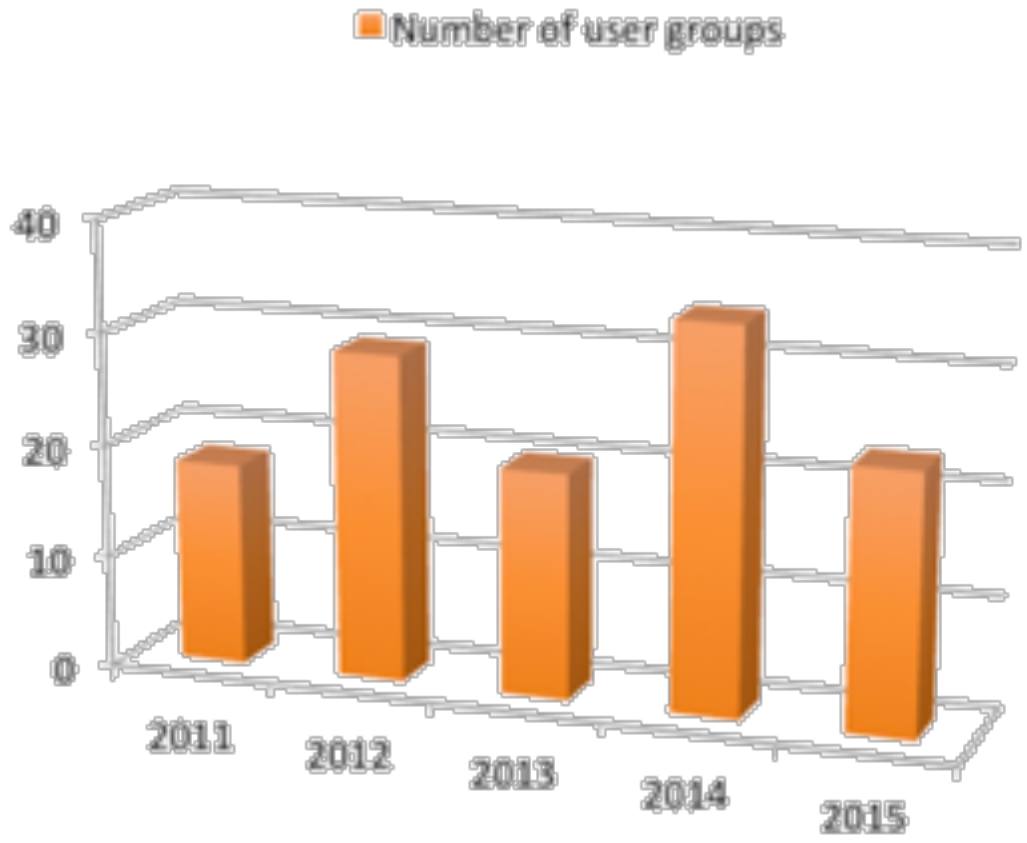

**Figure 1.25:** Total number of user groups accessing the BTF in the last five years of operation.

The overbooking factor, i.e. the ratio between the required and assigned beam time, has ranged among about 120% up to 150%, depending on the year.

The large majority of runs performed at the BTF were related to the high-energy physics and astro-particles communities, mainly with the purpose of testing, characterizing and calibrating particle detectors. As a consequence, most of the runs are performed at low intensity (few particles regime), exploiting the full range of beam energy and spot parameters listed in Tab. 1.2.

Practically all kind of detectors presently used in the HEP community and all major collaborations in this field have profited at some time of the BTF beam: calorimeters (tile, “spaghetti”, crystal, digital and double readout, etc.), scintillating detectors (segmented, timing detectors, fiber trackers), drift chambers, RPC, micro-pattern gas detectors (GEM, planar, mini-TPC and cylindrical, and MSGC), hybrid gas/Silicon detectors, diamond detectors, silicon pixels (also monolithic), silicon micro-strips, fluorescence detectors, nuclear emulsions, Cherenkov threshold and RICH, etc.

A smaller fraction of the beam shifts have been devoted to high intensity measurements or for small, dedicated experiments looking at specific electromagnetic processes, ranging from the measurement of the absolute fluorescence yield of air and other gases, to the detection of microwave emission in air showers, from the thermo-acoustic expansion of a super-conducting resonant antenna (see Fig. 1.28), to the characterization of innovative beam diagnostics (both position sensitive, like segmented diamonds, multi-channel plates, and proportional counters).

Another small, but significant fraction of the beam time has been devoted to the production of tagged photons, especially for the calibration of astro-particles detectors, with the very important example of the pre-launch calibration with photons of the payload of the AGILE gamma ray astronomy satellite (Fig. 1.29).

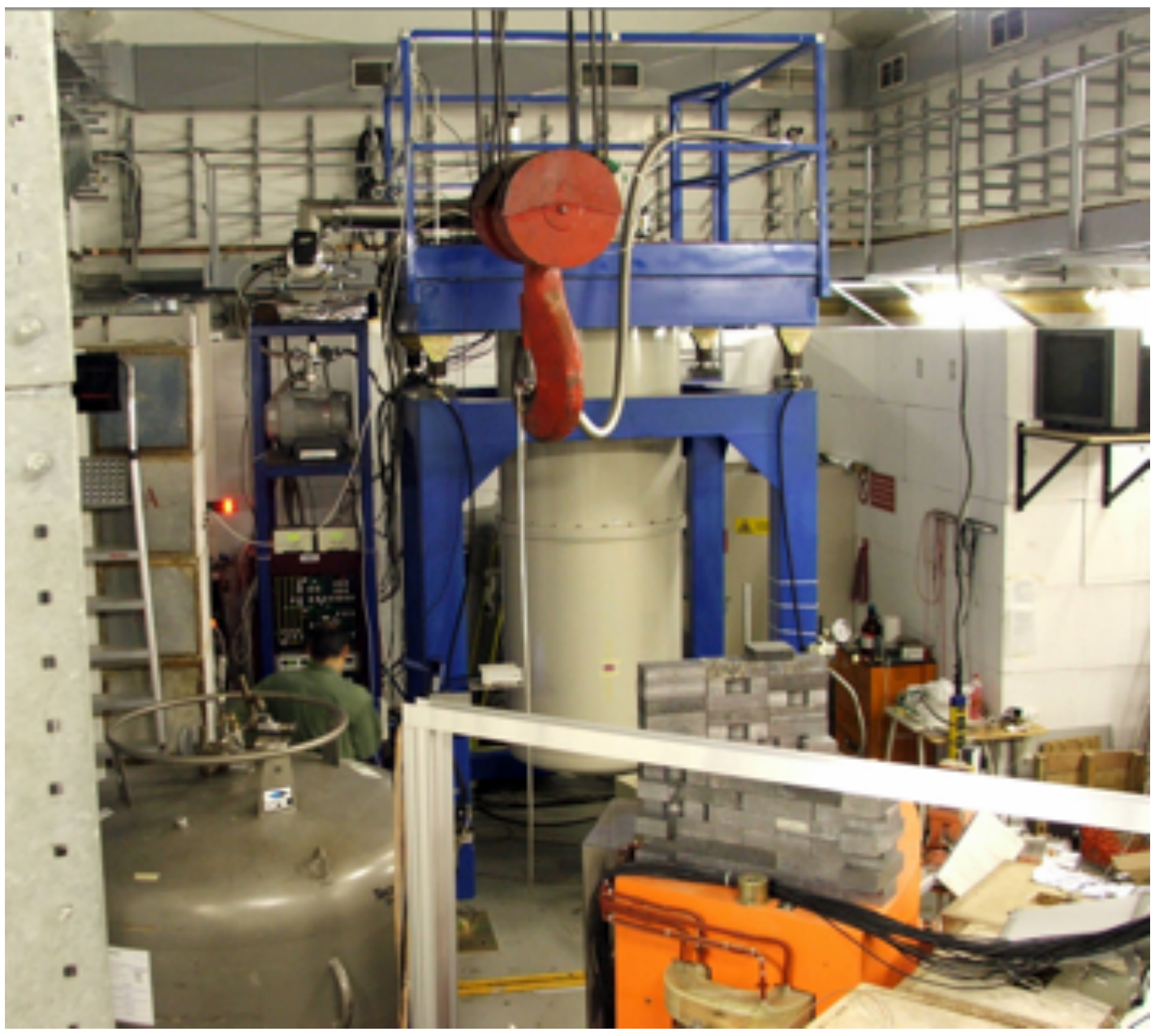

**Figure 1.28:** The RAP experiment took data at BTF with high intensity 510 MeV electrons from 2003 to 2007, measuring the effect of thermo-acoustic expansion on a super-conducting resonant antenna (inside the large dilution cryostat), made by an Aluminum cylinder,1:6 of the NAUTILUS gravitation antenna.

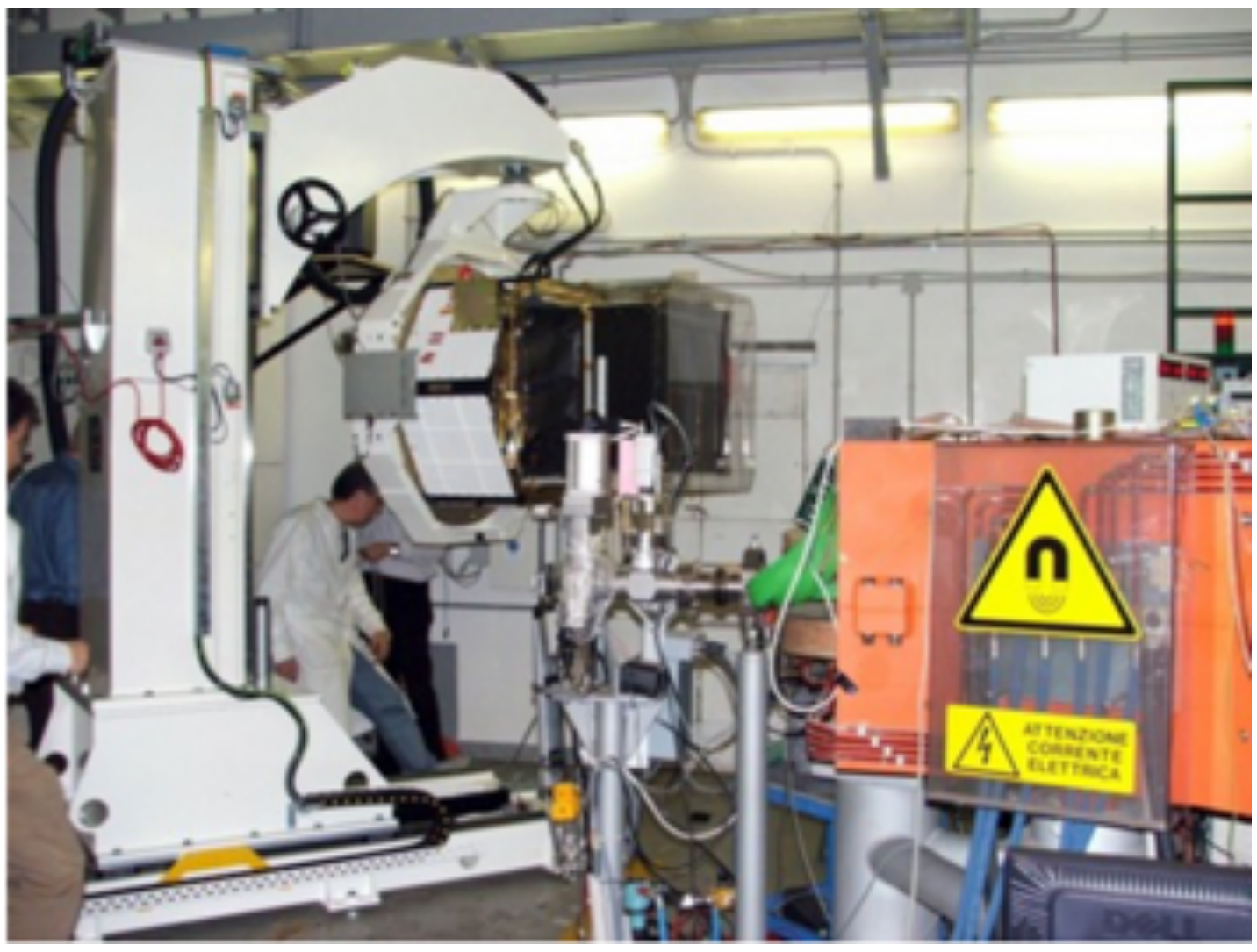

**Figure 1.29:** The AGILE gamma astronomy scientific payload installed on its dedicated movement system during the November 2005 dedicated calibration with the BTF tagged photon source.

Since 2008, one or two shifts/year have been devoted to the characterization and test of the BTF photo-production neutron source (n@BTF project, funded by the INFN Commissione V) [20]. Some more detail on the potential of such a neutron source driven by the BTF electron beam will be given in Sec. 1.7.

Finally, in the last few years, there has been a growing interest in using the BTF beam, especially selecting positrons, for detecting and measuring effects related to channeling in crystals, both bent and straight ones, at energy significantly lower with respect to the other facility used by this community (in Europe typically the CERN SPS at 450 GeV), or for the detection of parametric radiation from crystals, characterization of crystal ondulators, etc.

# 2 MOTIVATIONS AND REQUIREMENTS

In the previous sections, the good performance and the broad use of the Beam-Test Facility of the DAΦNE complex have been briefly illustrated. Apart from very limited periods, in the last 11 years the facility has been running in the shadow of the operation of the electron-positron collider (the so-called "parasitic" mode), taking advantage of the availability of the electrons or positrons from the LINAC not used for injection into DAΦNE.

Due to the constantly growing interest for the facility and the already broad community using the BTF beams, it is interesting to list all possible use cases. This not only in view of keeping the current level of activity, but also of extending the beam parameters and the capability in terms of users access.

The beam time in the last five years, as summarized in Sec. 1.3.5, has been devoted practically only for electron and positron running, with the limitations on the beam parameters due to the concurrent operation of the DAΦNE collider, and with scarce use of the tagged photon beam and of the neutron source, due to the limitations in their operation (see Sec. 2.2 and 2.3).

Taking into account the planned activities for fundamental physics (Sec. 2.4) and irradiation with electrons (2.5), in addition to the "usual" test-beam activities (Sec. 2.1) with electron and positrons, with improved photon (Sec. 2.2) and neutron (Sec. 2.3) sources, it is clear that an increase of the number of beam-lines is strongly motivated.

On another hand, planning a longer life for a facility with an even wider application range, certainly requires to consolidate the accelerator infrastructure and improve the LINAC beam parameters (in particular the maximum energy, for some applications, and the beam pulse duration and charge).

## 2.1 ELECTRON AND POSITRON BEAM-TEST ACTIVITIES

In the HEP and astro-particles communities testing and calibrating detectors is widely done using hadron beams, both protons extracted from accelerators (both linear and circulating) or secondary particles produced by protons on target, due to the properties of heavy charged particles, especially in terms of ionization. Electrons (and photons) are however needed when the response to an electromagnetic shower has to be carefully studied, the typical example being the calibration of a calorimeter. Even though using high energy and heavy charged particles is preferable when studying tracking detectors, due to the dependence of the Coulomb scattering on $v^{-1}p^{-1}$, electron facilities are widely used also for studying the performance of tracking detectors, as demonstrated by the number of this kind of users at the BTF (Sec. 1.3.5).

The availability of "high energy" electron beams, i.e. above the threshold of few MeV (the range of most energetic radioactive sources) and well beyond the GeV, is presently limited to a relatively small number of laboratories, generally dedicated to fundamental physics (especially HEP, but also some injector for photon science machines) or, much more rarely, extracted from industrial or medical accelerators.

Taking into account the European continent, the main electron facilities are CERN in Geneva, Switzerland (electrons and hadrons from PS and SPS, up to 450 GeV), DESY in Hamburg, Germany (1 to 6 GeV), MAX-LAB in Lund, Sweden (up to 3.7

GeV), MAMI in Mainz, Germany (up to 1.5 GeV). In Frascati, SPARC-LAB also provides electrons up to 150 MeV.

In the last five years the BTF user requests have been on average at the level of 150% of the facility capability (summarized in Fig. 1.23), with peaks in correspondence with shut-down periods of other European facilities, like DESY and CERN, in particular in 2013-2014 (due to LS1 of CERN accelerators, including beam-test facilities). This trend is also confirmed for year 2016.

The **main requirements** from the users concerning "traditional" detector testing beam-test activities can be easily summarized:

- Good quality beam, in particular from the point of view of beam size, divergence and background, down to the low end of the BTF energy range, i.e. few tens of MeV. This requirement is particularly difficult to match if the setup is (as in the great majority of the cases) in air, downstream of the 0.5 mm thick Beryllium window.
- Extending the energy range towards higher energies: tracking and efficiency studies suffer from the Coulomb scattering of electrons, which scales as $1/p$. Higher energies are also very useful for extending the range for the calibration of calorimeters.

## 2.2 TAGGED PHOTON SOURCE

The present BTF tagged photon source is constituted by two elements (see Fig. 2.1):

- an active target for the production of Bremsstrahlung photons from the (attenuated) electron beam, made by silicon micro-strip detector planes;
- a series of modules of silicon micro-strip detectors, placed in the final dipole (DHSTB02) pole gap, for the measurement of the radiating electron energy loss.

The system has been designed in collaboration with the AGILE (gamma astronomy satellite) team, with the purpose of calibrating the scientific payload before the launch, with a well defined energy and impact point photon beam [18].

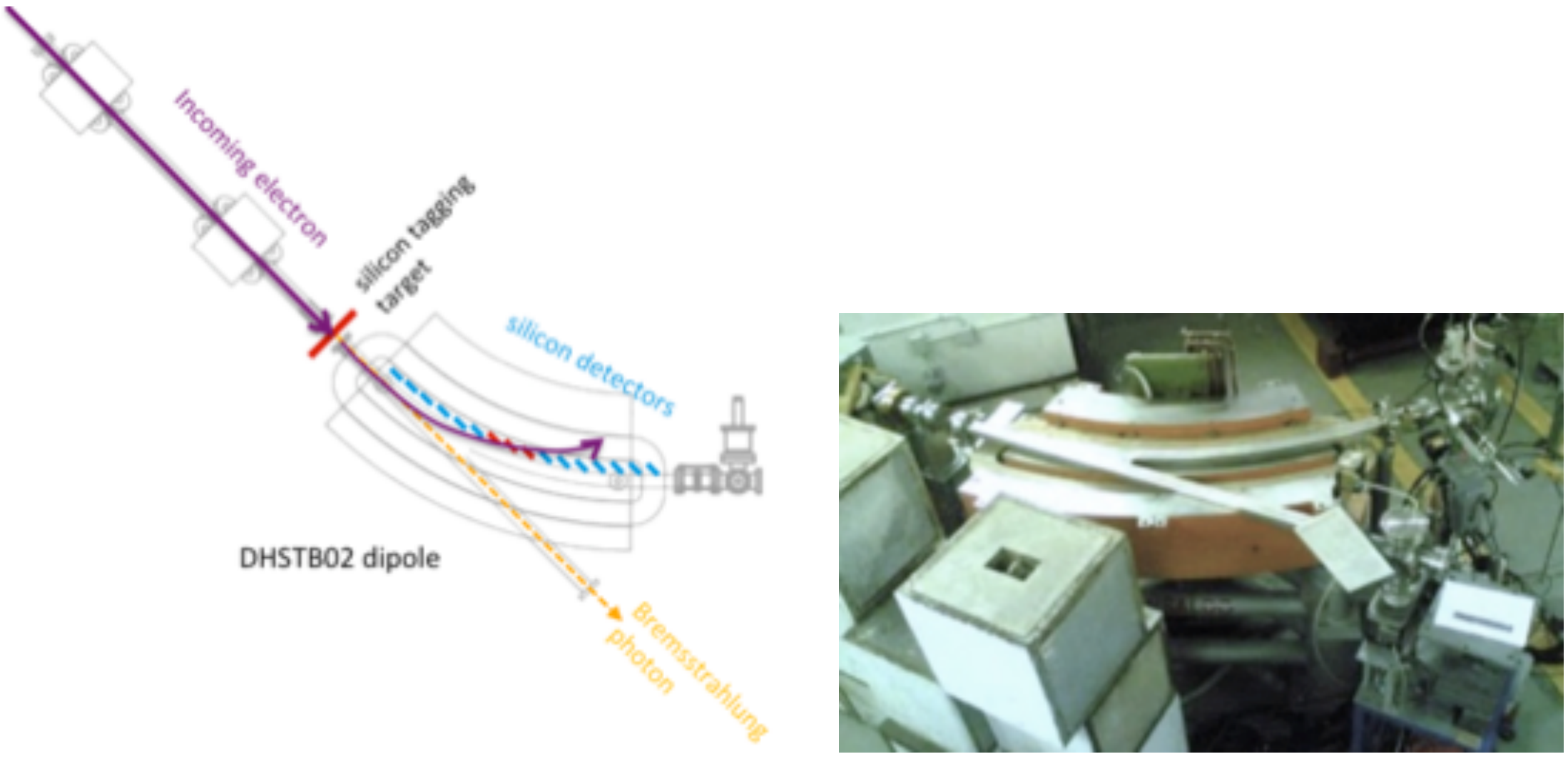

**Figure 2.1:** Left: the tagged photon source principle of operation; right: the vacuum pipe inside the DHSTB02 dipole (with the top pole and yoke removed).

Due to the fact that only possibility was to place tagging detectors just outside the vacuum pipe, a sub-mm spatial resolution was needed, so that Silicon micro-strip detectors have been realized (see Fig. 2.2). In order to minimize the effect of multiple scattering, the steel vacuum pipe has been replaced with a 3 mm Aluminium one.

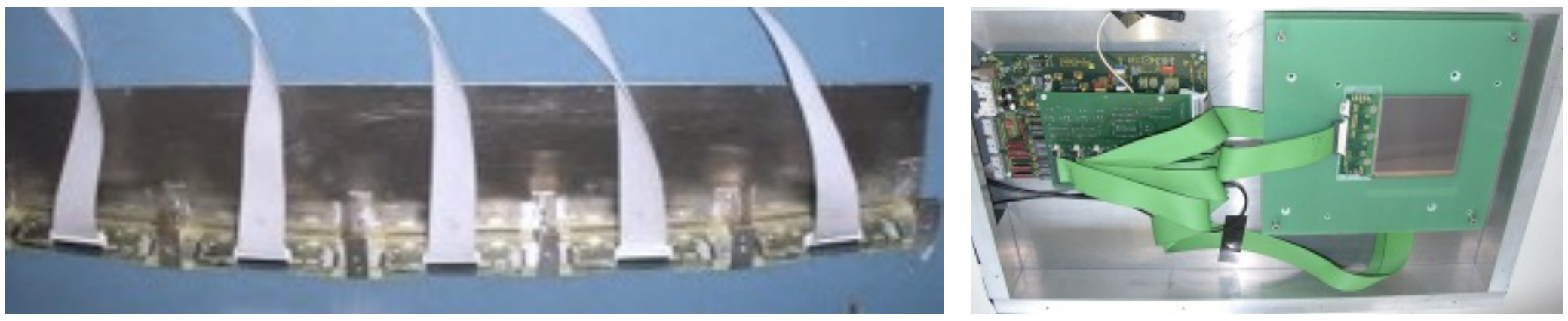

**Figure 2.2:** Silicon micro-strip tagging modules (left), placed along the curve beam-pipe, and active Silicon micro-strip target (right).

Still, the system suffers from spurious events [19] (hits without a corresponding radiated photon): this is mainly due to the fact that the tagging detector is placed very close to the nominal electron trajectory, giving rise to events with scattered electrons (mainly coming from the target) hitting the modules inside the magnet gap from outside.

The energy resolution quickly degrades for decreasing photon energy, as shown in Fig. 2.3, since low energy photons correspond to electrons close to the nominal trajectory inside the dipole, impinging the last tagging modules inside the magnet gap with a grazing angle.

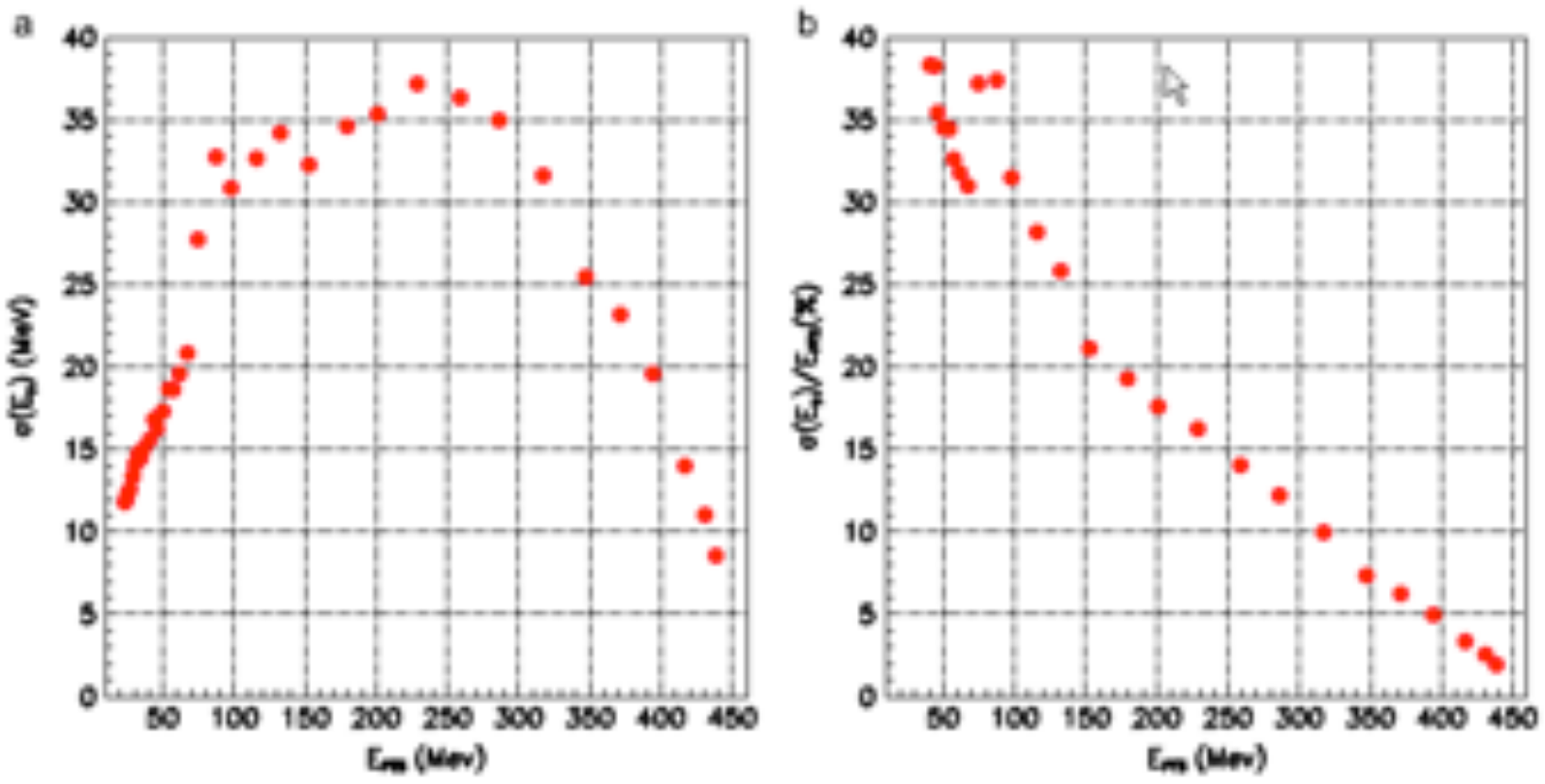

**Figure 2.3:** Absolute (left) and relative (right) energy resolution of the photon tagging system.

Finally, in order to have a sufficiently clean photon sample, stringent cuts on the number of tracks in the active target and on the track reconstruction quality are needed, thus requiring to keep the electron multiplicity close to one, and yielding a low tagged photon rate.

A re-design of the existing tagged photon source allows to greatly increase the efficiency and to improve the energy resolution by replacing the H-shaped magnet with a open-yoke one, allowing to place tagging detectors in vacuum and in the focal plane of the dipole. This allows to relax the requirements on the tagging detector: a 5 mm spatial resolution should ensure a tagging resolution at the level of 2% in the range between 10% and 70% of the incoming beam energy.

An improved tagged photon source would be extremely useful for the testing and calibration of calorimeters and a wide range of gamma ray detectors, of great interest for the large astro-particles scientific community. Finally, the possibility of placing the renewed photon tagging on a second beam-line would make its design and operation easier.

## 2.3 Neutron source

In the period 2008-2010 a feasibility study for an electro-production neutron source using the LINAC beam in the BTF line has been funded by the INFN "Commissione V" (n@BTF project) [20]. In this framework, an optimized Tungsten target and shielding, made of lead and polyethylene layers, have been simulated, designed, built and assembled in the BTF hall (see Fig. 2.4).

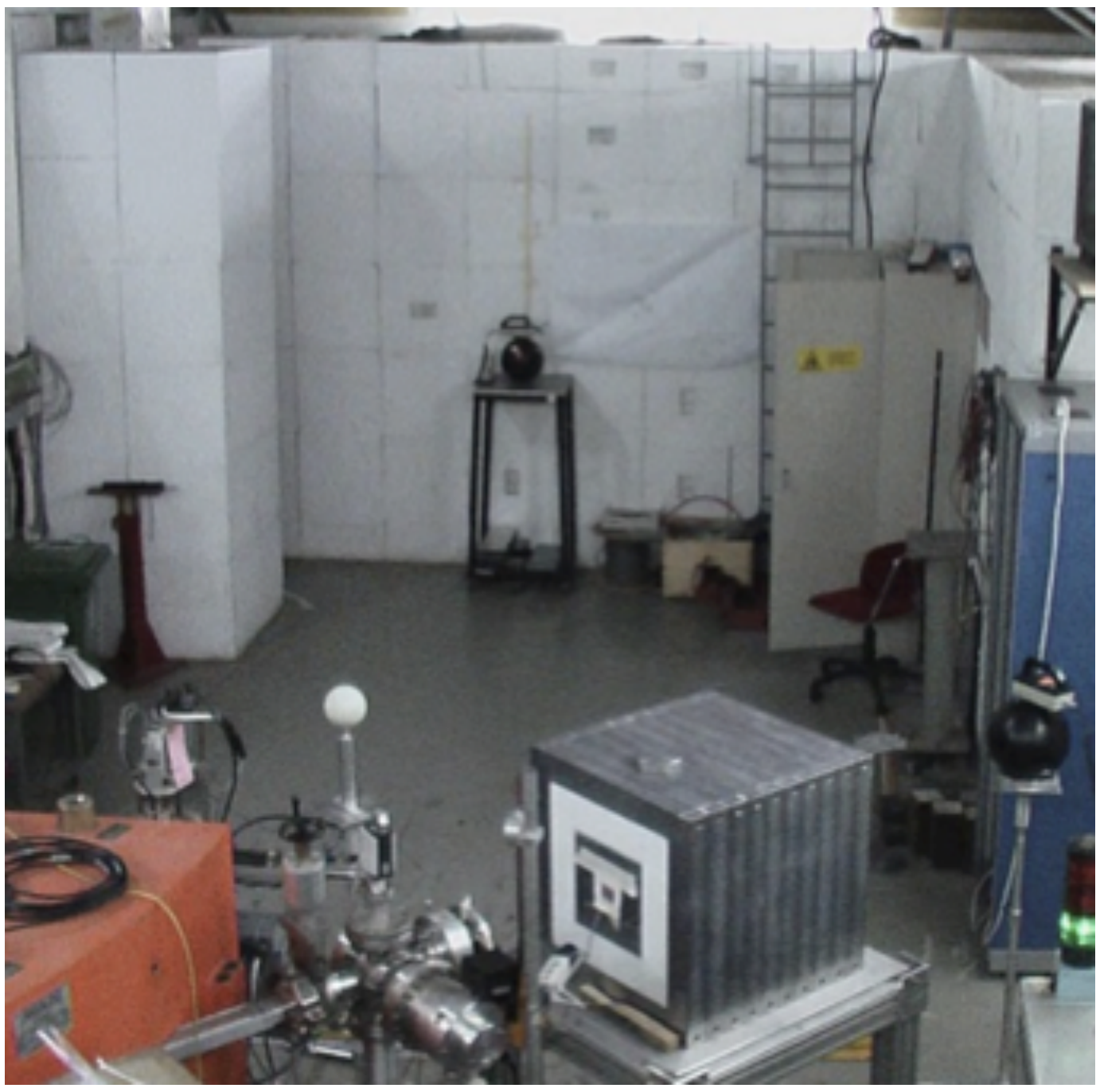

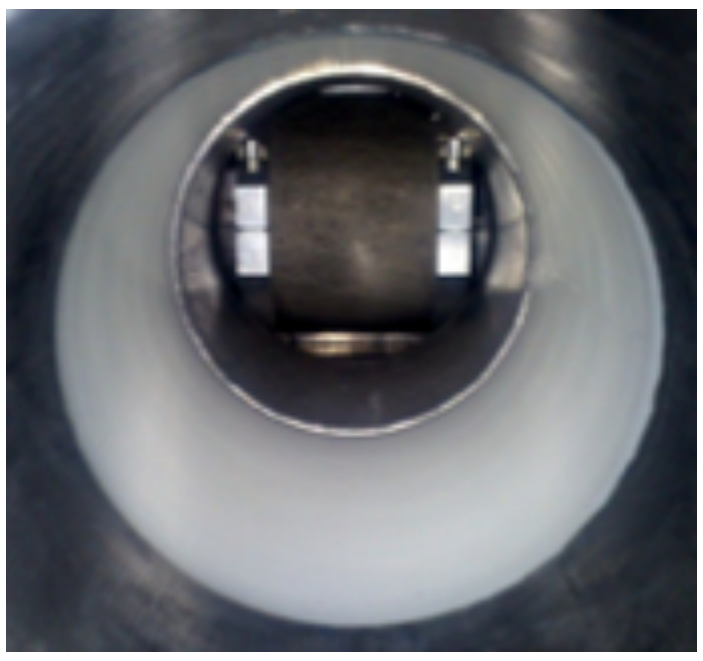

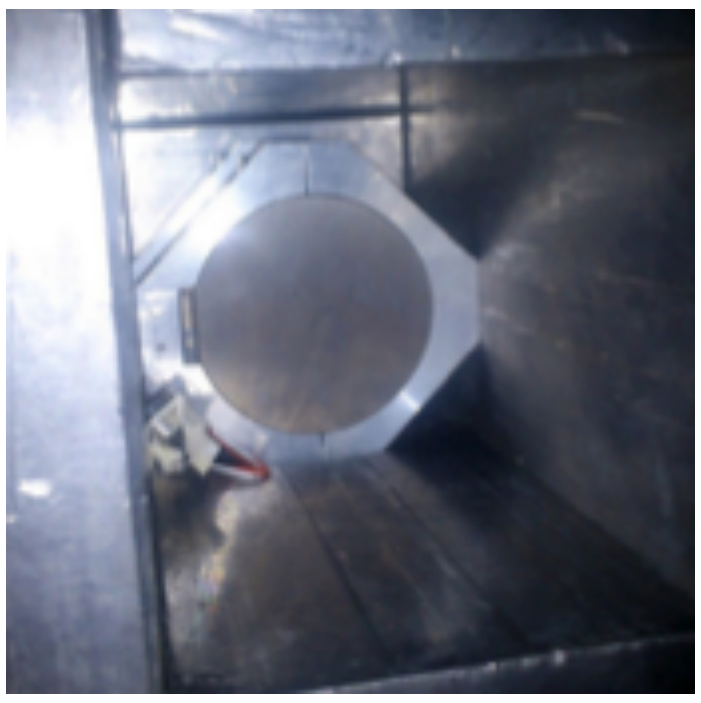

**Figure 2.4:** The n@BTF optimized target and shielding for photo-production of neutrons at the BTF: on the left the assembly (lead and polyethylene shielding hosting the target) mounted on the straight BTF exit, during a measurement of the neutron spectrum with Bonner spheres; right top and bottom, respectively: side view (from the neutron extraction channel) and front view (from the beam entrance channel) of the Tungsten target inside the shielding.

Since then the neutron spectrum emerging from the target has been characterized, showing the expected "evaporation" peak at about 1 MeV (see Fig. 2.5), but the source has been used only a few times, essentially due to the following limitations:

- The lack of on-line neutron fluence diagnostics: up to now, only user-provided neutron detectors have been used (apart from the radio-protection neutron dose measurement stations);
- The low neutron flux, given the electron beam intensity driven into the BTF hall: $4\times10^5$ n/cm$^2$/s on a 10 cm sphere at 1.5 m distance, reducing to about $8\times10^3$ n/cm$^2$/s with the present running mode. In order to avoid occasional safety interlocks to the LINAC, due to lost electrons at high intensity, we routinely run without BTF target at 1 Hz repetition rate (2 Hz during injections), instead that the nominal one, up to 49 Hz;
- The huge prompt photon background, only partially reduced by the choice of the 90° extraction channel from the lead/polyethylene shielding.

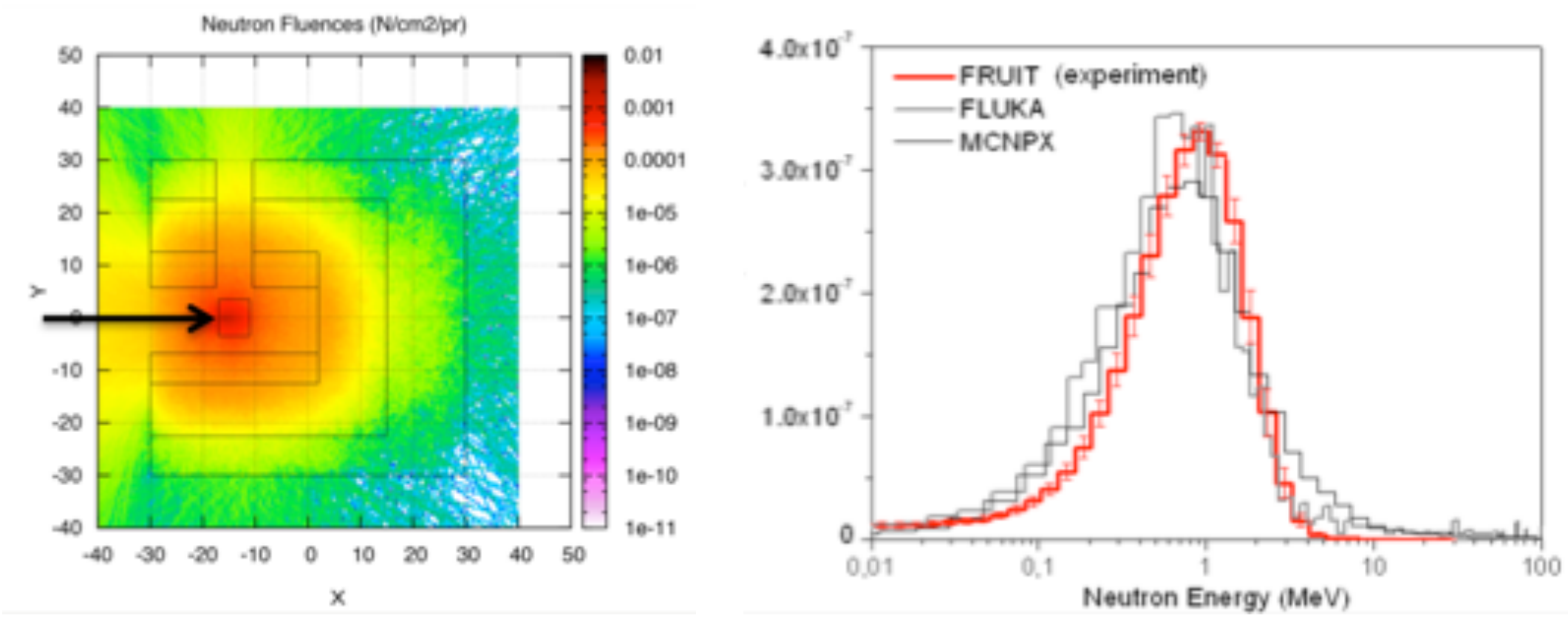

**Figure 2.5:** The n@BTF: simulated optimized shielding neutron fluence from the output channel, on the top of the picture, at 90° with respect to the incoming electron beam (left) and isolethargic neutron spectrum simulated and measured (right).

A number of improvements can be done in order to enlarge the use of the neutron source, in particular the realization and calibration of a suitable neutron diagnostics and the optimization of the neutron yield. Another interesting possibility for several users would be to moderate the produced neutron spectrum.

The use of the neutron source in any case will benefit of a dedicated running of the LINAC, both from the point of view of the maximum duty-cycle, and for the possibility of optimizing the electron energy, since increasing the primary energy slightly increases the neutron cross section and also enhances the tail of faster neutrons.

## 2.4 Electron and positron fundamental physics experiments

### 2.4.1 The PADME experiment

In the last few years the long-standing problem of dark matter has been addressed reviving the concept of “dark” or “hidden” sector: new particles are not directly connected with the Standard Model (SM) gauge fields, but only through mediator fields or ``portals", connecting our world with new hidden sectors.

This concept, which is quite natural in many SUSY models and appears in string theory, is more general and can be realized by adding a new gauge symmetry to the SM. One of the simplest models just adds an additional $U(1)$ symmetry, with its corresponding vector boson. All SM particles will be neutral under this symmetry, while the new field will couple to the charged particles of the SM with an effective charge ε, so that this new particle is often called ``dark photon" (also $U$ boson, or $A'$).

Additional interest arises from the observation that such a new particle, in particular a dark photon in the mass range 1 MeV/$c^2$ to 1 GeV/$c^2$ and coupling $\varepsilon \approx 10^{-3}$, would justify the discrepancy between theory and observation for the muon anomalous magnetic moment, ($g$-2).

Up to now, almost all direct searches have been performed in colliding-beams and fixed target experiments, by producing the dark photon in the analogous of the Bremsstrahlung process ($e^-(Z) \rightarrow e^-(Z)\ A'$) or of neutral meson electromagnetic decays ($\pi^0,\eta \rightarrow \gamma A'$), and then looking at its decay back to SM particles, i.e. to $e^+ e^-$ pairs and, for $M_{A'}$ greater than two muon masses, to $\mu^+ \mu^-$.

These experiments provide quite stringent limits (summarized in Fig. 2.6, left), in particular the muon ($g$-2) band has been practically ruled out. However, all these measurements rely on the strong assumption that no particle lighter than the $A'$ exists in the hidden sector. In this other more general – and probably more interesting case – the branching fraction to SM particles will be suppressed by a factor $\varepsilon^2$, and the dark photon would dominantly decay in ``invisible" particles in the hidden sector. The resulting, model-independent, exclusion plot is to date almost empty (Fig. 2.6, right).

An alternative approach, so far not exploited by any experiment, would be to use the lower cross-section annihilation process (analogous of pair production): $e^+ e^- \rightarrow \gamma A'$. In such a fixed target experiment, knowing the quadri-momentum of the incoming positron and by measuring the quadri-momentum of the SM photon, one can search for a non-zero missing mass.

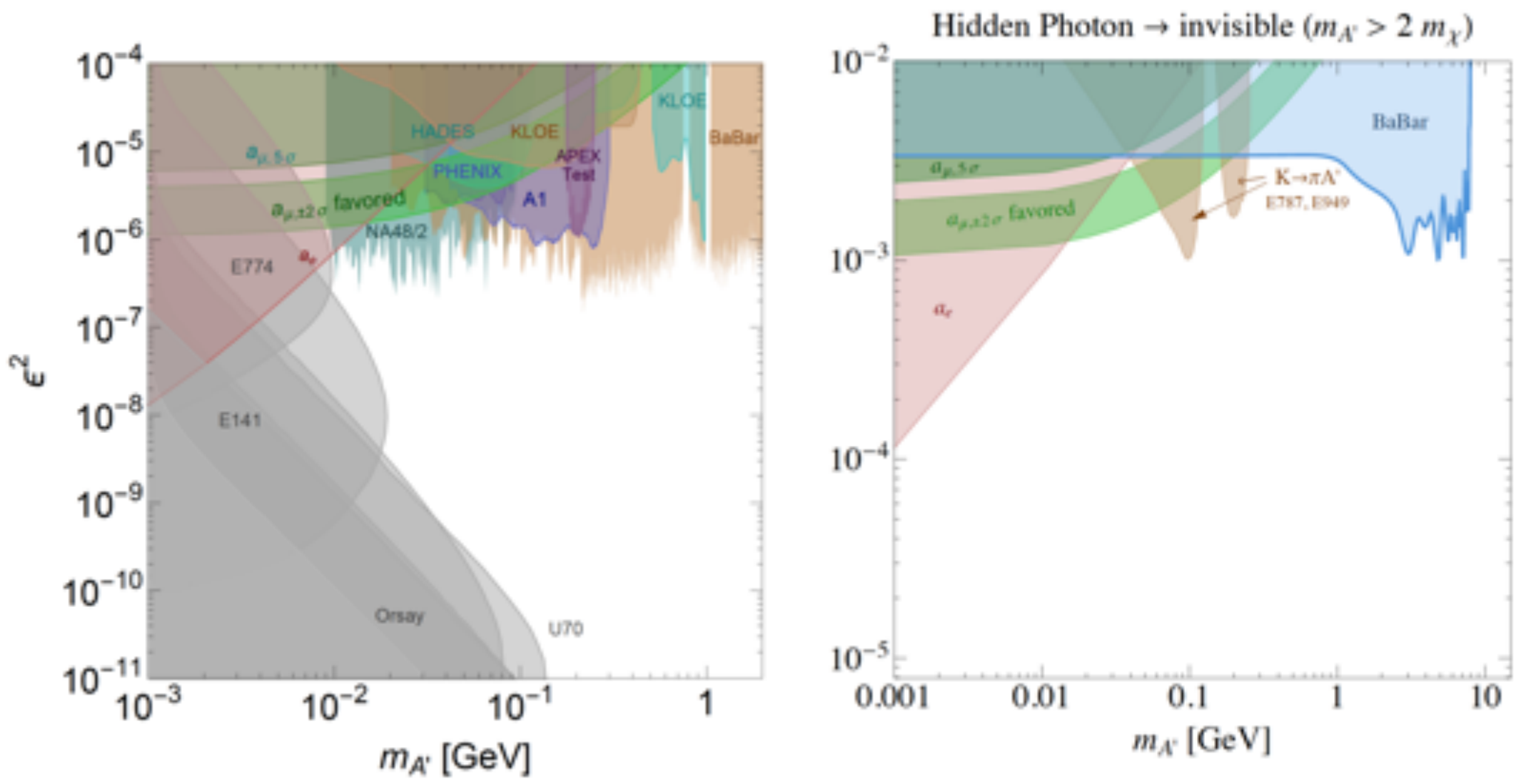

**Figure 2.6:** Exclusion plots in the $\varepsilon^2$ vs. $M_{A'}$ plane for a dark photon "visible" decays to lepton pairs (left) and for dominant "invisible" decay to light dark sector particles $\chi$ (right).

At the end of 2015 INFN has formally approved a new experiment, in the framework of the "What Next" program of new experimental initiatives for searching new Physics hints. The idea of the PADME experiment [21,22] is to exploit the DAΦNE LINAC and the BTF beam-line for producing a positron beam with well defined and easily tunable parameters, mainly with the objective of extending the exclusion – or even better, aiming at the discovery – of a dark photon in a relevant part of the mass and coupling range interesting for the muon ($g$-2) problem.

The experiment aims at collecting $10^{13}$ electrons on target, after a quick R&D and construction phase (approximately two years) of the setup, built around a large gap dipole magnet instrumented with segmented scintillator detectors, for measuring the momentum of positrons irradiating a background Bremsstrahlung; and a high-performance electromagnetic calorimeter, made of BGO crystals (reused from the former L3 experiment), for measuring the angle and energy of photons produced at the target, for the rejection of the $e^+ e^- \rightarrow \gamma\gamma$ and $\gamma\gamma\gamma$ backgrounds. A sketch of the PADME experimental setup is shown in Fig. 2.7.

Positrons will be annihilated on a thin, active diamond target (50-100 μm thick) with readout strips for the monitoring of the beam position.

The requirements from the point of view of the beam are:

- Highest possible momentum positrons, in order to cover the larger possible dark photon mass range;
- Longest possible beam pulse, in order to keep the pile-up probability in the calorimeter as low as possible, given its granularity and the beam intensity;
- Beam spot below $\sigma$=1 mm, divergence below 1 mrad;
- Beam momentum spread below 1%;
- Active diamond target in vacuum, remotely controlled for removing it from the beam-line.

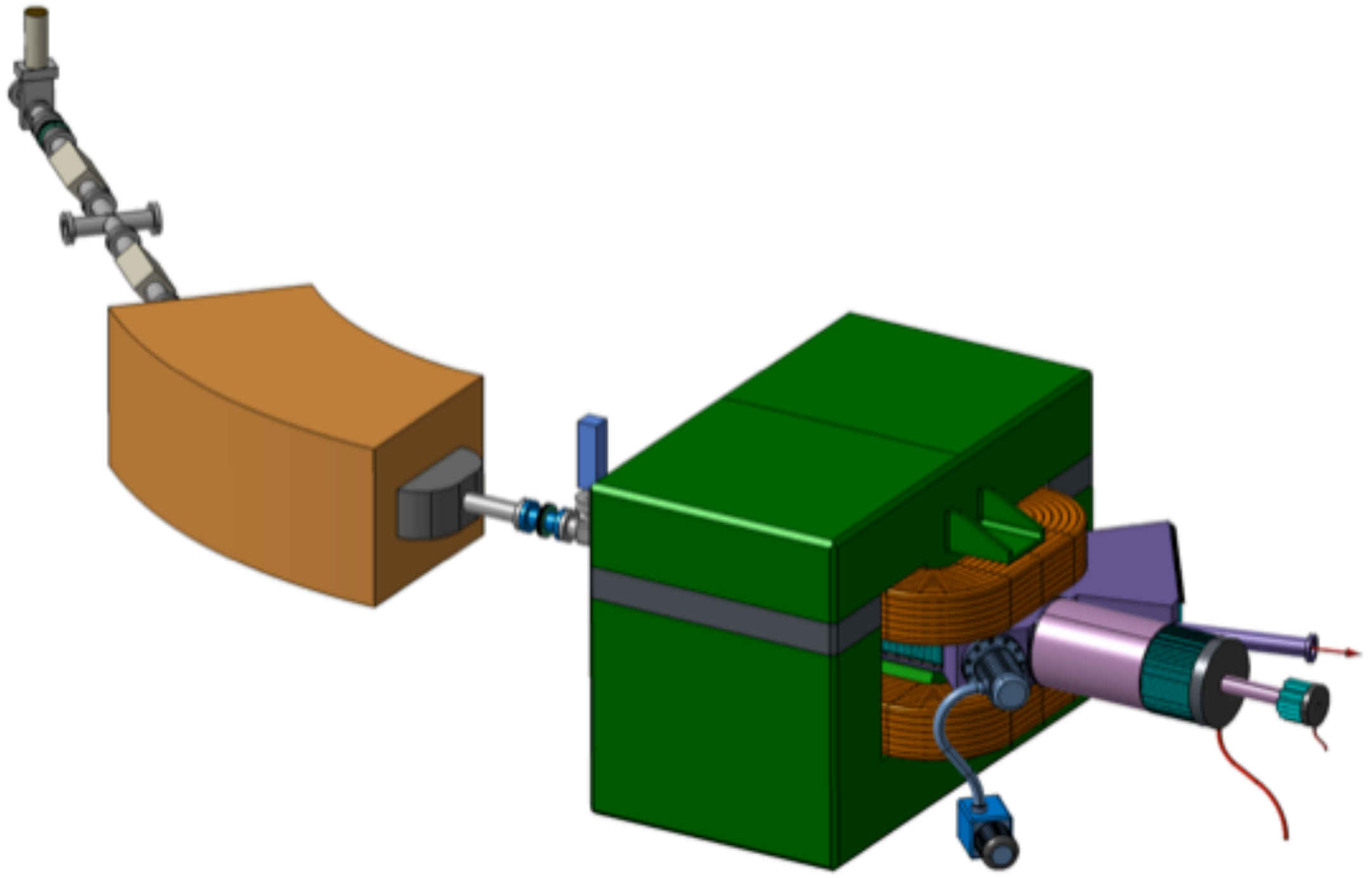

**Figure 2.7:** PADME setup sketch, installed downstream of the DHSTB02 last BTF dipole (orange): inside the poles of the large (green) analysing dipole a vacuum pipe with positron veto scintillating detectors is connected to the downstream vacuum decay volume, ending with the BGO crystal calorimeter. A central hole allows forward Bremsstrahlung background photons to hit a fast small angle calorimeter. The primary positron beam is swept away from the calorimeter by the same dipole field.

Given the present 40 ns maximum length of pulses from the LINAC, and the baseline choice for the geometry of the calorimeter (656 cells, 2×2 $cm^2$, arranged in a cylinder of 30 cm radius at 3 m distance from the target, i.e. covering ≈100 mrad), the single positron pulse cannot exceed $10^4$ particles, so that a dedicated running of the BTF for PADME would require **from one to two years** of data taking in dedicated mode, i.e. without the DAΦNE collider operation. This is necessary both for using the extended beam pulse (40 ns or more) and for running the LINAC at highest energy.

In the present configuration of the facility, a run of PADME would of course exclude the use of the BTF for beam tests or any other activity for several months. In addition, to the running time a not negligible amount of time for the preparation of the BTF infrastructure (e.g. magnet installation, vacuum integration), experiment mounting and commissioning on the beam-line has to be considered.

Running for $10^7$ seconds with a pulse longer than 40 ns (and thus with more than $2 \cdot 10^4$ positrons, keeping the same pile-up probability) would increase the sensitivity of the experiment. Preliminary estimates with a full GEANT4 Monte Carlo, accounting for all background sources, gives the exclusion power shown in Fig. 2.8 for $10^{13}$ positrons of 550 MeV (solid red curve). Improving the background rejection with a further optimization of the setup would push the exclusion closer to the single event sensitivity (i.e. zero background) curve (orange in Fig. 2.8), while any upgrade of the beam energy $E_b$ would of course extend the exclusion at higher dark photon masses as $\sqrt{(2 m_e E_b)}$.

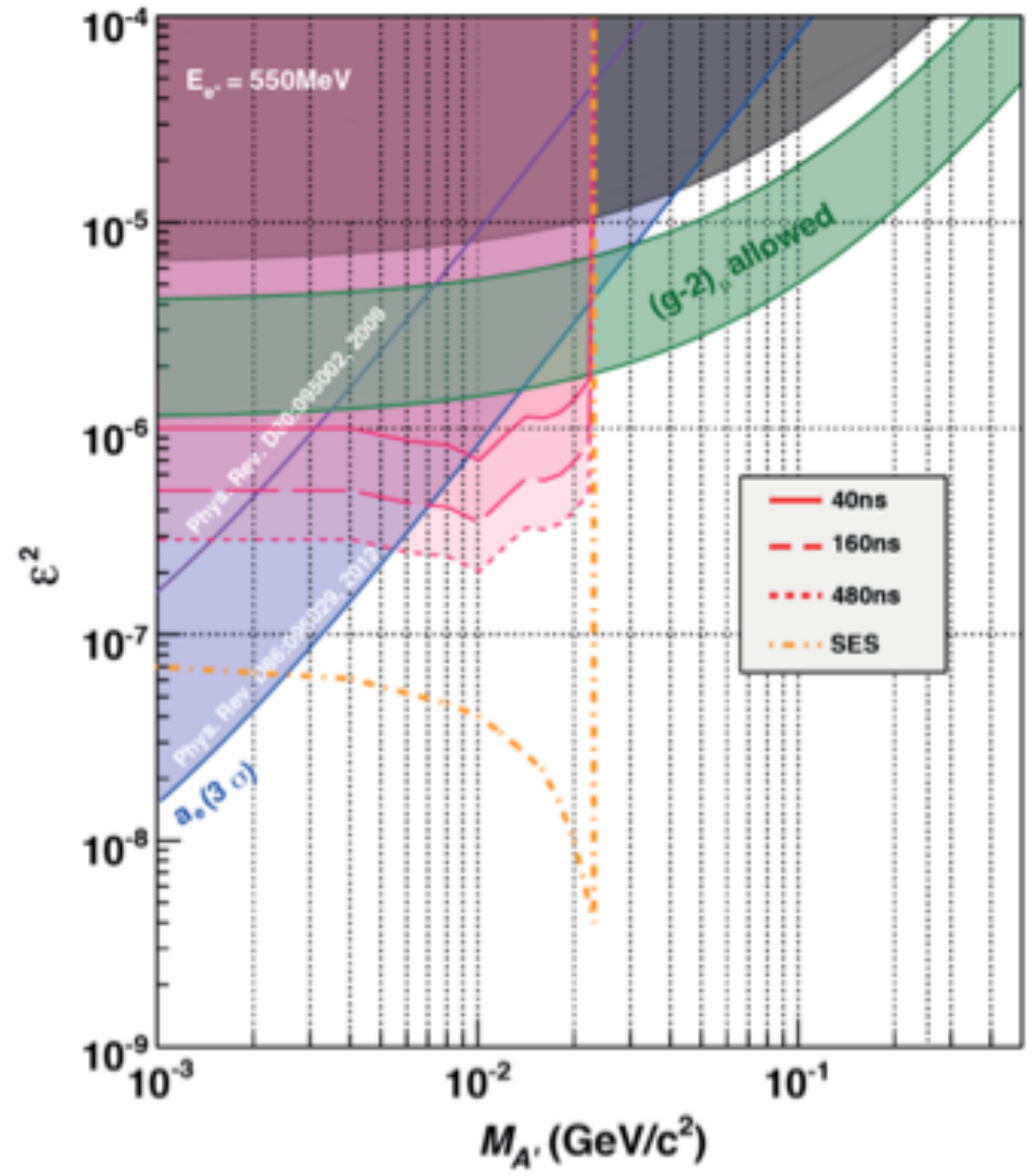

**Figure 2.8:** PADME sensitivity for dark photon exclusion in the $\varepsilon^2$ vs. $M_{A'}$ plane.

### 2.4.2 CHANNELING EXPERIMENTS

When a charged particle beam impinges the atomic planes of a crystal at a sufficiently small angle, the particles are channeled between the crystalline planes. In 1976, Tsyganov proposed the usage of channeling in bent crystals for the steering of beams in accelerators. For many years the UA9 collaboration has investigated this possibility, successfully achieving the steering of the protons both in the SPS accelerator (450 GeV) and in the LHC (at the record energy of 6.5 TeV). Indeed, connected interesting phenomena have been investigated by a number of experimental groups in the world: axial channeling, volume reflection, volume capture, emission of parametric radiation, etc.

For all this experiments the basic requirement is to have a beam divergence better than the typical deflection angles. The main parameter for channeling is the Lindhart angle: $\theta_c=(2U_0/pv)^{1/2}$, being $p$ and $v$ the particle momentum and velocity, and $U_0$ the depth of the potential well, of about 20 eV for Silicon (the most used material) so that $\theta_c$ is 280 μrad at 500 MeV, to be compared with 9.42 μrad at 450 GeV. A few hundreds μrad divergence is almost an order of magnitude better than what can be achieved with the present BTF collimation scheme, also considering that this kind of experiments should be performed in vacuum, thus avoiding multiple scattering both in air and on the beam-pipe window.

The emission of radiation due to the channeling on the planes of single crystal of positrons in the energy range of the BTF beam would result in the interval from few tens of KeV to ≈2 MeV. The requirement of this kind of experiments is then to keep the low energy photon background to an acceptable level, in order not to cover the channeling radiation peak.

## 2.5 Electron Irradiation

The effect of ionizing radiation on electronics, optical and opto-electronics readout is a concern for a large range of applications: not only devices inside nuclear or accelerator plants, but also avionics and aeronautical apparata are sensitive to the enhanced rate of high-energy cosmic rays in the upper part of the atmosphere, satellites and their payload and all components in general travelling in space. Of course the main effects are related to heavy charged particles (protons, ions) and neutrons, but the growing interest for the exploration of Jupiter and its surroundings in the last years has also increased the relevance of the damage induced by a high fluence of electrons, due to the high density of relatively low energy electrons in the magnetosphere of the giant planet: indeed the trapped and accelerated particles create radiation belts thousand of times stronger than Van Allen ones around Earth.

The main requirements for running the BTF as an electron irradiation facility are the following:

- High fluence of electrons of well defined energy (10%) over a wide range: down to few tens of MeV and up to the GeV range;
- Uniformity of irradiation over a $O(\mathrm{cm}^2)$ surface within 10%;
- Determination of the dose with an uncertainty better than 10%.

The present BTF beam quality is adequate, provided that a better and calibrated fluence and uniformity diagnostics is provided to the irradiation users, as required, for instance, by ESA ESCC (European Space Component Coordination) guidelines. Preliminary studies of the defocussed beam, in order to control the uniformity at the required level, demonstrate that BTF can be qualified quickly as an irradiation facility, suitable for space qualification (Fig. 2.9).

Preliminary contacts with the Italian Space Agency (ASI) have been established, in order to both qualify the BTF as a facility recognized by ESA, and to include it in a network of Italian irradiation facilities for space (ASIF).

With similar purposes, BTF is also included in the NASA-SSERVI solar system exploration program, in collaboration with the American space agency and the Frascati spatial qualification laboratory (SCF-LAB).

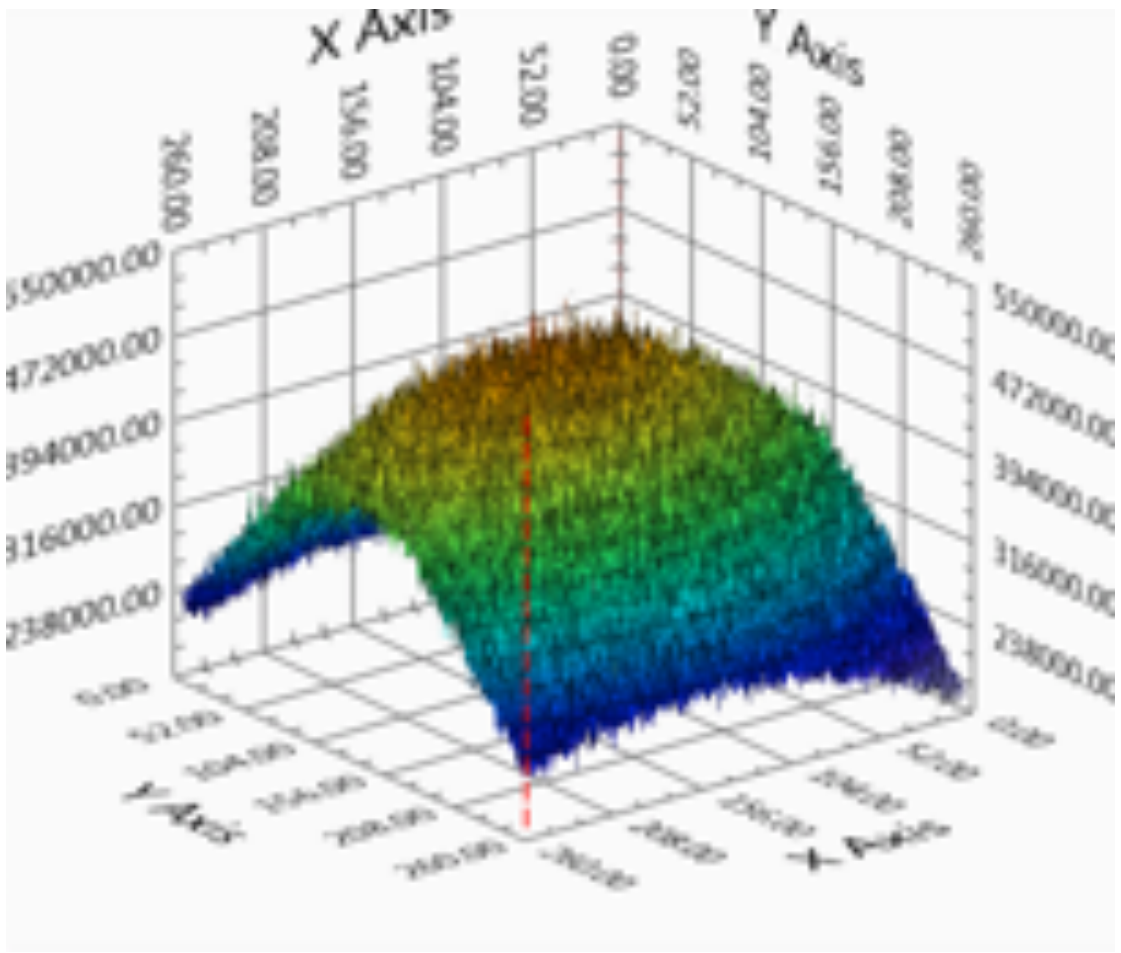

**Figure 2.9:** Spatial distribution, measured with Medipix detector over a 1.45×1.45 $cm^2$ sensitive area, of 50 MeV electrons, $3.5\times10^3$/pulse, during a test BTF irradiation run.

The recent improvements on the BTF hall shielding (see Fig. 2.10), with the new arrangement of the concrete block-house, would allow running at the nominal maximum intensity of $3.125\cdot10^{10}$ particles/s, thus allowing to reach higher fluences also at lower energies (in case by using an optimized target).

Also in this case, extending the energy range towards 1 GeV would increase the potential of the facility, and the availability of a second experimental hall would be fundamental in order to fully exploit all those activities.

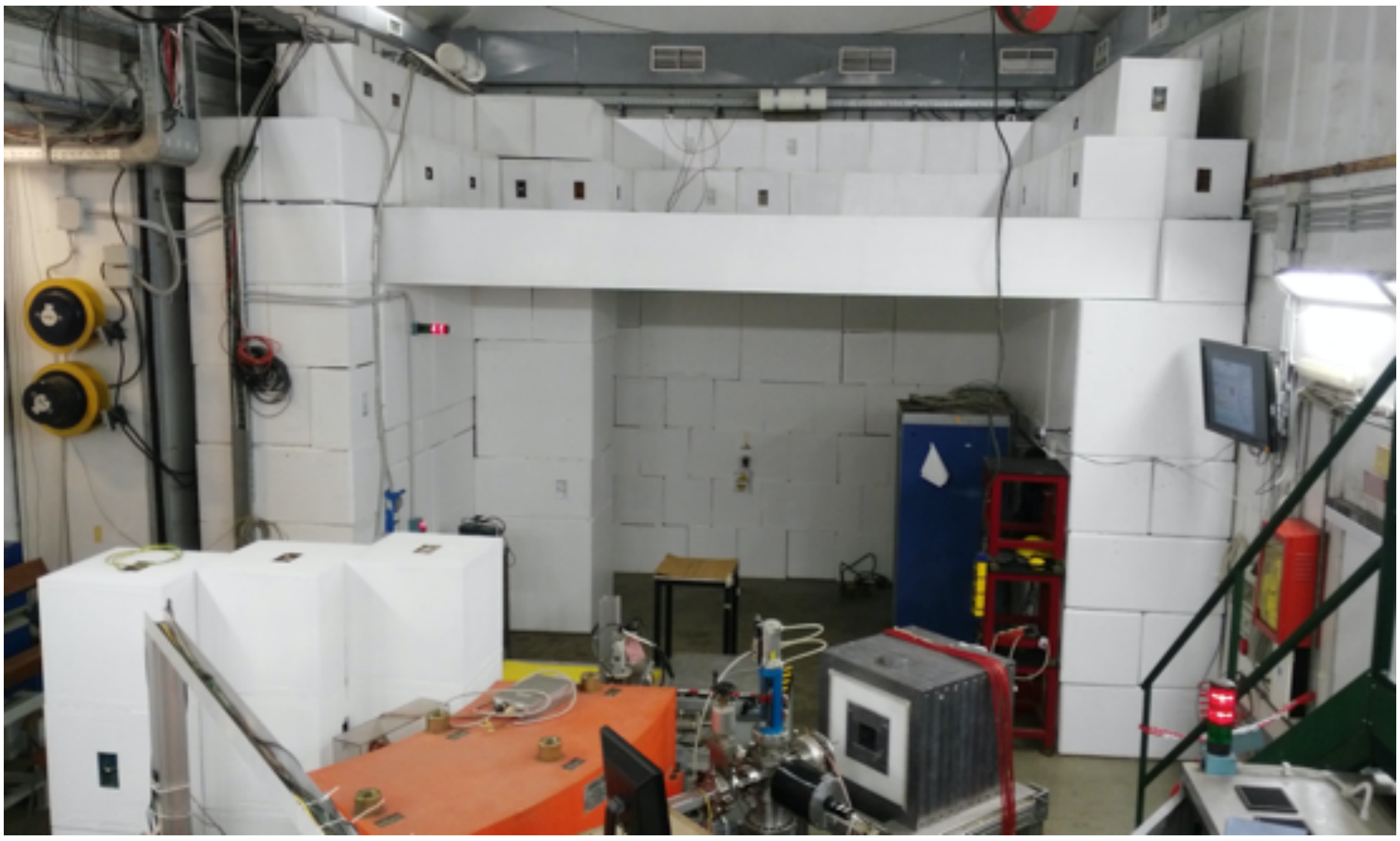

**Figure 2.10:** During the July-September 2015 shutdown the concrete block shielding of the BTF experimental hall has been dismounted and reinstalled in an improved configuration, adding a concrete ceiling (50 cm concrete).

# 3 PROJECT DESCRIPTION

The project aims to a significant upgrade of the LINAC and BTF infrastructure, with the motivations and requirements detailed in the previous chapter 2, and can be divided in three work packages:

- LINAC modulators consolidation and realization of a complete test RF station. This two different activities are strongly related, since the additional RF power station is a fundamental element for the preparation and testing of the components for the consolidation of the existing modulators, thus avoiding a very long shut-down of the LINAC (WP1).
- Energy upgrade of the LINAC up to 1 GeV, by adding four more accelerating sections, supplied by a fifth RF station (WP2).
- Doubling of the beam-lines, re-organization of the experimental hall and realization of an additional experimental hall (WP3).

In the following sections these packages are described, starting from the present status of the infrastructure.

Even though the project has been built developing these three activities in an integrated way, it would be possible to implement only a part of the planned improvement program. In particular, the consolidation of the LINAC (WP1) and the splitting of the BTF lines (WP3), can reasonably proceed as stand-alone projects.

On the contrary, an energy upgrade of the LINAC can hardly be performed without planning a consolidation and improvement program for the modulators and RF power distribution.

## 3.1 LINAC MODULATORS CONSOLIDATION

The DAΦNE LINAC has been briefly described in Sec. 1.2; in particular this S-band, travelling wave linear accelerator has been modelled on the SLAC design. The 16 (including pre-buncher and buncher), 3 m long, SLAC-type accelerating sections are fed by four RF stations, with the power distribution scheme shown in Fig. 1.2: the first part of the LINAC is “high current”, the second part is the “high energy”, and in particular the last 4+4 accelerating sections, out of 10 downstream the positron converter, have an identical configuration in which the RF from each of the two klystrons is split twice with 3 dB couplers, providing an approximate peak power of about 40 MW per section.

The scheme of a single RF station is shown in Fig. 3.1: the 200 W RF from the 2856 MHz source, pre-amplified by the RF driver, is amplified up to the maximum power by the klystron (fed by the modulator), and compressed by the SLED (up to a factor 3.5 at the top), prior to be distributed by the waveguide network to the accelerating sections.

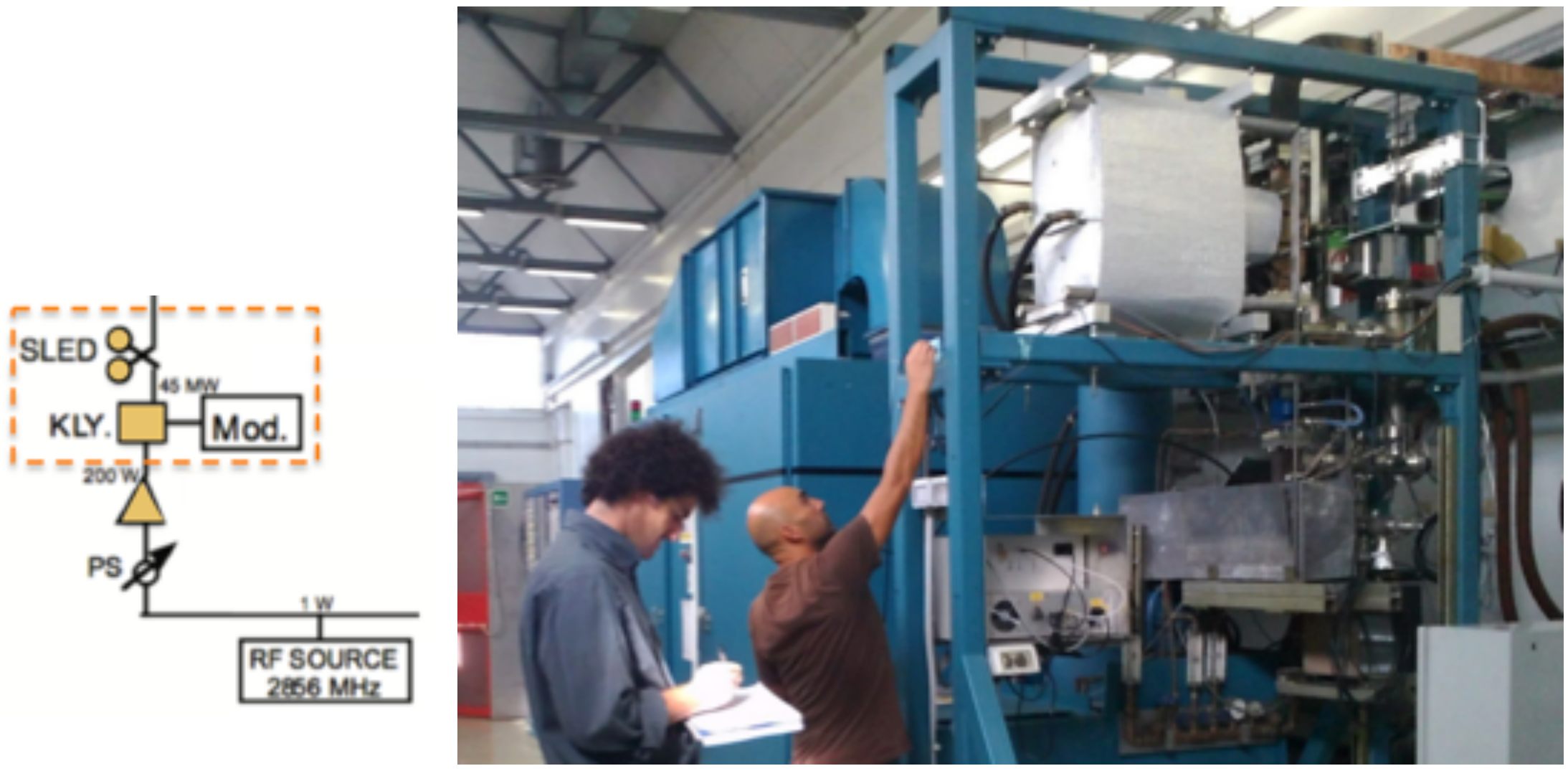

**Figure 3.1:** The RF power station (left, logical scheme; right, photograph): the 1 W RF from the 2856 MHz source, pre-amplified to 200 W by the RF driver, is amplified up to the maximum power by the klystron, fed by the modulator, and compressed by the SLED, prior to be distributed by the waveguide network to the accelerating sections.

Each modulator is composed by the rectifying transformer, the charge power supplies, charging up to 50 KV the pulse forming network (PFN), composed by 9 LC cells; then the charge pulse through a thyratron (using hydrogen gas as switching medium, E2V model CX2668A) is transferred to the 12× up-scaling pulse transformer (installed inside the cooling oil-filled tank hosting the klystron), which gives the high-voltage pulse (up to 310 KV) to the klystron.

The overall functional scheme is shown in Fig. 3.2, while pictures of the main components inside the modulator racks and the klystron tank are shown in Fig. 3.3 and 3.4.

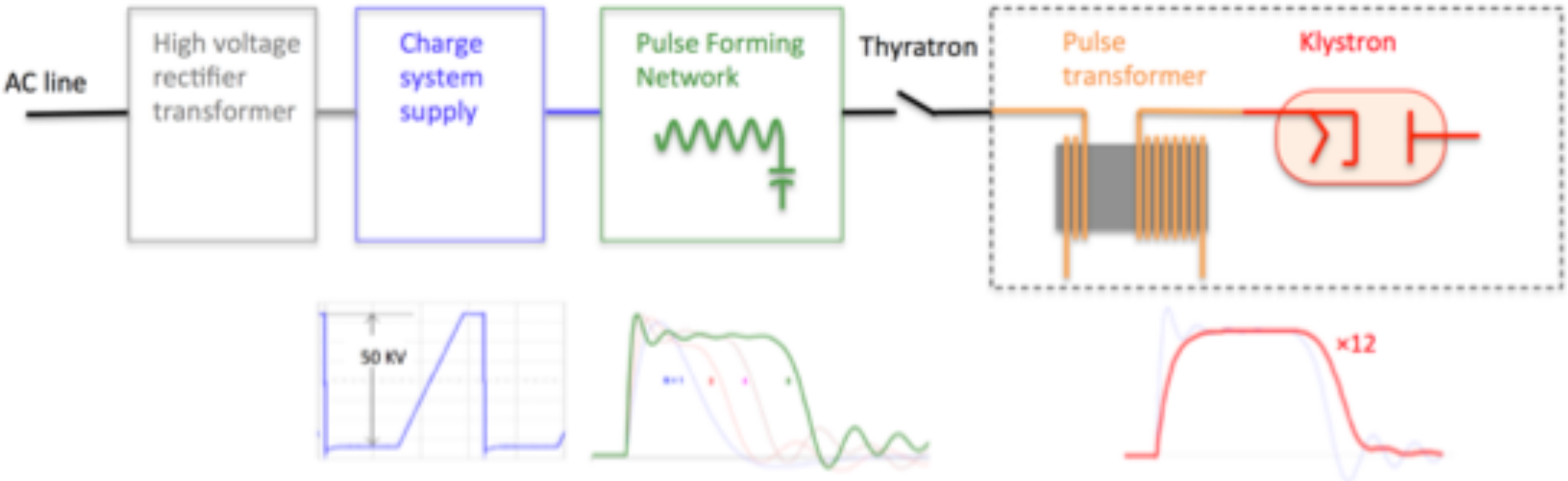

**Figure 3.2:** Logical scheme of the modulator, thyratron and pulse transformer supplying the klystron.

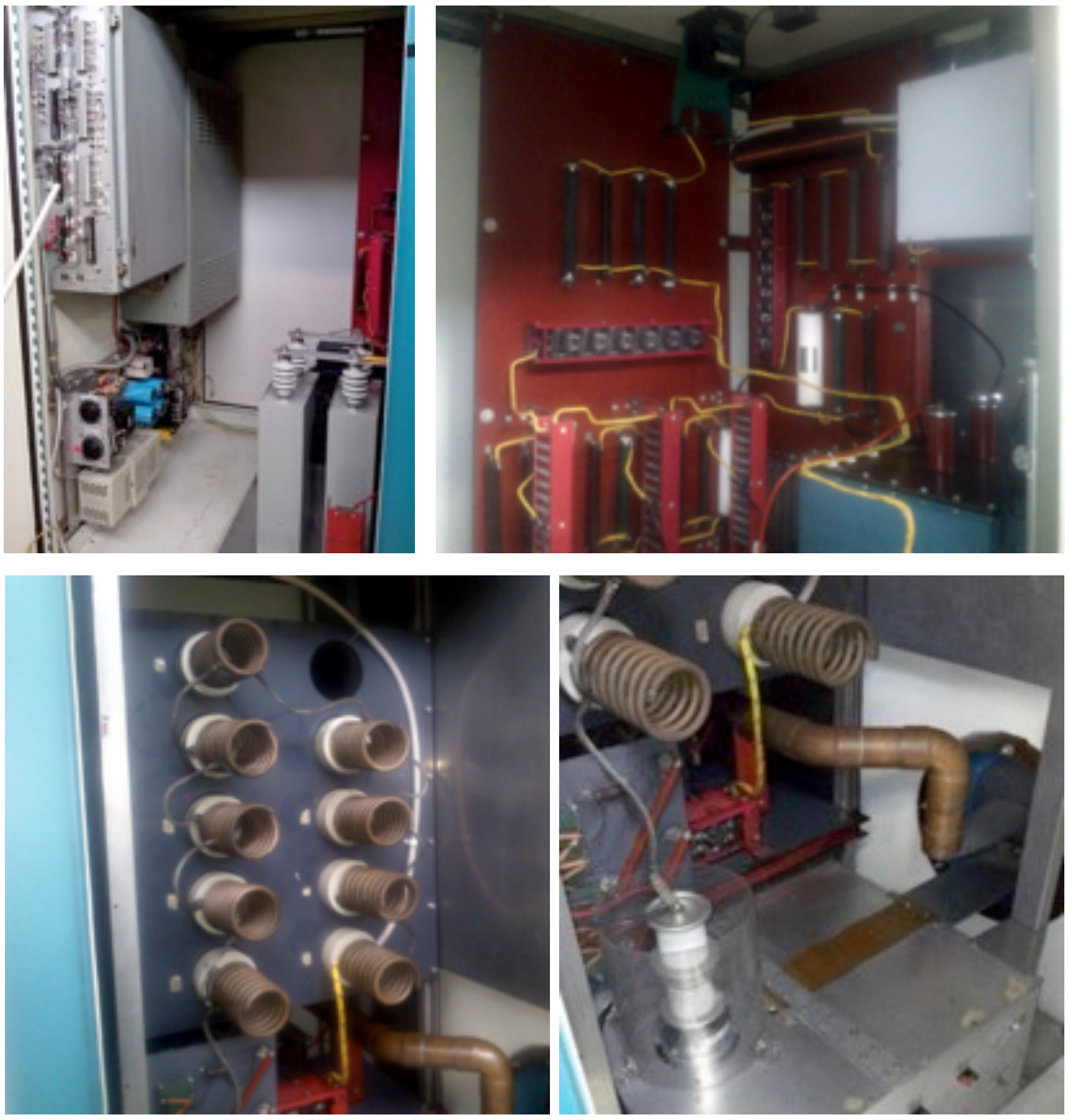

**Figure 3.3:** The SCR rectifier, the core bias and filament power supplies are placed inside one door of the modulator racks (top left), together with the diode arrays, the large charge capacitor and filters of the charging circuit (top right). In the other rack door: the 9 LC cells PFN (bottom left), the Hydrogen gas thyratron and the conductor towards the pulse transformer (bottom right).

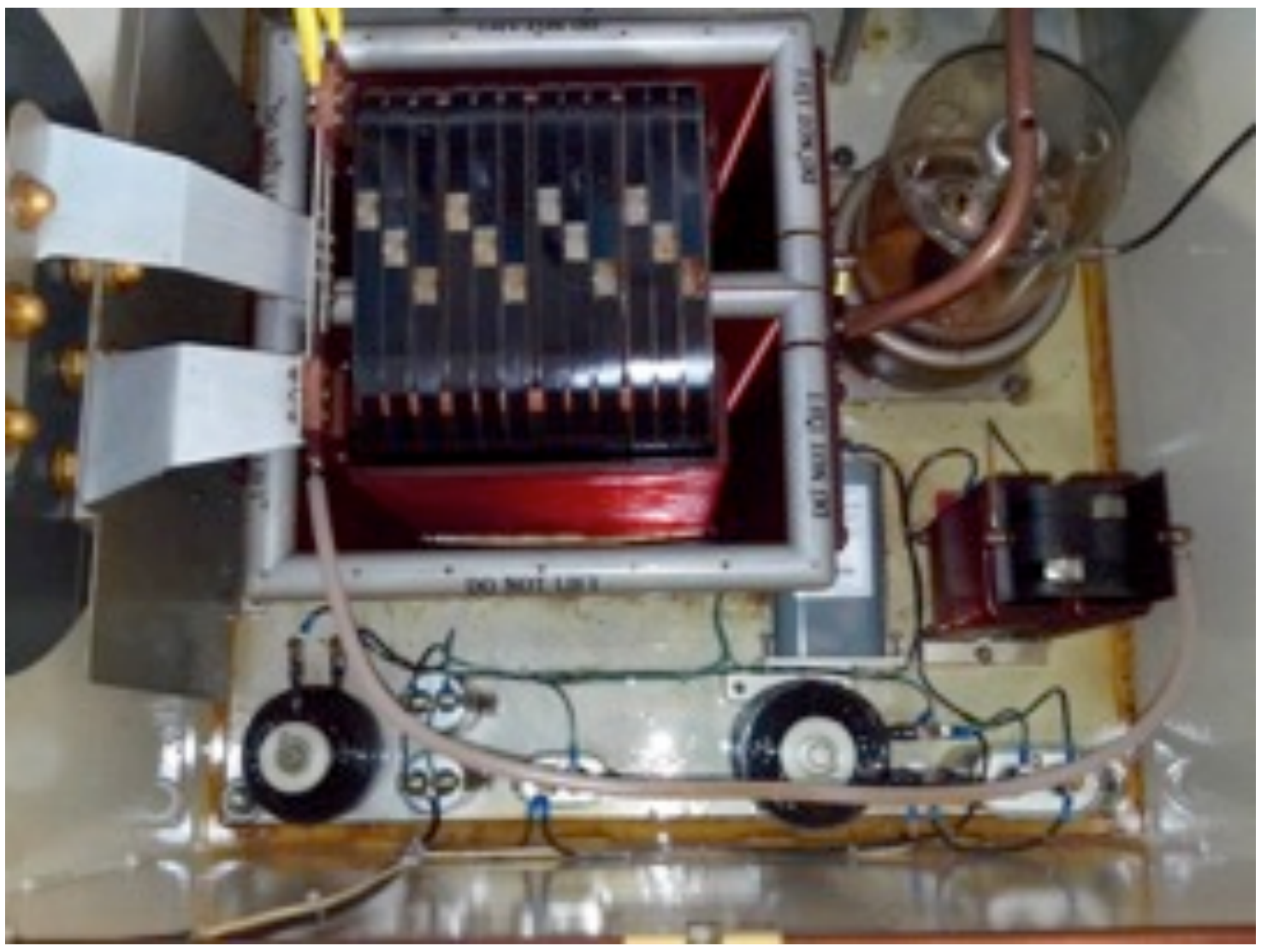

**Figure 3.4:** The klystron tank, hosting the pulse transformer and the capacitive divider 10,000:1 (for the measurement of the input klystron potential, $V_k$, on the top right of the picture).

The main parameters of the klystron, Thales model TH2128C (shown in the pictures of Fig. 3.5), are listed in Tab. 3.1.

| **Parameter** | **Unit** | **Value** |
|---|---|---|
| Center Frequency | MHz | 2856 |
| Peak output power | MW | 45 |
| Peak average power | KW | 10 |
| RF pulse width | µs | 4.5 |
| Peak beam voltage | kV | 315 |
| Peak beam current | A | 360 |
| Micro-perveance | µA/V3/2 | 2.0 |
| Heater voltage | V | 20~30 |
| Heater current | A | 20~28 |
| Focusing currents | A | 40 |
| Peak driver power | W | 200 |
| Gain | dB | 54 |
| Efficiency | % | 43 |
| Pulse repetition rate | Hz | 50 |

**Table 3.1:** Main parameters of the klystron.

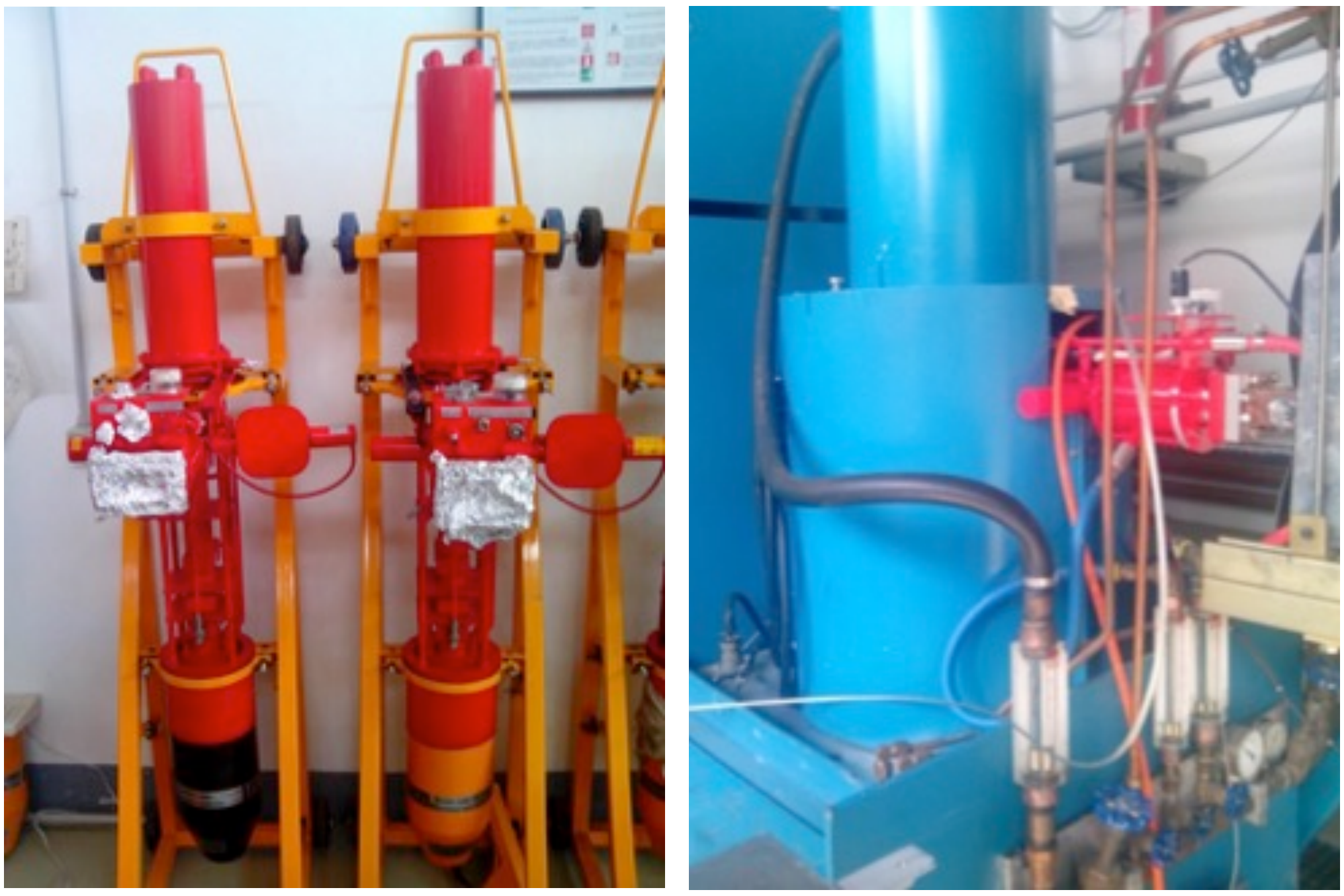

**Figure 3.5:** Left: two klystrons Thales TH2128C (exausted); right: the klystron installed inside the cooling oil tank and surrounded by its focalizers.

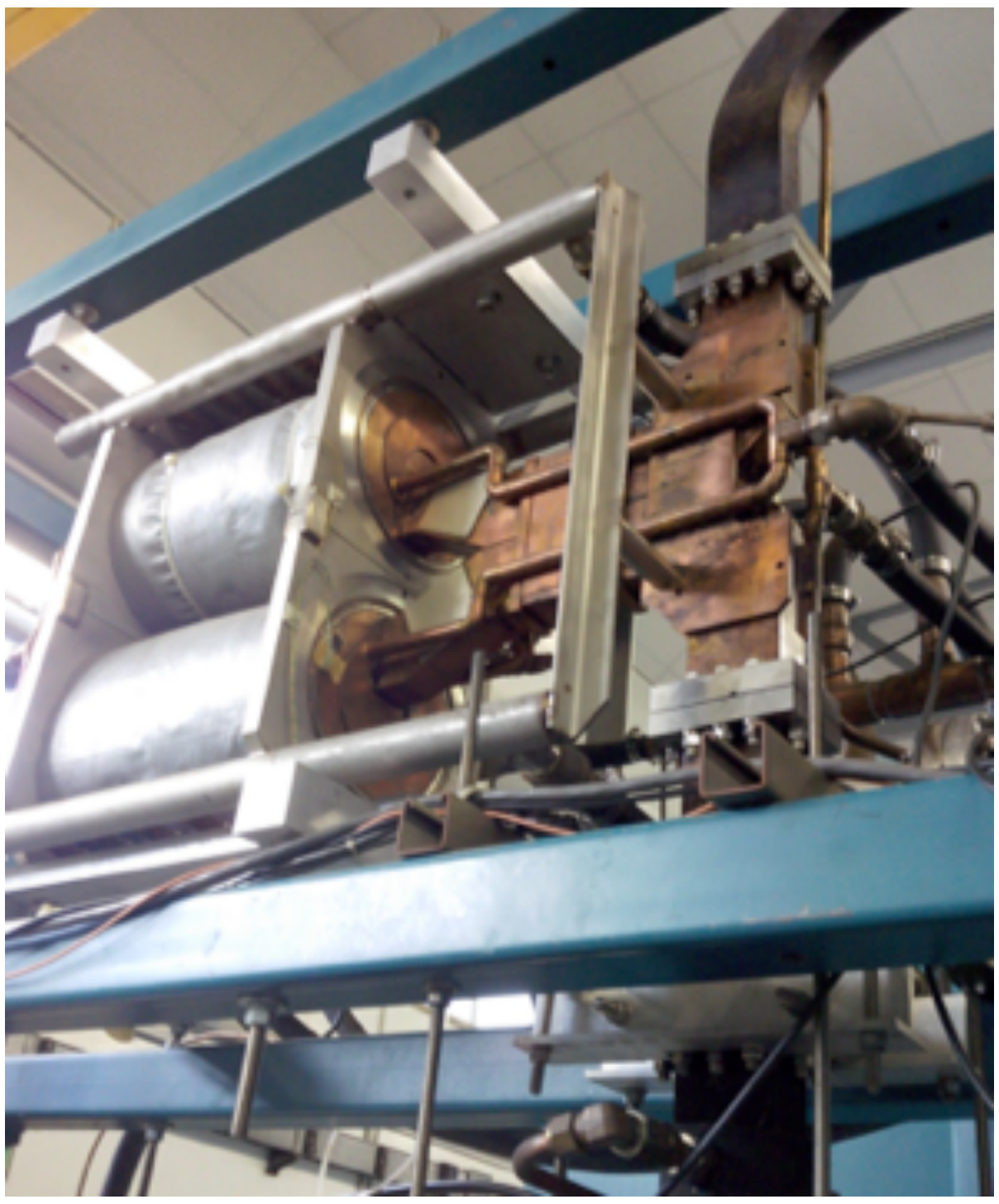

**Figure 3.6:** The SLED device is composed by two (cylindrical) resonant cavities, connected by a 3 dB hybrid coupler, taking as input the power from the klystron (just below, not shown in the picture) and output to the waveguide network towards the accelerating sections.

The 45 MWp RF power from the four klystrons is a square wave, 4.5 μs long, that is compressed by means of the two resonant cavities, connected by a 3 dB hybrid coupler, constituting the SLED device (see Fig. 3.6): the shapes of the power out of the klystron and exiting from the SLED are shown in Fig. 3.7 (left). The power reflected back to the klystron should be kept at a minimum fraction, as shown in the example in Fig. 3.7 (right).

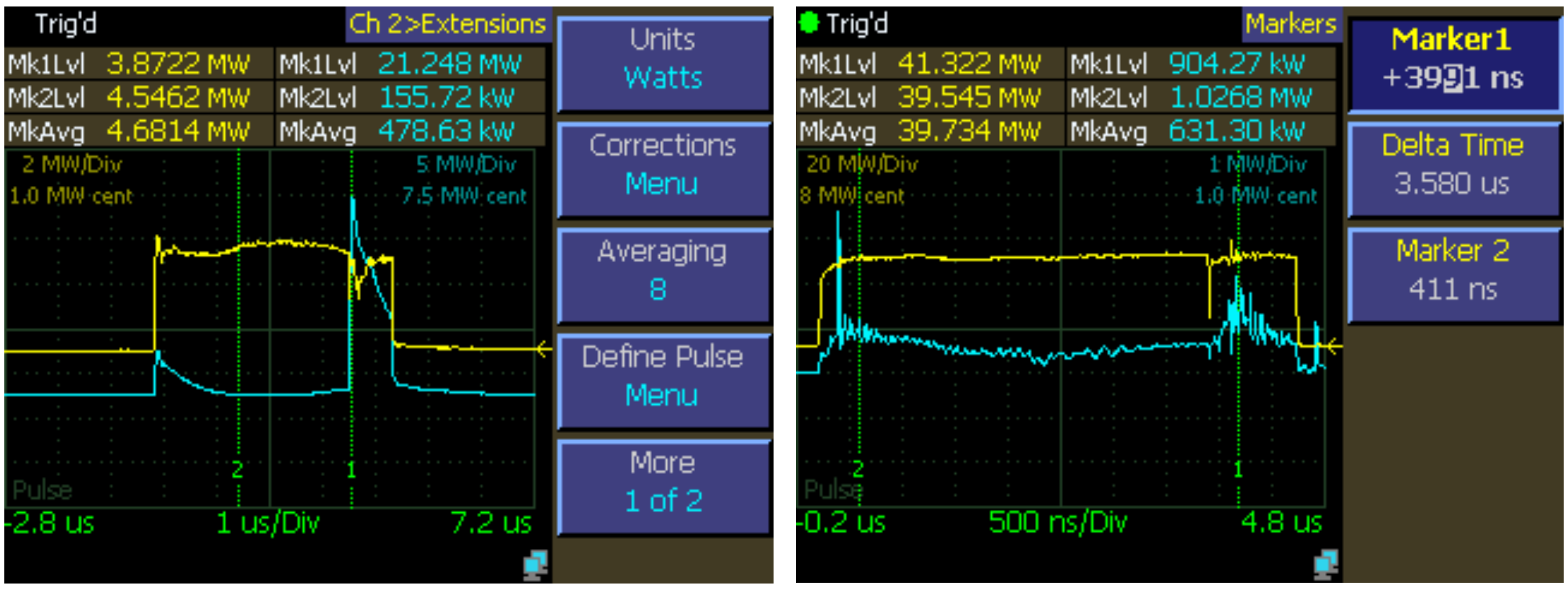

**Figure 3.7:** Left: measured forward klystron power (yellow) and SLED output power (cyan) for one of the four RF stations. Right: measured forward (yellow) close to the maximum klystron power ≈41 MW, and reflected (cyan) ≈1 MW.

During the last long shutdown (January-May 2013), an extensive maintenance program has been performed on all the LINAC components, in particular on the four RF power stations [23].

Several components such as filter capacitors, thyratrons and high power pulse discrete elements have been replaced, and a newly designed RF driver system has been installed, in order to achieve a better stability of the delivered power. In critical parts, like waveguides downstream the SLEDs, additional pumps with higher pumping speed have been added, in order to reduce discharge occurrences. All the ceramic windows, placed downstream the klystron ones to decouple the LINAC vacuum, have been substituted.

The LINAC control system has been revised in order to be compatible with the new network infrastructure. A new electron gun system has been also developed and put into operation. Also the control of the cooling has been upgraded to a new PLC-based system, together with a revision of the water ducts, flux-meters and water pumping system of the primary cooling system (at 30° C).

The water system for the SLED and waveguide network (the so-called secondary circuit, at 45° C) has to be completely revised, in order both to improve the stability of the temperature control, and to allow the increase of 20% of the cooling power, with the addition of a fifth RF station.

The most critical parts of the LINAC sub-systems, however, are the modulators, and their control and safety interlocks. Indeed, all consolidation and maintenance activities in the last years concerned mainly the RF stations, but still a number of critical components are the original parts installed by TITAN BETA more than 20 years ago, namely:

- The high voltage rectifier transformers;
- The charge power supplies;
- The PFN inductances and capacitors;
- The core bias and filament power supplies;
- All the safety interlocks from the LINAC subsystems (modulator signals, water flow, temperature and pressure meters, vacuum pressure gauges, pump controls) hard-wired in custom-built logics boards, readout by a CAMAC-bus system (see Fig. 3.8);
- The phase switchers, klystron power controls, also performed with digital I/O controlled by CAMAC-bus;

Realizing a complete RF station of course requires also the installation of the klystron, and thus the tank, the pulse transformer, the thyratron, the PFN, and all the other needed components. Building an additional RF station, not connected to the accelerating sections, would also be extremely useful in order to perform routinely tests and maintenance activities on all the RF components, without interfering with the operations of the LINAC.

A summary of the components needed for this work package (WP1), is shown in Tab. 3.1, together with the estimated costs.

| Item | N. | Unit cost (k€) | Cost (k€) |
|---|---|---|---|
| New supplies for PFN charging | 3 | 40 | 120 |
| New supplies for PFN charging (existing modulators) | 8 | 40 | 320 |
| New readout system for LINAC signals | 1 | 50 | 50 |
| New modulator control system | 1 | 100 | 100 |
| New readout and DAQ for BPMs | 1 | 50 | 50 |
| New gun pulser 1 ns - 4.5 us | 1 | 70 | 70 |
| Secondary cooling upgrade for RF structure and SLED | 1 | 100 | 100 |
| Modulator | 1 | 600 | 600 |
| Klystron filament and core bias supplies | 10 | 6 | 60 |
| **Total WP1** | | | **1470** |

**Table 3.1:** LINAC modulators consolidation and upgrade costs (WP1).

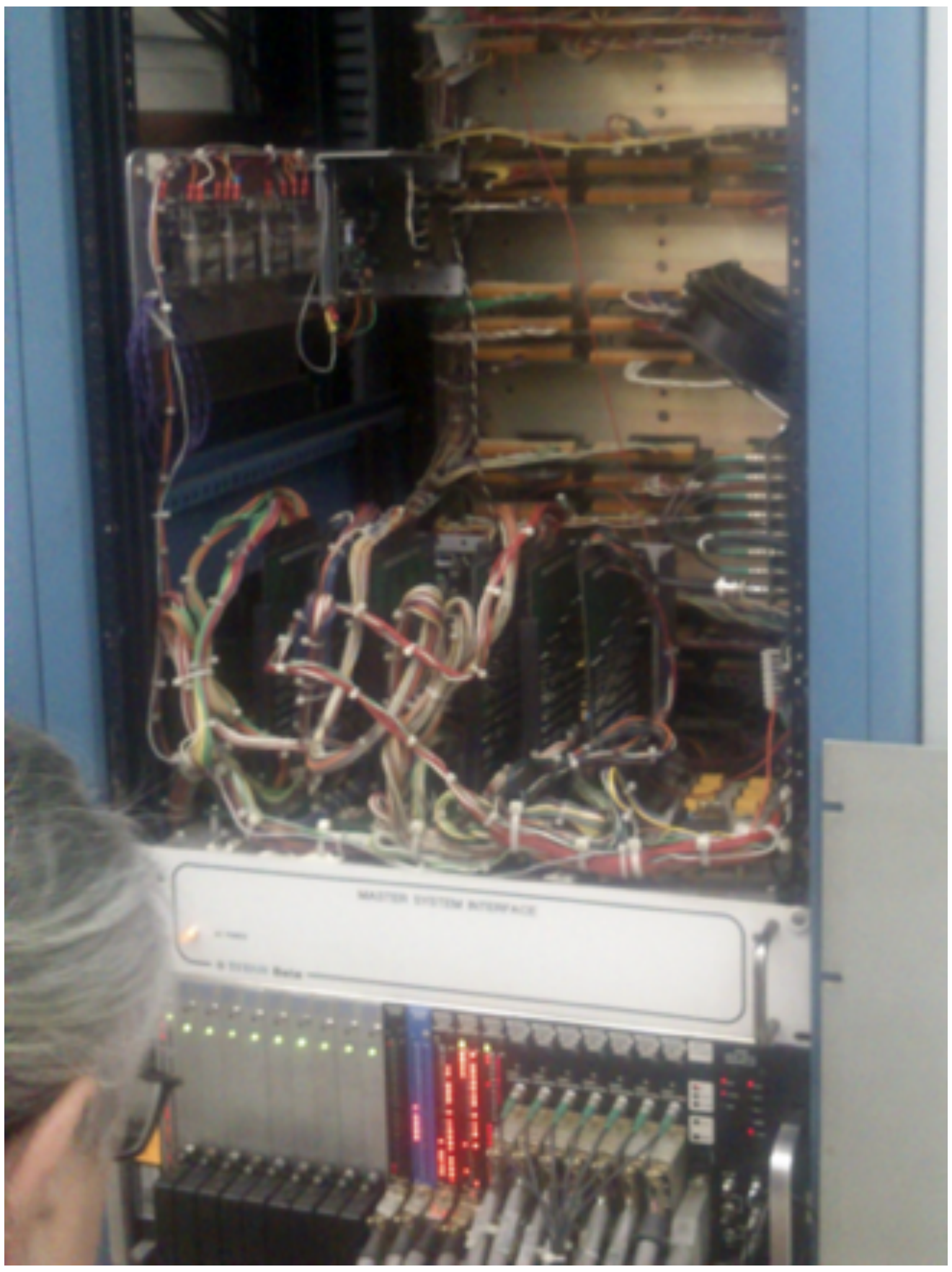

**Figure 3.8:** Interlock analog signals routed to the control custom-built board. All signals are readout by a CAMAC-bus system (visible at the bottom of the picture).

## 3.2 LINAC ENERGY UPGRADE

The ~60 m long DAΦNE LINAC, already described in Sec. 1.2, comprises 16 S-band (2856 MHz) constant gradient, travelling wave, 2/3π phase advance, SLAC type accelerating sections, made up of 86 cells, 3.5 cm long (for a total length of 304.8 cm), loaded with disks with variable iris radius, from 1.31 down to 0.96 cm. In Fig. 3.9 a schematic view of the sections structure is shown, while the main parameters are listed in Tab. 3.1.

Each section is fed from the upstream end with RF from the power stations (four, with the configuration shown in Fig. 1.2) and the power is dissipated onto a water-cooled load on the downstream end of the section (as shown in Fig. 3.10, right).

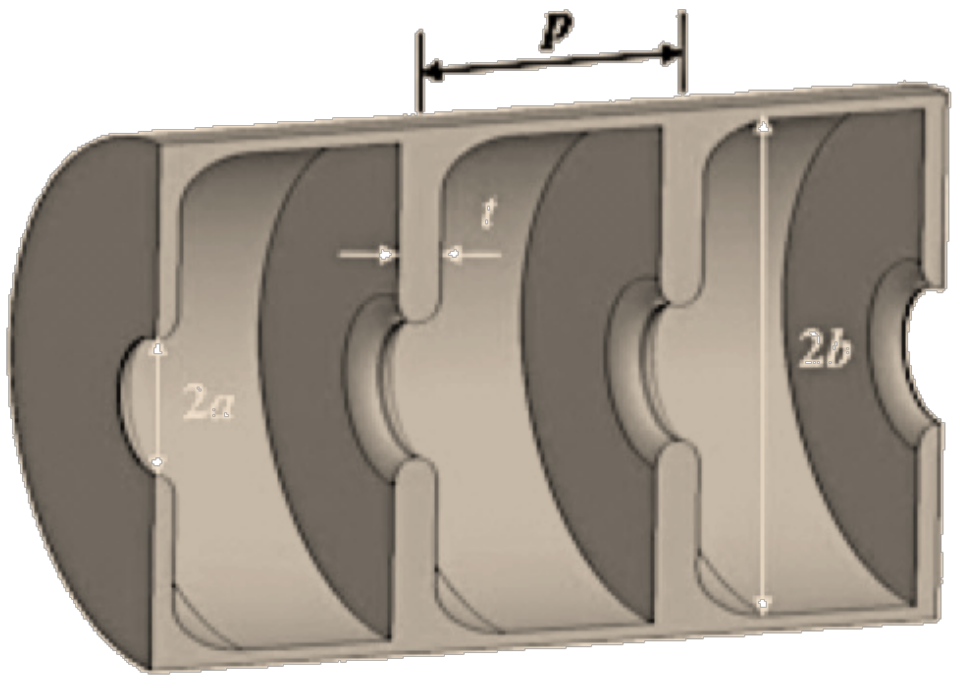

**Figure 3.9:** Scheme of the disk-loaded accelerating structure of radius *b* and cell length p; disks have thickness *t* and internal radius *a*.

| Parameter | Symbol | Unit | Value |
|---|---|---|---|
| Frequency | $\omega/2\pi$ | MHz | 2856 |
| Length | $L$ | m | 3.048 |
| Cell Radius | $b$ | cm | 4.17 – 4.09 |
| Iris Radius | $a$ | cm | 1.31 – 0.96 |
| Cell Length | $p$ | cm | 3.50 |
| Phase Advance per Cell | $\psi$ | - | 2π/3 |
| Disc Thickness | $t$ | cm | 0.584 |
| Quality Factor | $Q$ | - | 13,000 |
| Shunt Impedance per unit length | $r_l$ | MΩ/m | 52 – 60 |
| Filling Time | $t_f$ | ns | 830 |
| Group Velocity | $v_{gr}$ | % c | 2.0 – 0.65 |
| Attenuation | $\tau$ | “nepers” | 0.57 |
| Typical Unloaded Gradient | $G_0$ | MV/m | 21 |
| Typical Input Power | $P_0$ | MW | 45 |

**Table 3.2:** SLAC type accelerating section main parameters.

Different solutions for increasing the total energy of the LINAC have been studied in the last years, in particular in the framework of the proposal of the energy upgrade of the DAΦNE collider; a comprehensive summary is discussed by R. Boni in Ref. [24].

In this document, a simpler scheme is proposed, which just exploits the extra-length in the LINAC tunnel, of about 15 m, between the last accelerating section at the end of the LINAC and the three-way switch-yard (towards the damping ring, the BTF and the spectrometer line, at 0°, 3° and 6° respectively), presently used as a drift space, with four quadrupoles for preserving the beam optics.

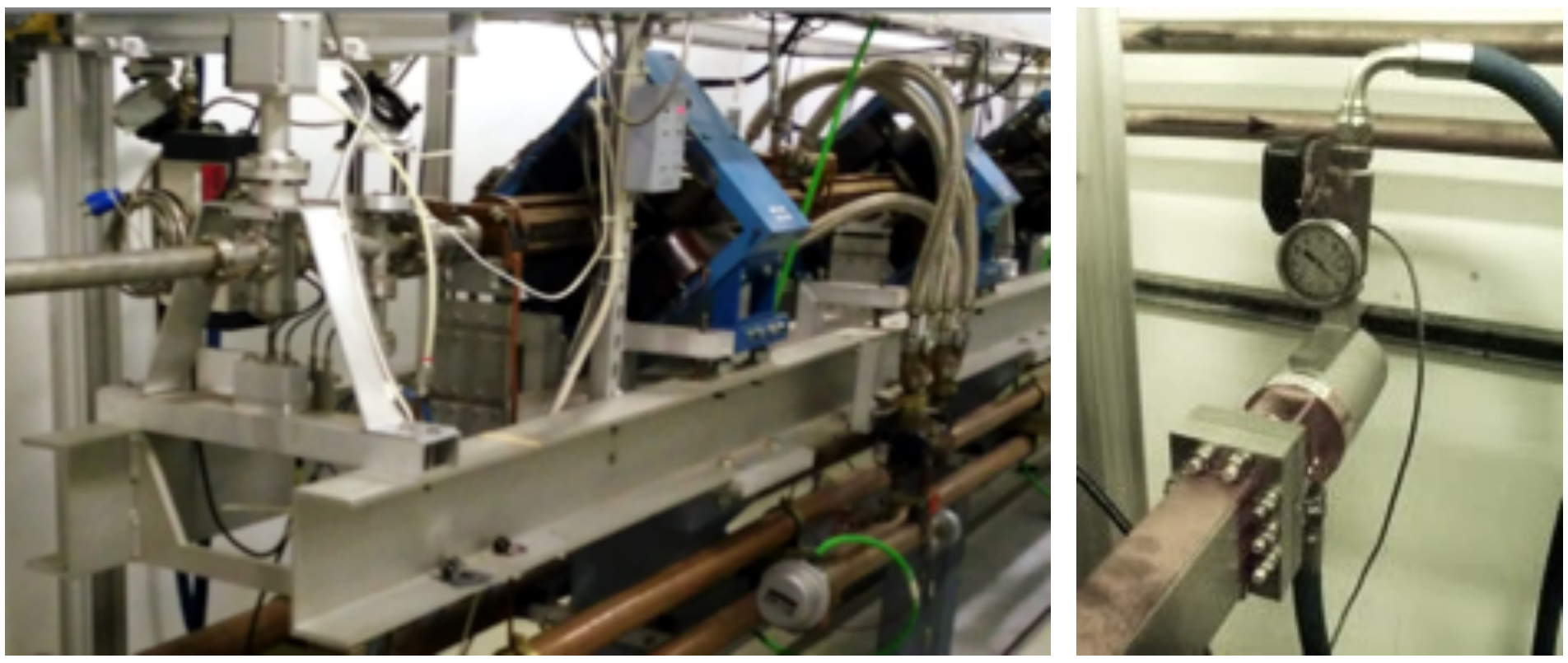

**Figure 3.10:** Left: the last SLAC type accelerating section at the end of the present LINAC is shown followed (from right to left) by the load, the diagnostics and vacuum pump; the two wrap-around quadrupoles (blue) and the cooling circuit pipes (bottom right) are also visible. Right: detail of the load with its cooling system.

This segment of the LINAC is shown in Fig. 3.11. By replacing this drift space with four 3 m long accelerating sections, identical to the other 16 ones, the total energy at the end of the LINAC would be increased, by an amount depending on the accelerating gradient, in turn determined by the RF power fed into those additional sections.

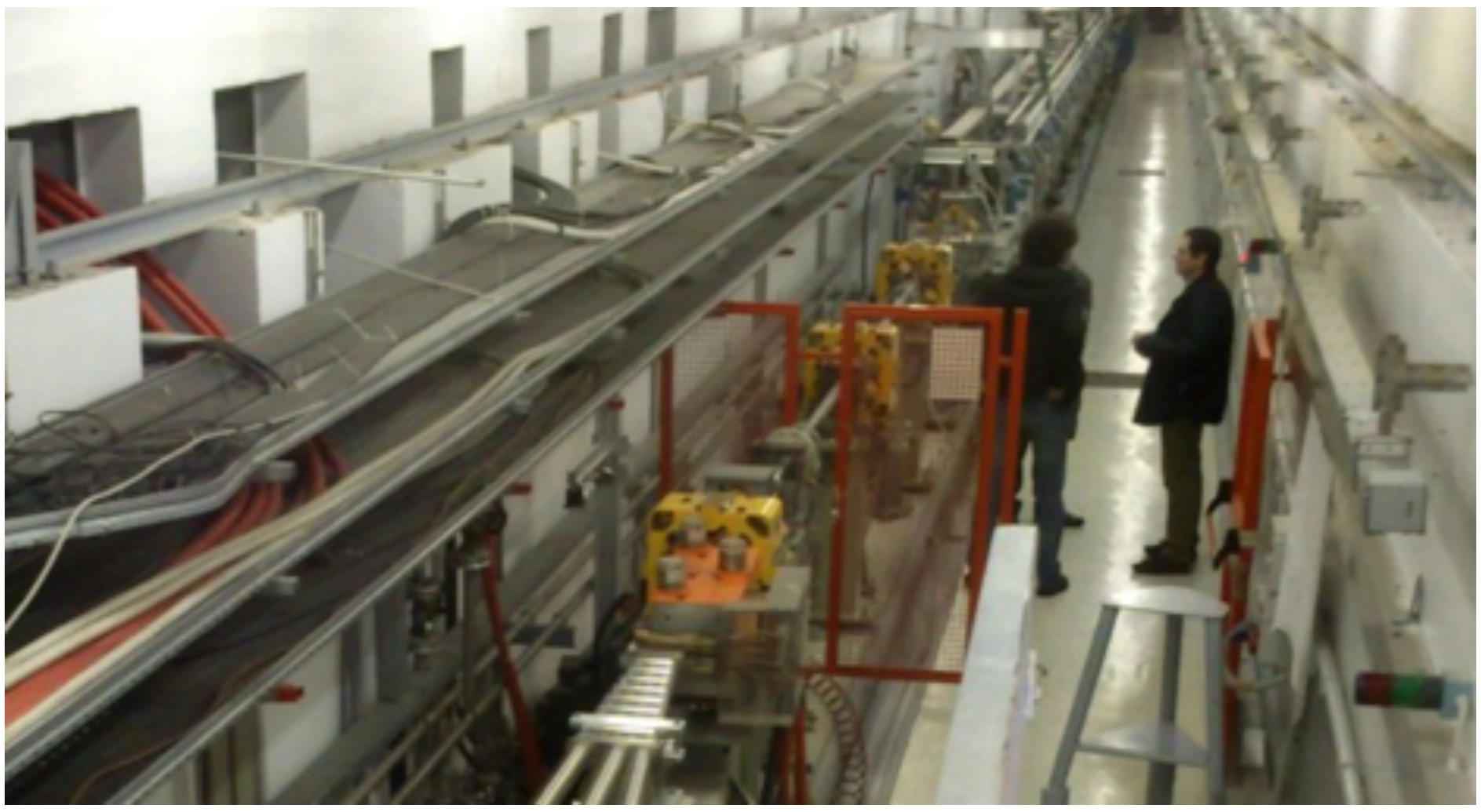

**Figure 3.11:** View of the LINAC tunnel from the three-way switch-yard, with the two pulsed magnets driving the beam towards the BTF line (3°, yellow dipole, more precisely a modified quadrupole) and the spectrometer line (6°, orange dipole), towards the LINAC (top of the picture). In between, a ≈15 m drift, with four quadrupoles (yellow). In the upper part of the picture, the passageways towards the parallel modulator tunnel for cables, pipes and waveguides, is visible.

The expression of the energy gain $U_0$ in travelling wave, constant gradient accelerating structures, can be derived from [25], as a function of the section parameters (using the symbols in Tab. 3.1) and the input power $P_{in}$:

$$U_0 = (1\text{-exp}(-2\tau))^{1/2}\cdot(P_{in}\, r_l\, L)^{1/2}$$

Substituting the values from Tab. 3.1, a practical formula can be used (the same used in [24]): $U_0 \cong 10.4\cdot P_{in}$ [MW].

The most conservative assumption for the accelerating field is to use the same RF configuration as in the last eight sections of the present LINAC: the power from each of the two klystrons (identical to the rest of the LINAC, i.e. Thales TH2128C), once compressed by the SLED up to 160 MW, is split twice with 3 dB couplers, in order to feed each section with about 40 MW, corresponding to a gradient of about 17 MV/m and an energy gain of 65 MeV/section. The additional four sections, fed by one more (a fifth) RF station (identical to the already existing four: klystron, powered by its modulator, plus SLED), would in this simple configuration give 260 MeV more, reaching 1 GeV for electrons and about 800 MeV for positrons.

The overall RF layout would be the one depicted in Fig. 3.12.

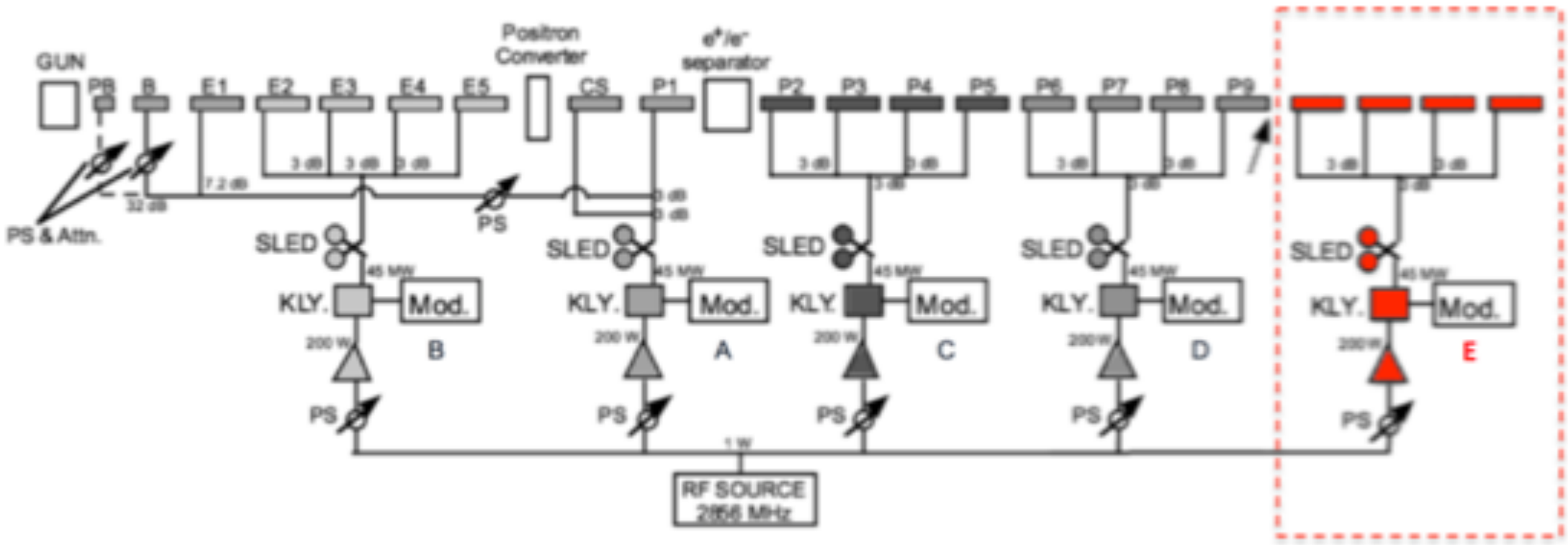

**Figure 3.12:** Proposed RF distribution scheme for the upgraded LINAC. The fifth RF station (modulator and klystron E) has a configuration identical to the one of the previous two stations: the additional modulator and SLED-ed klystron feeds the four new accelerating sections with a double 3 dB split of the power.

Two FODO wrap-around quadrupoles for each section, with the same parameters of the existing ones, should be realized: 25 cm magnetic length for a gradient of 610 G/m. The overall transport optics should be of course optimized including the additional portion of the LINAC. Also the diagnostics and services for the additional four sections should be of course installed and commissioned.

A more ambitious upgrade can be easily conceived by doubling the number of additional RF stations (again identical to the original four) and halving the RF power out of each SLED only once with a 3 dB splitter, in order to get an accelerating field similar to the one in the capture section, immediately downstream of the positron converter (see Fig. 1.2), of about 26 MV/m. In this case, the energy gain should reach 90 MeV/section.

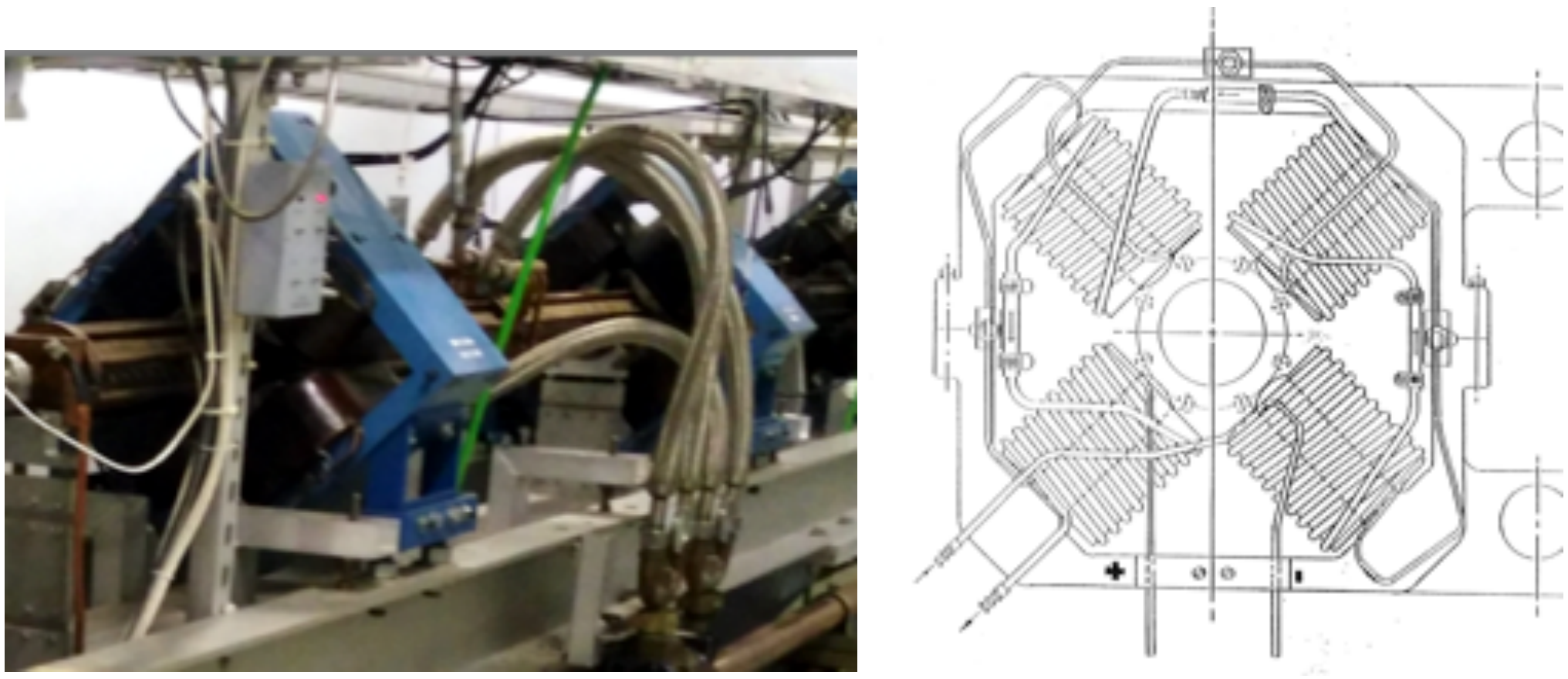

**Figure 3.13:** Two FODO wrap around quadrupoles are foreseen for each of the 3 m SLAC accelerating sections: picture of existing ones (left) and mechanical design (right).

The needed components for the baseline solution (WP2), i.e. maximum energy 1 GeV, four more sections, one additional RF station) are listed in Tab. 3.3, together with the cost estimate.

| **Item** | **N.** | **Unit cost (k€)** | **Cost (k€)** |
|---|---|---|---|
| SLED | 1 | 200 | 200 |
| Klystron | 1 | 200 | 200 |
| Accelerating sections | 4 | 100 | 400 |
| Waveguide network | 1 | 150 | 150 |
| Quadrupoles (LINAC) | 4 | 75 | 300 |
| Correctors (LINAC) | 8 | 10 | 80 |
| **Totale WP2** | | | **1330** |

**Table 3.3:** LINAC energy upgrade costs (WP2).

To the cost of the modifications we should add the resources necessary for the adaption of the dipoles (at least) driving the beam from the end of the LINAC down to the BTF experimental hall. All the DAFNE transfer lines, indeed, have been designed for a maximum energy of 800 MeV, so that some modification or replacement has to be taken into account in order to reach ≈1 GeV, in case the required magnetic field would bring the iron at saturation and/or the coils and power supply above the maximum power.

The minimum set of magnets to be checked working up to 1 GeV, and in case adapted or replaced, is (referring to the schematic layout in Fig. 1.12:

- Pulsed magnet at 6° DHPTS01 and 60° dipole DHSTS01, for the LINAC energy measurement
- Pulsed magnet at 3° DHPTB101 and 45° dipoles DHSTB01 and DHSTB02.

We are currently evaluating the possibility of modifying the larger dipoles, DHSTB01 and DHSTB02, slightly reducing the present gap h=42 mm down to the range around 30 mm, in order to reach a high enough field and, as a consequence, the required bending power at 1 GeV. This would require to modify the beam pipe, reducing the height inside the gap of those dipoles, but this should not impact the transport of the beam, since the optics of the BTF line can be adjusted in such a way that the vertical dimensions of the beam are reduced at the two dipoles.

Studies on the modification of the poles of the two magnets are currently under way, but preliminary results show that soft iron extensions of the order of 5 or 6 mm on each pole, properly machined and fixed, should be adequate for reaching the required field (see Fig. 3.14).

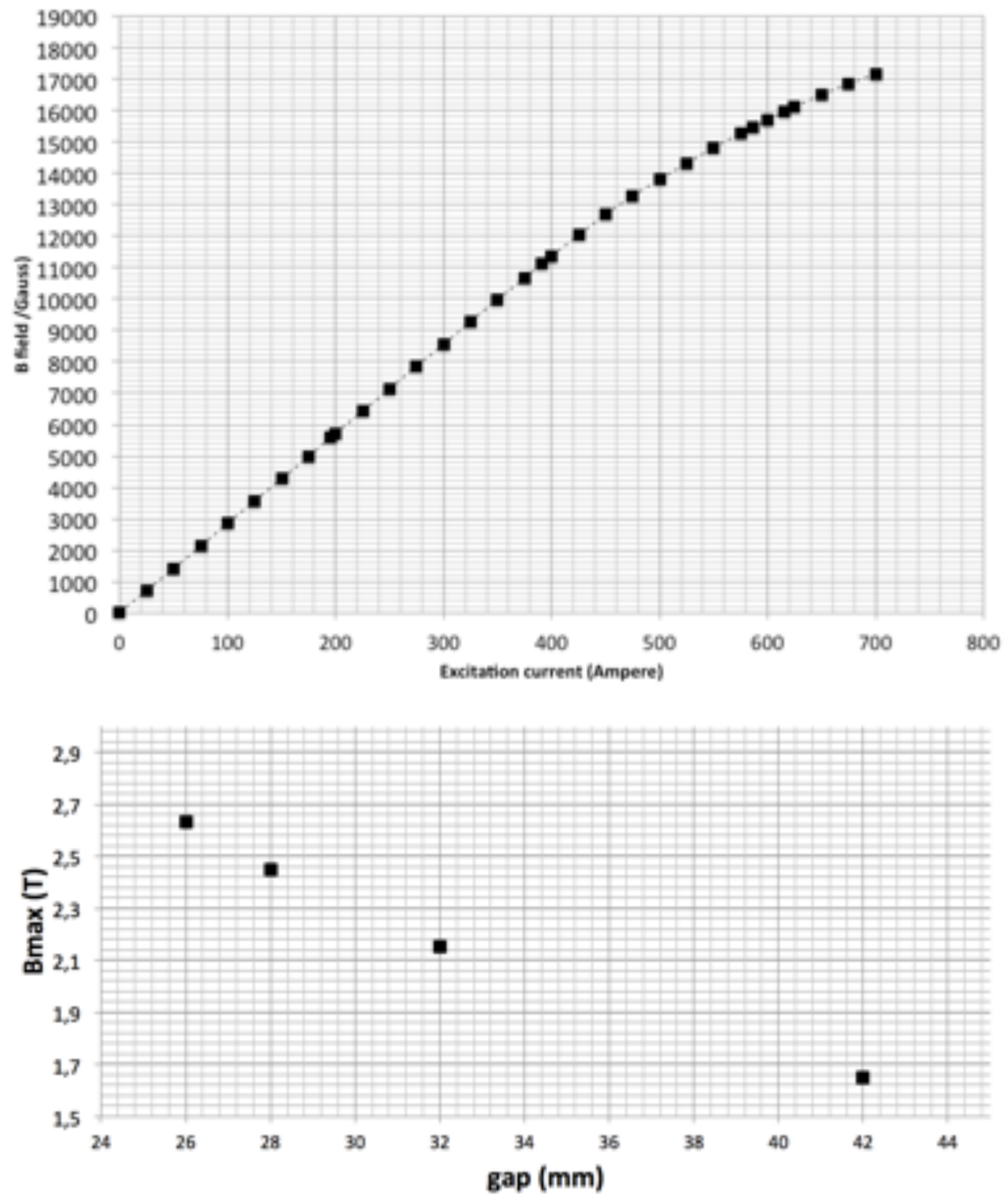

**Figure 3.14:** DHSTB dipole excitation curve (top); calculation of magnetic field as a function of the pole gap (bottom).

After having verified the overall layout, and having duplicated the RF station from one of the two in the “high energy” part of the LINAC, the dismounting of the 15 m drift will be followed by the installation of the four accelerating sections, that should be then connected to the waveguide network. The four new accelerating sections will also need:

- Mechanical supports;
- Vacuum and cooling of sections, waveguide network and loads;
- Controls;
- Quadrupoles and power supplies;
- Diagnostics (BPM, flag).

The extended LINAC will then need a dedicated period for the commissioning and optimization at higher energy.

Concerning the vacuum system, a preliminary layout of the extended LINAC, including the new pumps, valves and connections for the fifth RF station and the new accelerating sections is schematically shown in Fig. 3.15.

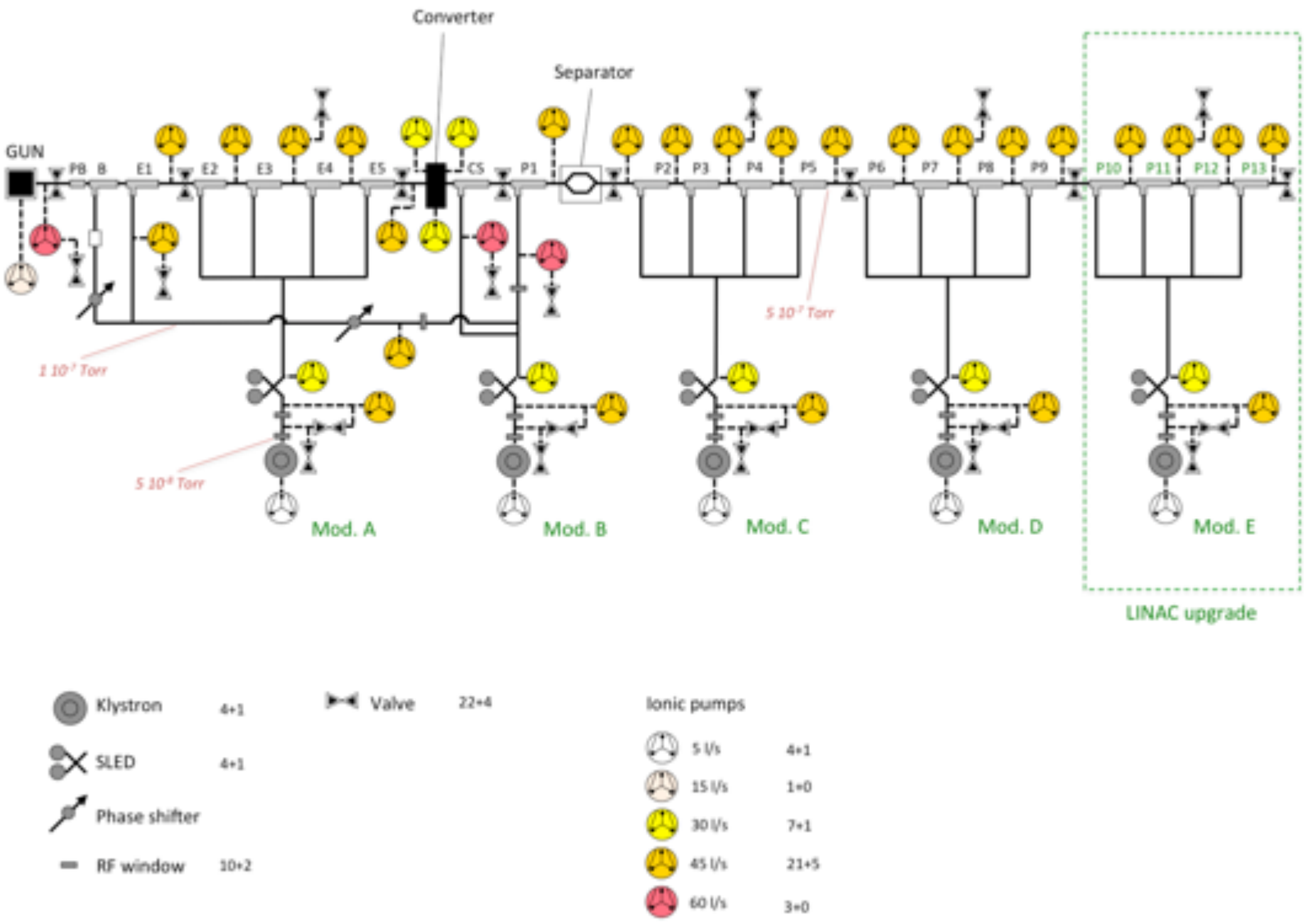

**Figure 3.15:** Extended LINAC vacuum system preliminary layout.

## 3.3 NEW BEAM LINES

The third area of intervention is the arrangement of the BTF beam-line. In particular WP3 has the two-fold objective of having:

- A second useful beam-line;
- A second experimental hall;

The present configuration, shown in Fig. 3.16, practically makes available only a useful beam exit, along the main axis of the unique BTF experimental hall. The useful experimental area, marked with the dashed purple rectangle in Fig. 3.15, is the one presently used for the beam-tests, and equipped with the remotely controlled trolley, Medipix and GEM diagnostics, scintillator hodoscope, etc. (see Fig. 3.16).

The exit at the end of the straight beam line, pointing towards the lateral wall of the hall, is currently used for high-intensity electron runs, typically with the neutron production target (described in Sec. 2.3), and as photon exit with the tagged photon source (described in Sec. 2.2). A picture of the present BTF hall showing these two beam exits is shown in Fig. 3.17.

During the 2015 summer shutdown the movable blocks shielding the experimental hall have been dismounted and repositioned in a different configuration, in order to allow mounting a series of concrete bin blocks to improve the shielding in the upward direction. The new shielding configuration of the BTF hall is shown in Fig. 2.10.

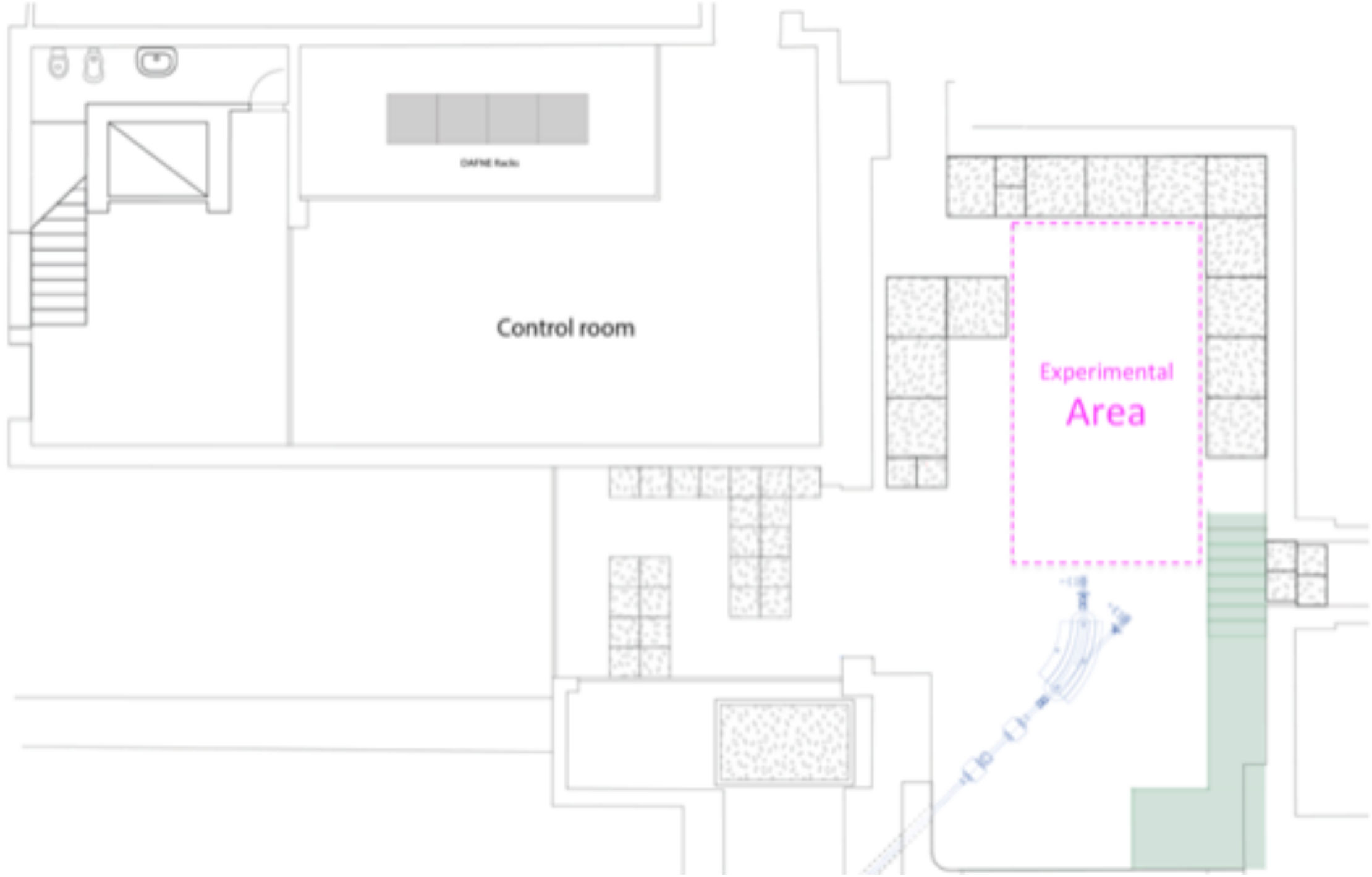

**Figure 3.16:** Present layout of the BTF beam-line, experimental hall and control room.

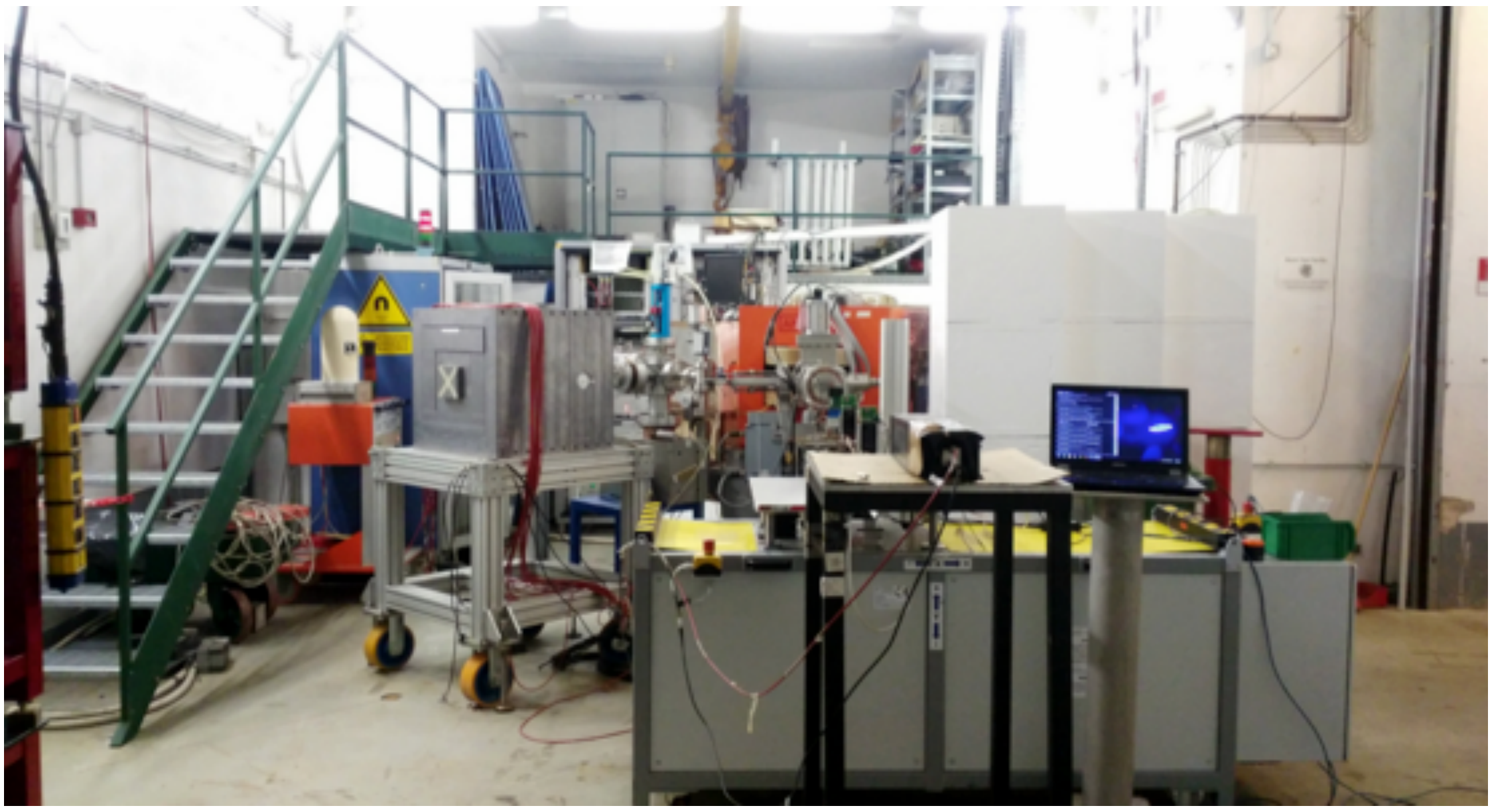

**Figure 3.17:** Picture of the present BTF beam-line and experimental, showing the useful user area at the exit of the DHSTB02 45° dipole (orange), equipped with diagnostics and remotely controlled trolley. The lead/polyethylene cubic shielding of the neutron production target is also visible at the exit of the straight beam pipe (on the left).

The idea for the new layout is schematically shown in Fig. 3.18: a beam-splitting dipole, wrapped around a double-exit pipe, can drive beam pulses from the upstream BTF beam-line alternatively to the two new lines. In case, the dipole can be connected to a pulsed power supply for a fast switch between the two lines.

In order to have sufficient space and to allow bending the beam towards the former control room, this dipole has to be placed as close as possible to the entry point of the BTF vacuum-pipe in the present experimental hall.

One line will be turned by 45°, driving the beam along the main axis of the room, using the existing DHSTB02, in a configuration very similar to the existing BTF line. The available space will be only slightly reduced in the transverse direction with respect to the beam ("Area 1" in Fig. 3.18), to a useful surface of about 3×6 $m^2$.

The second, new line, will be further bent in order to enter the former control room close to the intersection of the two perpendicular walls of the control room. A hole in the wall separating the two rooms should be realized (at a 45° angle). Since the entire area has a kind of L-shape, an additional angle will be needed in order to complete the 90° turn, thus directing the beam along the main axis of the second hall ("Area 2" in Fig. 3.18).

In order to preserve the beam quality, to this basic configuration quadrupoles and correctors should be added. In order to compute and optimize the optics of the two new lines, we have assumed to realize magnets essentially with the same parameters of the existing ones: for dipoles, a magnetic field in the same range as the existing DHSTB01 and DHSTB02 (identical) 45° bendings; for quadrupoles, the new elements have been modelled on the QUATB01-02 and QUAT03-04 doublets.

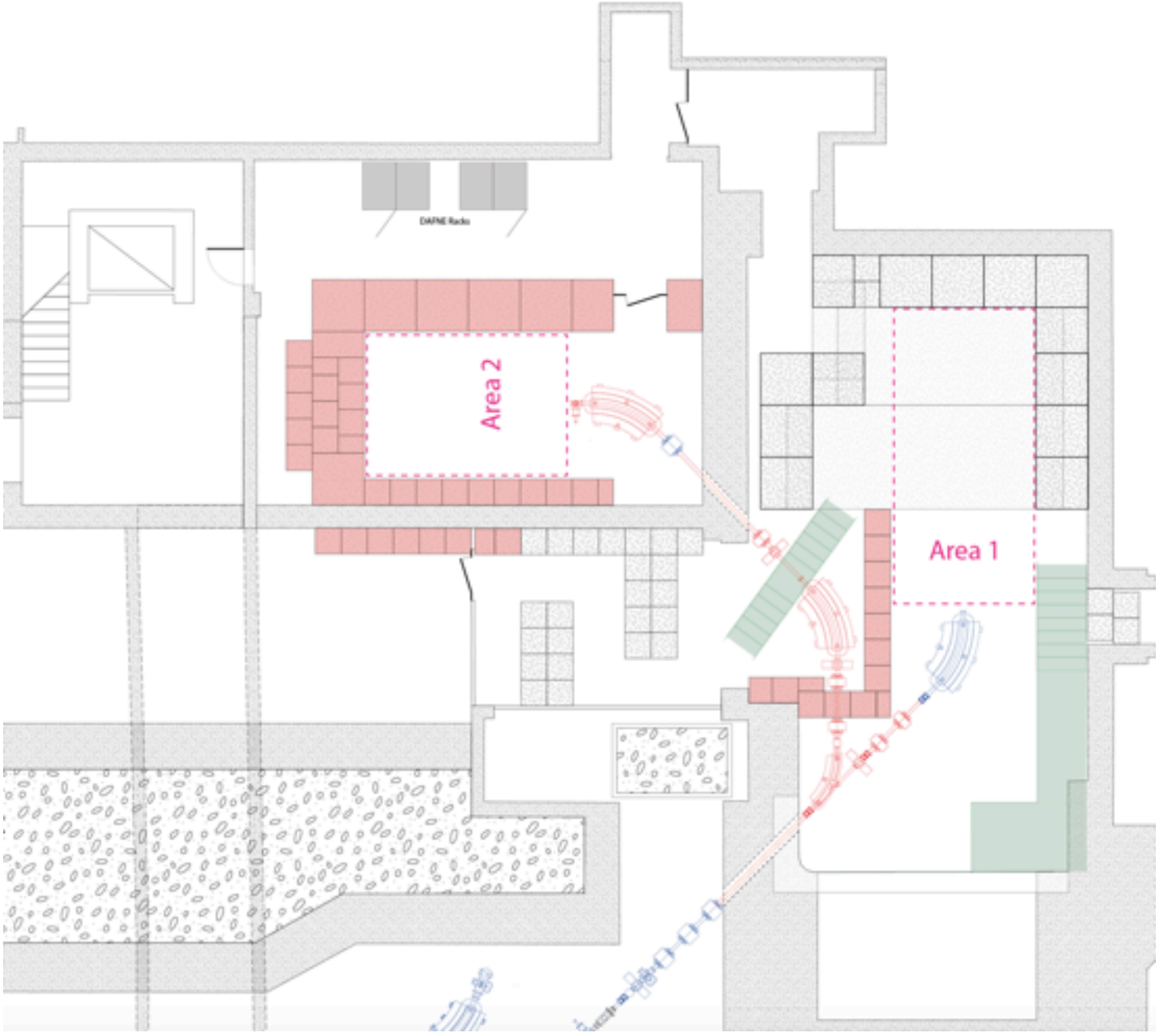

**Figure 3.18:** Proposed layout of the BTF beam-line, modified to have two lines, one in the existing experimental hall, and one serving the second area in the former control room.

In order to use the new experimental area, the present control room should of course be dismantled and moved in some other place, and an adequate shielding should be foreseen. Taking into account that the new room has to be shielded with concrete blocks as the existing one, and considering a depth of 1 m, the useful surface in this new hall will be of about 3×5 $m^2$. The new elements are shown in red in the schematic layout of Fig. 3.18.

One open issue is the presence of four racks hosting the readout of damping ring diagnostics, damping ring and BTF timing boards, network switches, power supplies and other ancillary systems of DAΦNE, that should be either moved in another room (with substantial re-cabling work) or protected with adequate shielding in order to allow access to the apparata. The resulting layout without those racks of course leaves more space in the new experimental shielded area, as shown in Fig. 3.19.

Another point is the possibility of enlarging the access to the new experimental hall, by modifying the stairs connecting to the upper floor of the building, as shown in Fig. 3.20.

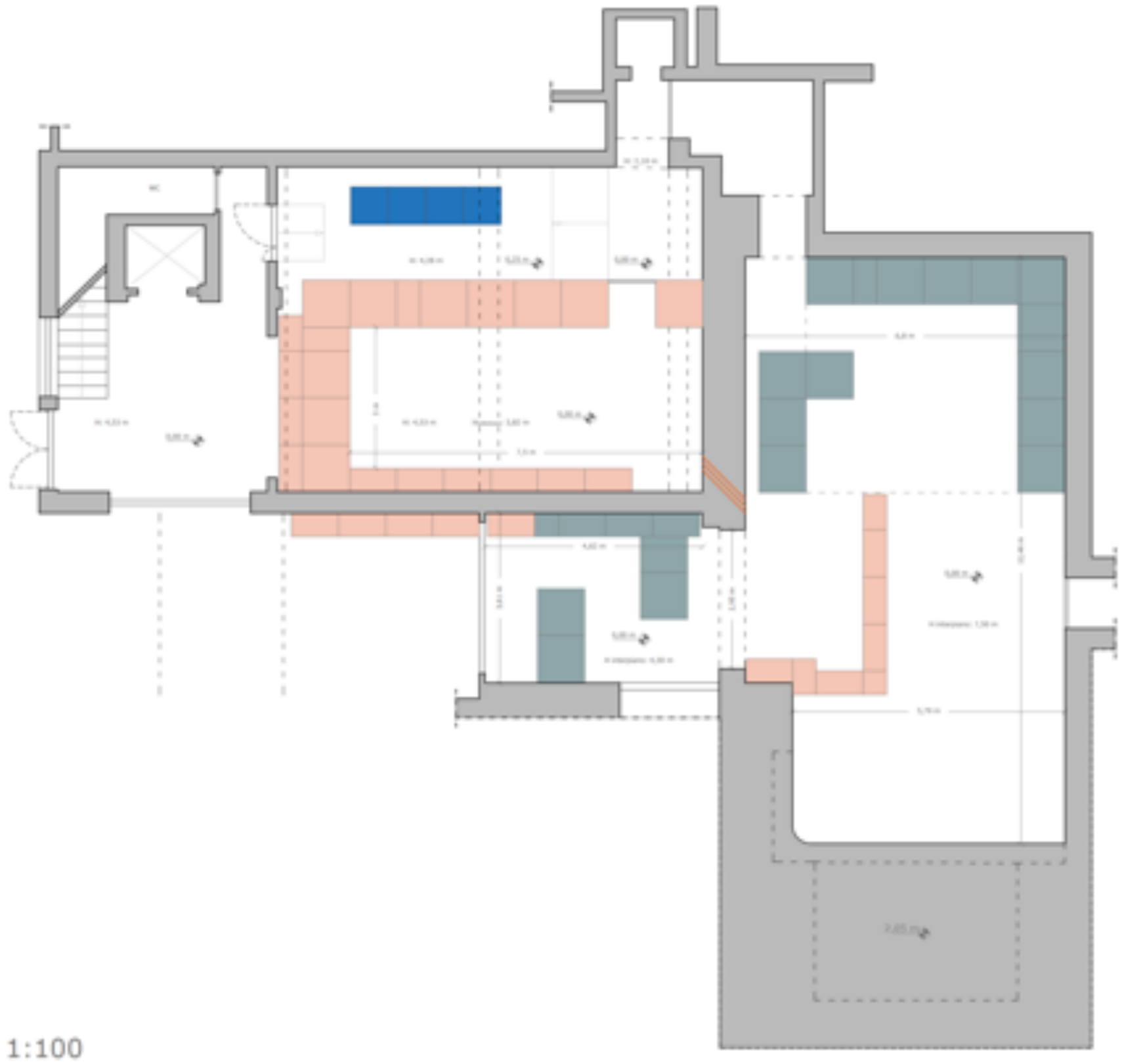

**Figure 3.19:** Proposed layout of the two new experimental areas, without moving the four DAΦNE instrumentation racks and without modifying the access to the building.

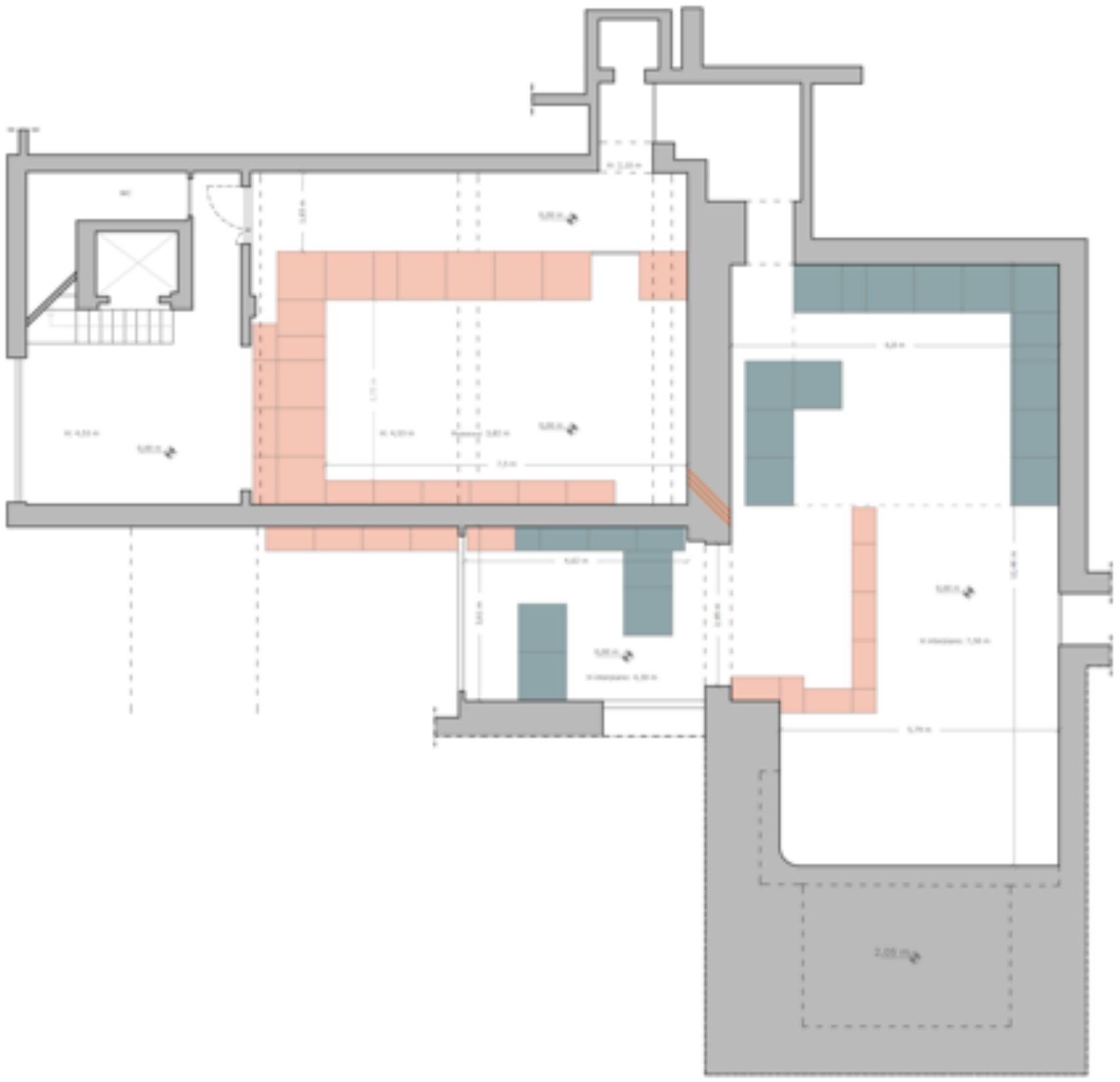

**Figure 3.20:** Proposed layout of the two new experimental areas, in the hypothesis of moving the four DAΦNE instrumentation racks and with enlarged access to the building.

Actually, a new control room for the BTF has been prepared and is ready to be used, refurbishing and adapting the control room in the past used for the ADONE accelerator before and for the FINUDA experiment, after the start of the DAΦNE project. This new control room, shown in Fig. 3.21, is on the upper floor with respect to the BTF area, very close to the main DAΦNE complex control room.

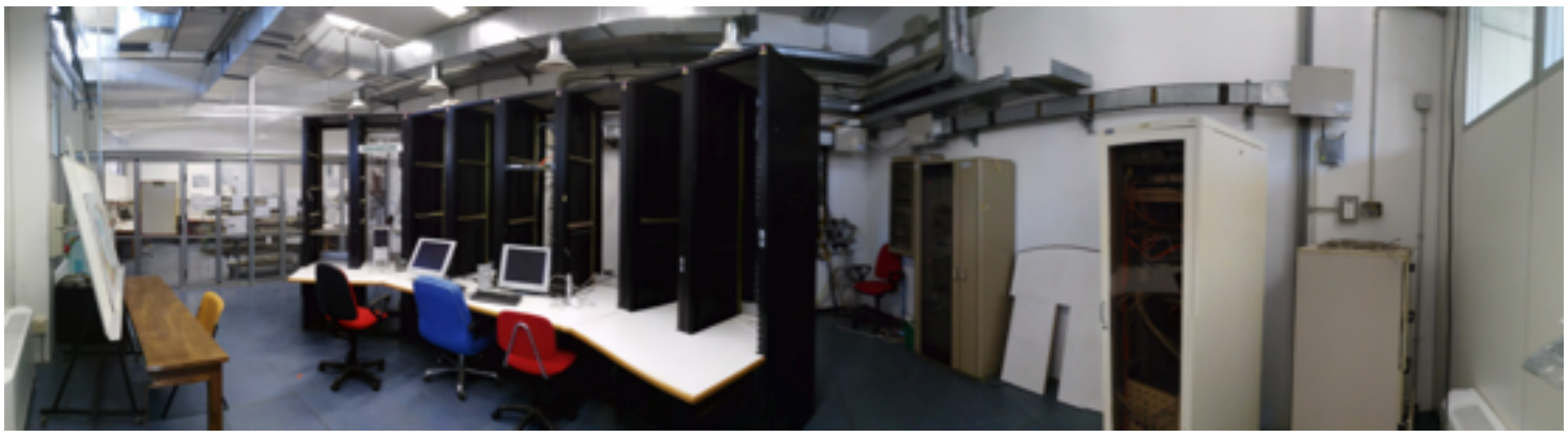

**Figure 3.21:** New BTF control room.

Even though this kind of layout is not the optimal one with respect to the use of the existing surface or from the point of view of the minimum amount of magnetic elements, the main advantages of this kind of layout are the following:

- It allows having two separate lines in the two separate rooms;
- The shielding the existing control room and a the same time, leave access (even with the beam) to the DAΦNE racks, hosting several diagnostics cabling and the main timing signals from/to subsystems of the accelerator.

- Allow using “Area 1”, under the new concrete bin ceiling, even when “Area 2” is open for letting the users access their setup.

Given the schematic layout shown in Fig 3.19 and assuming the same specifications as the existing magnets for the new elements, we have calculated and optimized the optics and the detailed layout of the new lines, using both the G4-beamline tracking code and the MAD-X software.

The simulated layout is shown in Fig. 3.22. The elements upstream the SLTB04 collimator (i.e. the last horizontal collimator after the energy selection) reproduce the present situation. All modifications are applied to the elements downstream.

We have indicated:

- with NEWD1 the new 45° bending downstream of a new quadrupole triplet substituting the two doublets (QUATB01-04);
- with BTF1 the line developing in the original experimental hall (very similar to the existing one), reusing the DHSTB02 dipole at 45°;
- with BTF2 the line going inside the former control room, after a 90° bending by means of the new dipoles NEWD2 and NEWD3.
- Additional collimators NEWSLT1 and NEWSLT2 have been also foreseen, in order to optimize the beam spot in the two new lines.

The resulting beam spot and divergence at different points along the original line and the new two lines are computed, in particular five intermediate positions and the two output ones have been considered (DIAG1 to DIAG5 and OUT1 and OUT2 in Fig. 3.22). Some typical results of G4-beamline simulations are shown in the following Fig. 3.23.a-d in these seven positions, for a given dataset for the magnetic and collimator configurations.

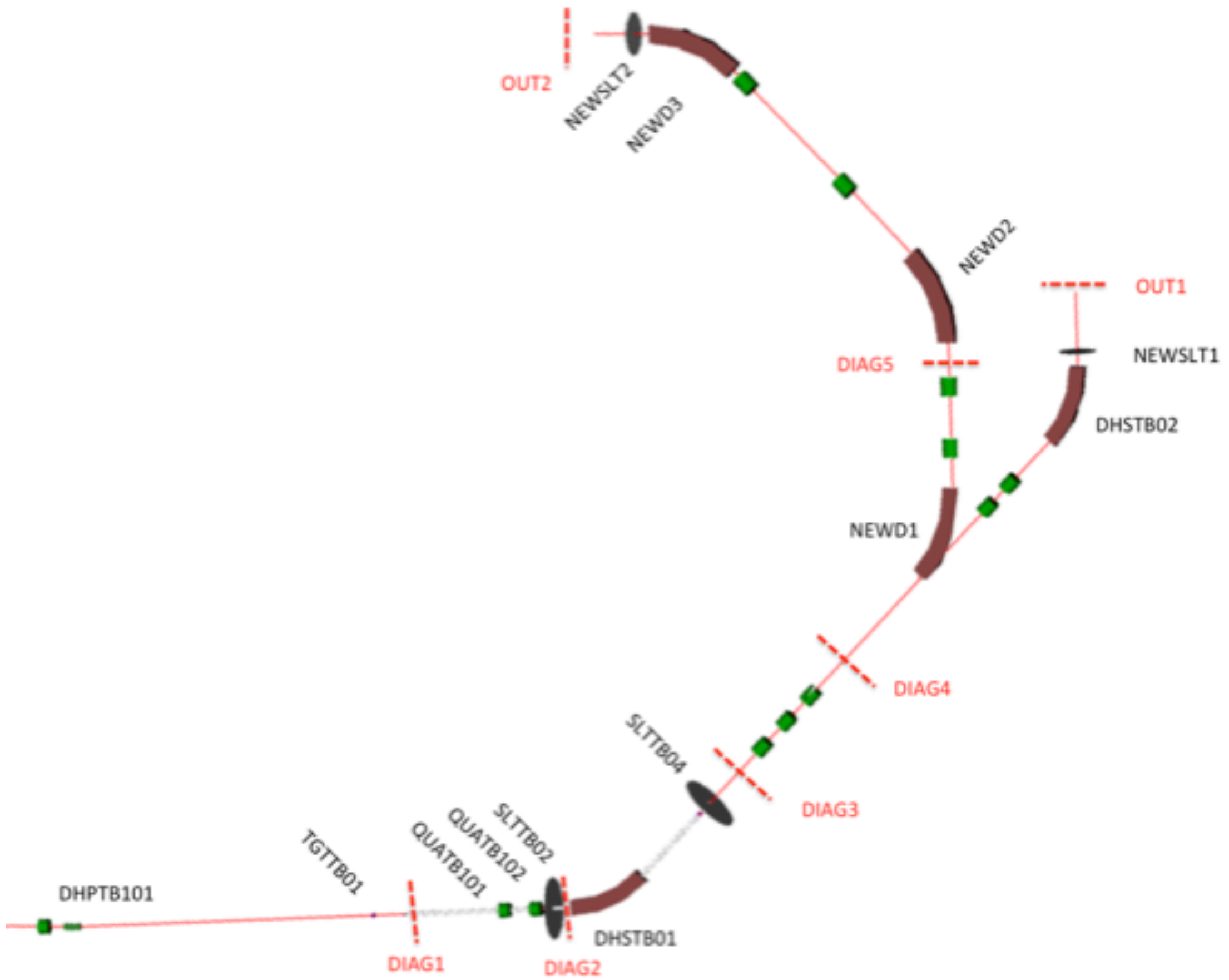

**Figure 3.22:** Graphical display of the G4-beamline simulation of the new beam-lines, showing the new elements and the detecting volumes (DIAG1 to DIAG5). The line coincides with the present line up to the DIAG3 position. The two final detecting volumes are OUT1 (BTF1 line, in the present area) and OUT2 (BTF2 line, in the new hall).

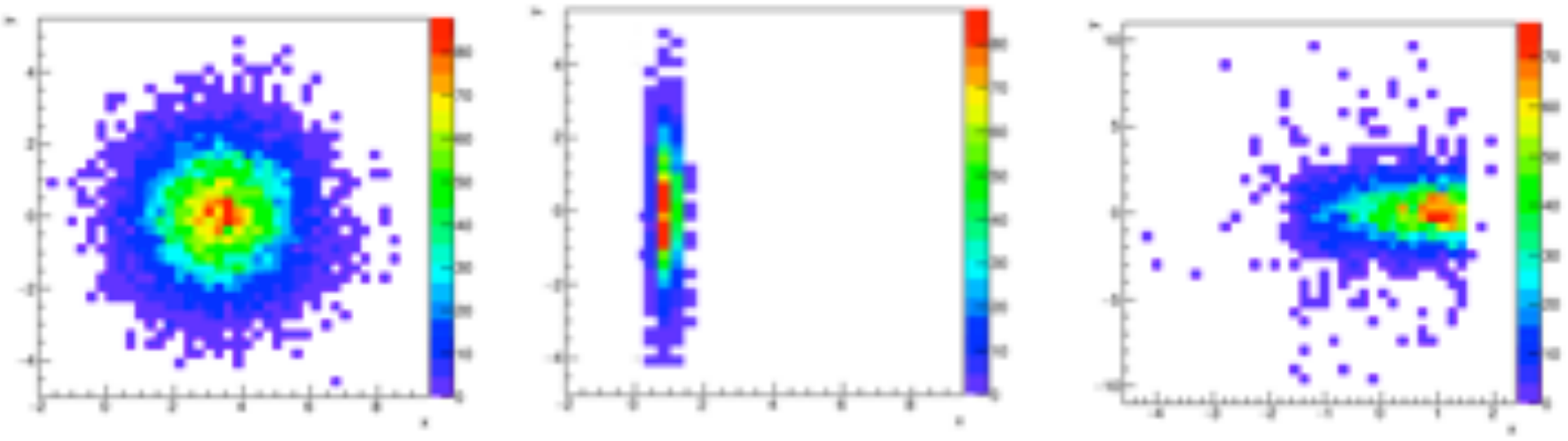

**Figure 3.23.a:** Example of transverse beam spots for 510 MeV electrons in the first part of the BTF line at the first three diagnostics positions: (from left to right) immediately after the TGTTB01 target (DIAG1) with a widespread and fully symmetric *x-y* distribution, immediately downstream of the first horizontal collimator SLTTB02 at the entrance of the energy selecting dipole DHSTB01 (DIAG2), and downstream of the second horizontal collimator SLTTB04 (DIAG3), showing the dispersive effect of the bending.

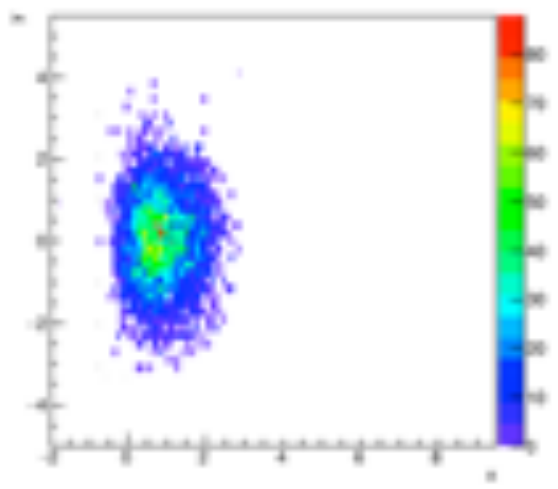

**Figure 3.23.b:** Beam spot for 510 MeV electrons immediately upstream of the splitting of the two BTF lines and downstream of the new quadrupole triplet (DIAG4), using the same dataset of Fig. 3.21.a.

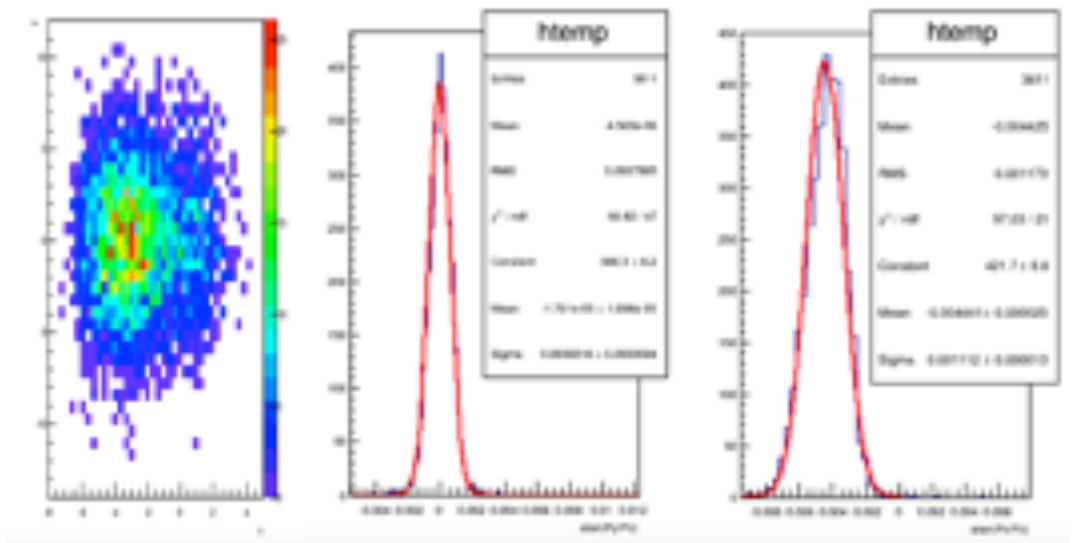

**Figure 3.23.c:** Example of transverse beam spots for 510 MeV electrons at the BTF1 line exit (OUT1 position), and beam divergence in $x$ ($\sigma_x$=0.7 mrad) and $y$ coordinates ($\sigma_y$=1.1 mrad), with the same dataset as Figg. 3.21.a and 3.21.b.

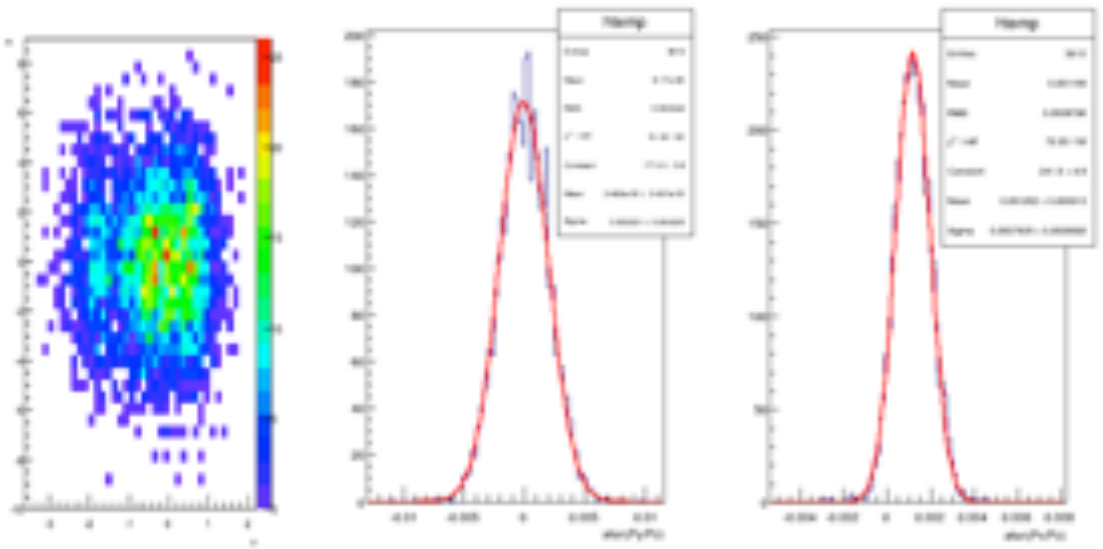

**Figure 3.23.d:** Example of transverse beam spots for 510 MeV electrons at the BTF2 line exit (OUT2 position), and beam divergence in $x$ ($\sigma_x$=2 mrad) and $y$ coordinates ($\sigma_y$=0.8 mrad), with the same dataset as Figg. 3.21.a and 3.21.b.

In order to validate the G4-beamline simulation of the new lines, the same framework has been used for implementing the present situation of the BTF, i.e. the layout shown in Fig. 1.12, applying a real configuration of magnetic fields and collimators opening, and then comparing the resulting beam parameters with the measured ones.

The beam spot, measured by means of the usual BTF Medipix detector, for 450 MeV electrons is compared with the G4-beamline simulation result, using the present BTF configuration and providing as input the settings of all elements. The simulated spot and measured one are shown in Fig. 3.24.

For the final optimization, and also as a cross-check of the G4-beamline results, the present BTF line and the two new beam-lines have been also simulated using the MAD-X code. The β function, dispersion and envelope in the three configurations are shown in Figg. 3.25.a and 3.25.b, for the usual reference energy of 510 MeV, showing consistent results with the G4-beamline output: the beam at the exit of the BTF1 and BTF2 new lines can be delivered with similar or event better parameters with respect to the existing line.

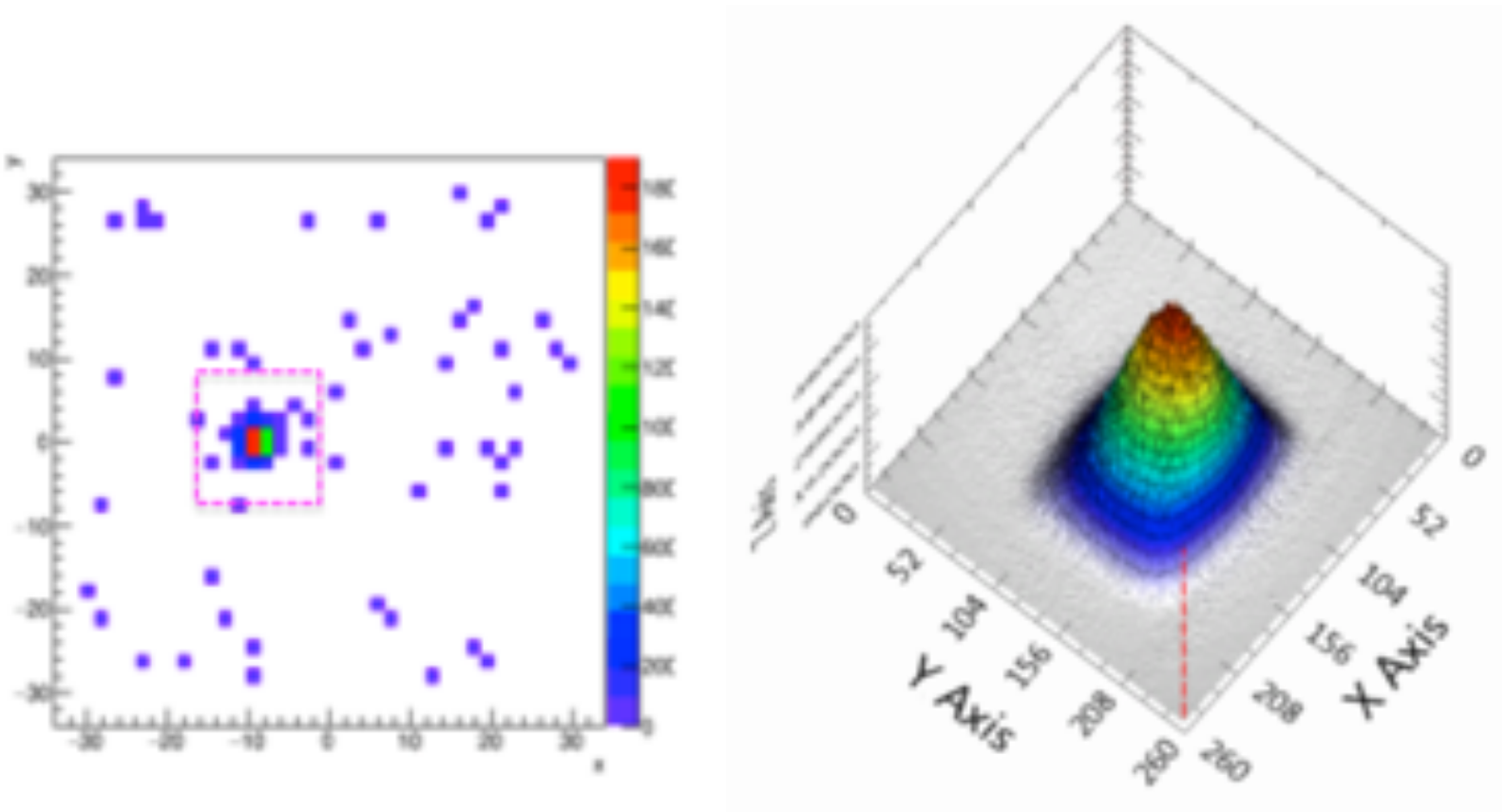

**Figure 3.24:** Left: beam spot simulated with G4-beamline for a 510 MeV positron beam and a real magnetic and collimator configuration, an area 14.5×14.5 mm$^2$, corresponding to the Medipix active area is shown by the purple square. Right: real measurement performed by means of the Medipix detector. In the two cases, similar values of about σ=2 mm in both coordinates are found.

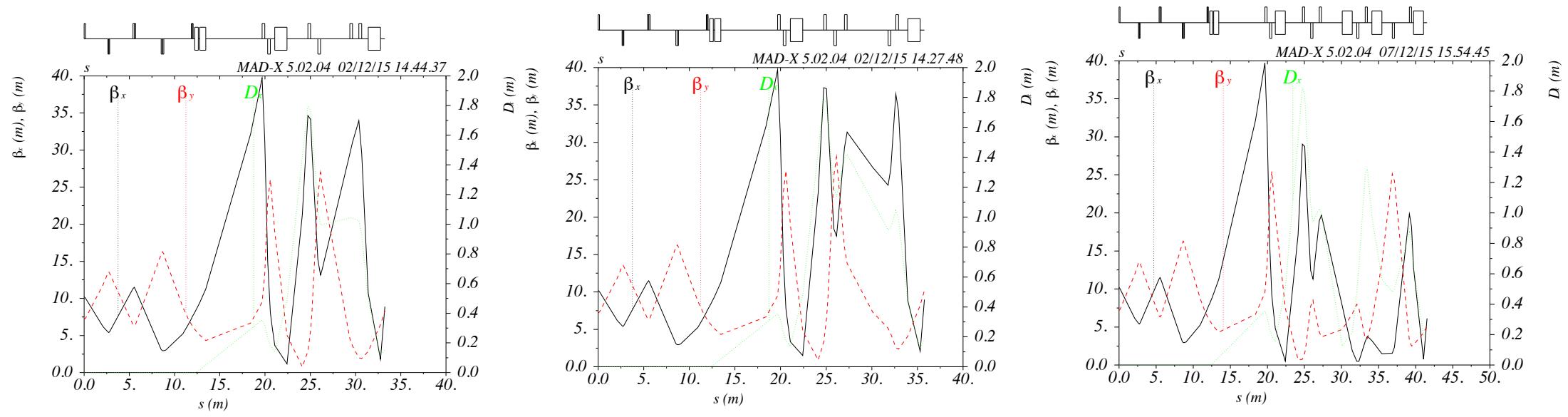

**Figure 3.25.a:** From left to right: MAD-X simulation results for the β function and dispersion present BTF, new lines “BTF1” and “BTF2” respectively.

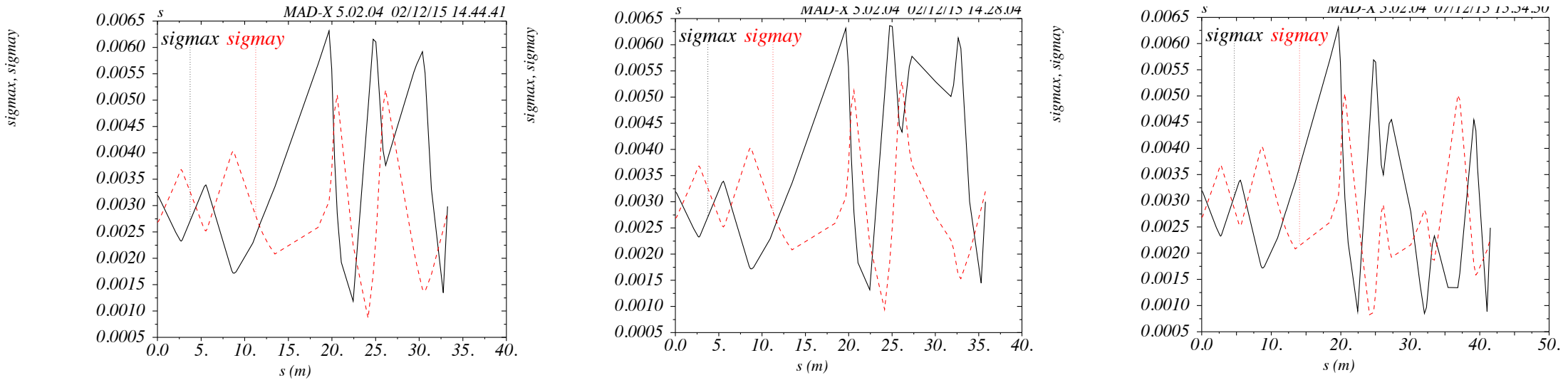

**Figure 3.25.b:** From left to right: MAD-X simulation results for the envelope ($\sigma_x$ and $\sigma_y$) of present BTF, new lines “BTF1” and “BTF2” respectively.

Once having validated the optics for the new lines configuration, a detailed layout should be designed, taking into account all the other elements of the beam-lines: vacuum pipe, mechanics and supports, vacuum valves and pumps, walls of the building and shielding blocks, etc.

A first sketch of the layout of the new BTF lines, including the existing elements is shown in Fig. 3.26.

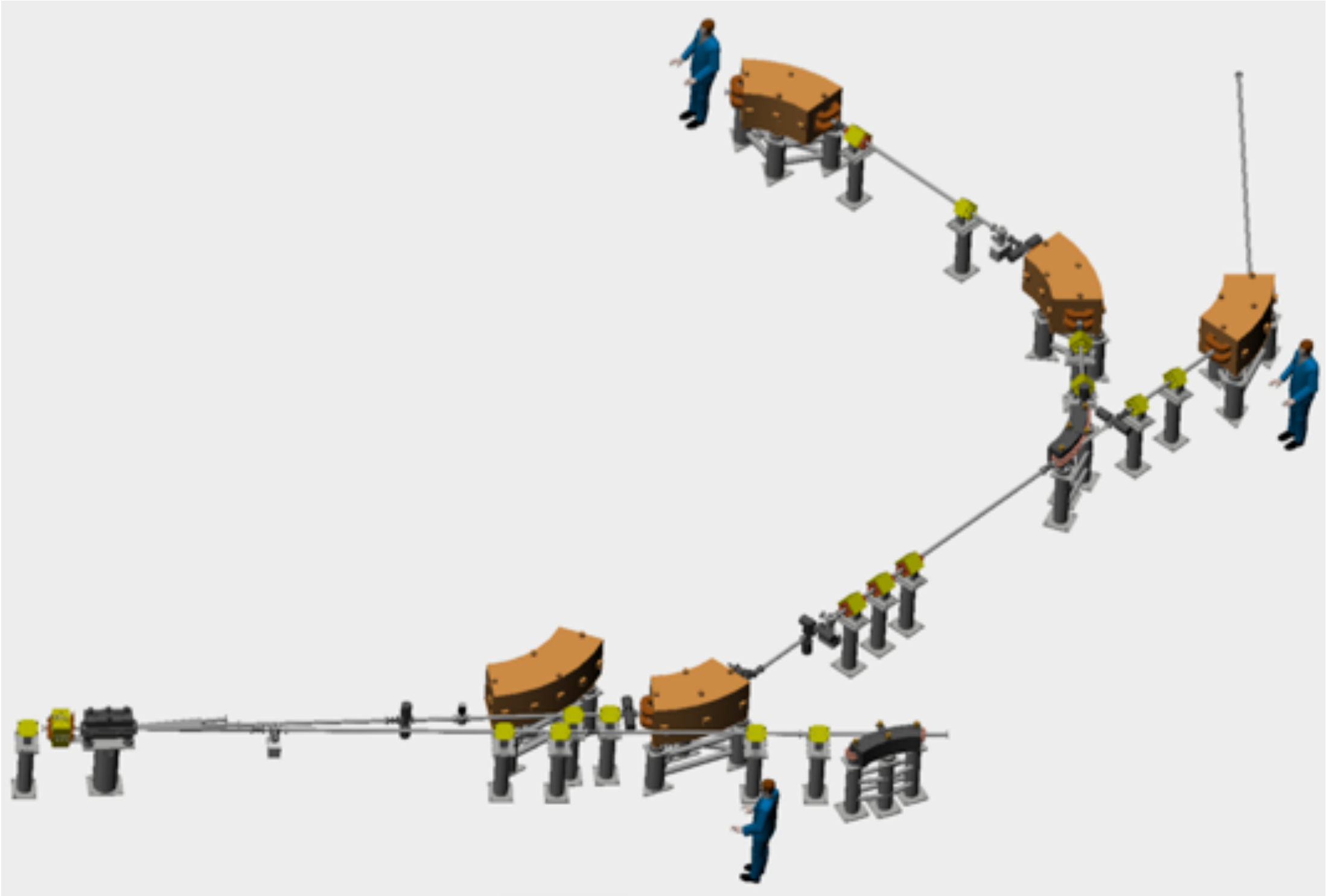

**Figure 3.26:** Proposed 3D sketch of the layout of the two new BTF beam-lines, walls and shielding removed.

| Element | | N. | Power (kW) |
|---|---|---|---|
| **Transfer lines** | QUATM01,02,03,04 (end of LINAC) | 4 | 4 |
| | DHPTS01 (pulsed) | 1 | 8 |
| | DHSTS01 (spectrometer) | 1 | 7,2 |
| | QUATM05,06,07 (transfer line) | 3 | 3 |
| | | | **22** |
| **BTF extension** | DHSTB01 | 1 | 20 |
| | QUATB101,102 (transfer line) | 2 | 2 |
| | QUATB01,02,03,04 (BTF) | 4 | 4 |
| | DHSTB02 | 1 | 20 |
| | | | **46** |
| **BTF 2nd line** | DHSTB03,04,05 | 3 | 60 |
| | QUATB05,05,07,08 | 4 | 4 |
| | PADME dipole | 1 | 35 |
| | | | **99** |

**Table 3.5:** BTF new lines power budget.

Even trying to reuse all the existing elements, new dipoles and quadrupoles (with their power supplies, cables, supports and services), and new vacuum pipe segments (even though for a moderate additional length of about 10 m), should be designed, procured, installed and tested. In addition, the power and cooling services should be upgraded in order to cope with the additional magnets (see Tab. 3.5).

### 3.3.1 Radioprotection studies

A preliminary study of the radio-protection issues has started, in order to assess the prescription, in terms of shielding and allowable beam intensity in the new hall. New possible layouts of the experimental hall and shielding walls have been simulated using FLUKA. In particular, a baseline solution of 1 m concrete shielding all around the room (see Fig. 3.27), with additional concrete in the forward direction.

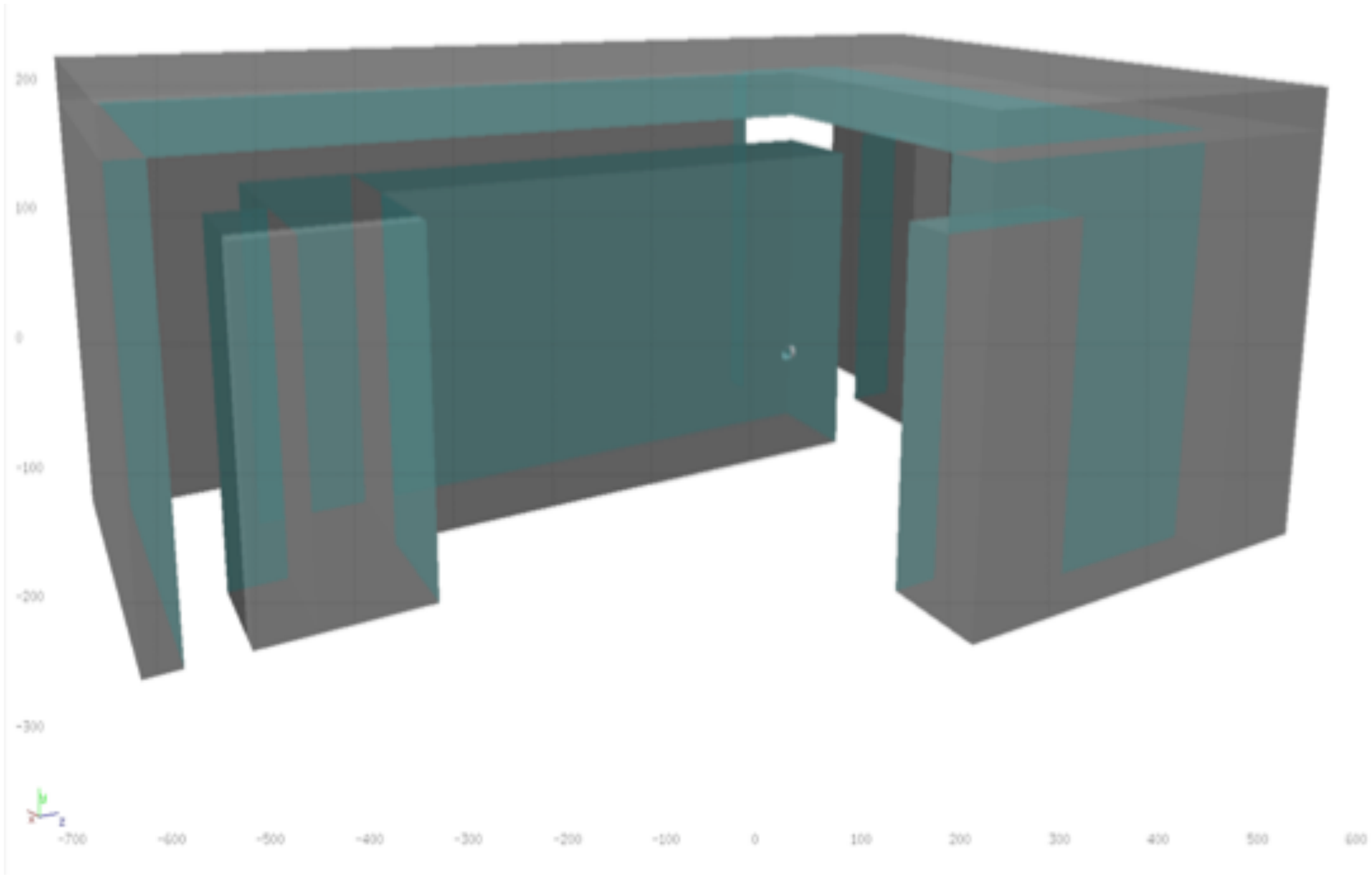

**Figure 3.27:** 3D layout of the two new BTF experimental hall walls and shielding (FLUKA simulation).

Preliminary results for the 3D dose rate maps are shown in Fig. 3.28.a, b and c for 550 MeV, 800 MeV and 1 GeV primary electrons respectively.

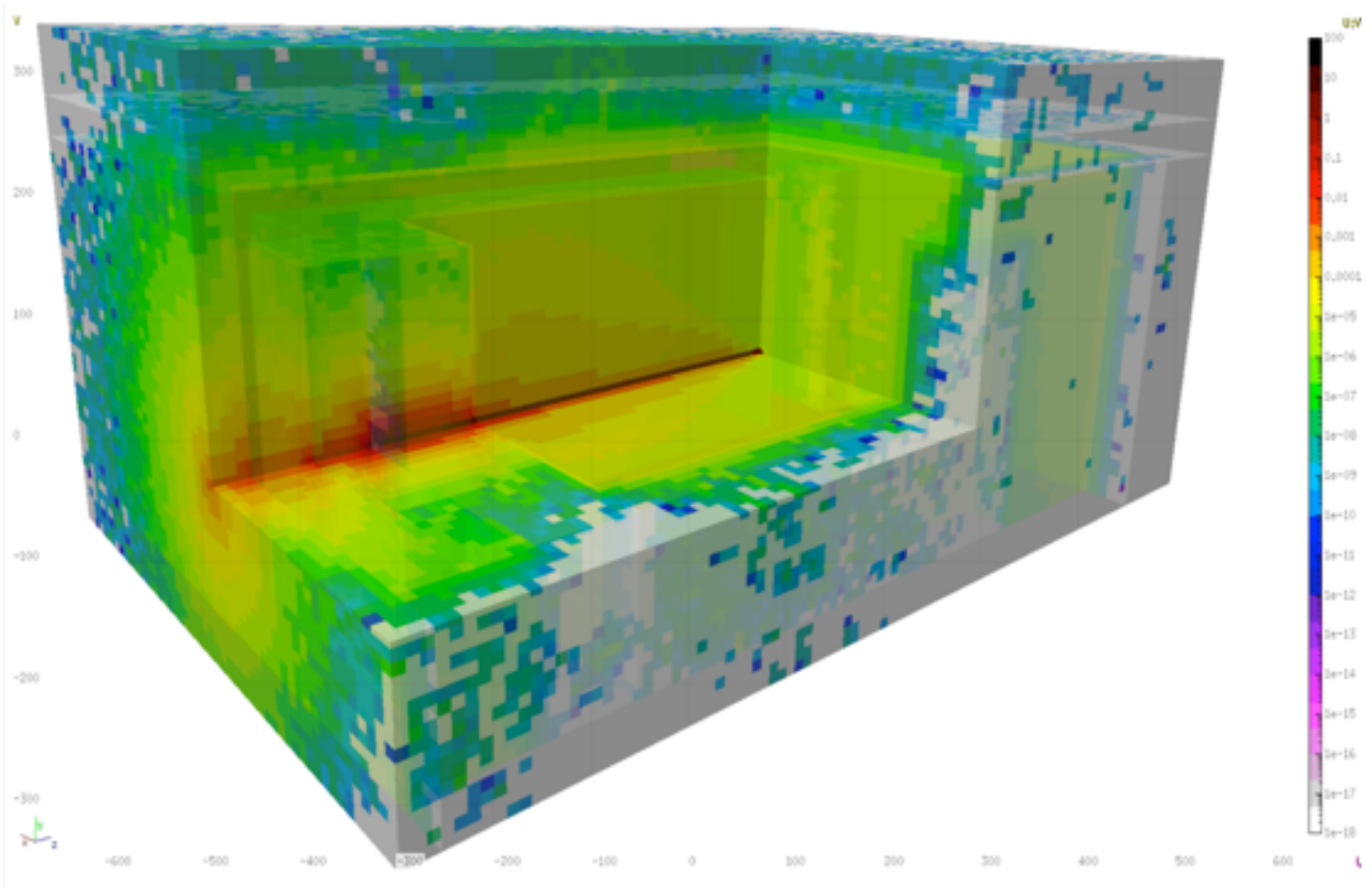

**Figure 3.28.a:** 3D dose rate map (pSv/primary) for the new BTF experimental hall, simulated in FLUKA, electron beam, 550 MeV energy.

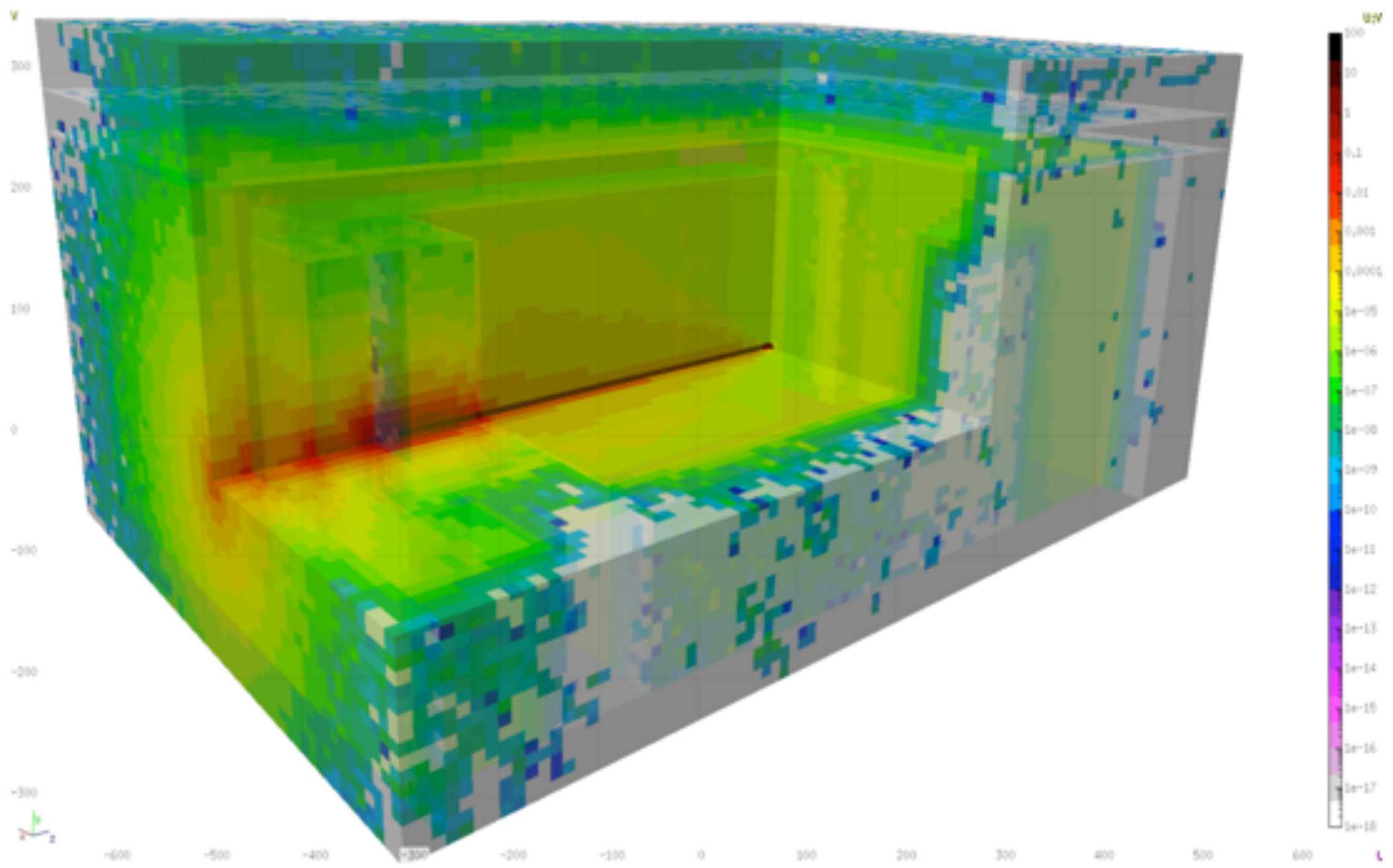

**Figure 3.28.b:** 3D dose rate map (pSv/primary) for the new BTF experimental hall, simulated in FLUKA, electron beam, 800 MeV energy.

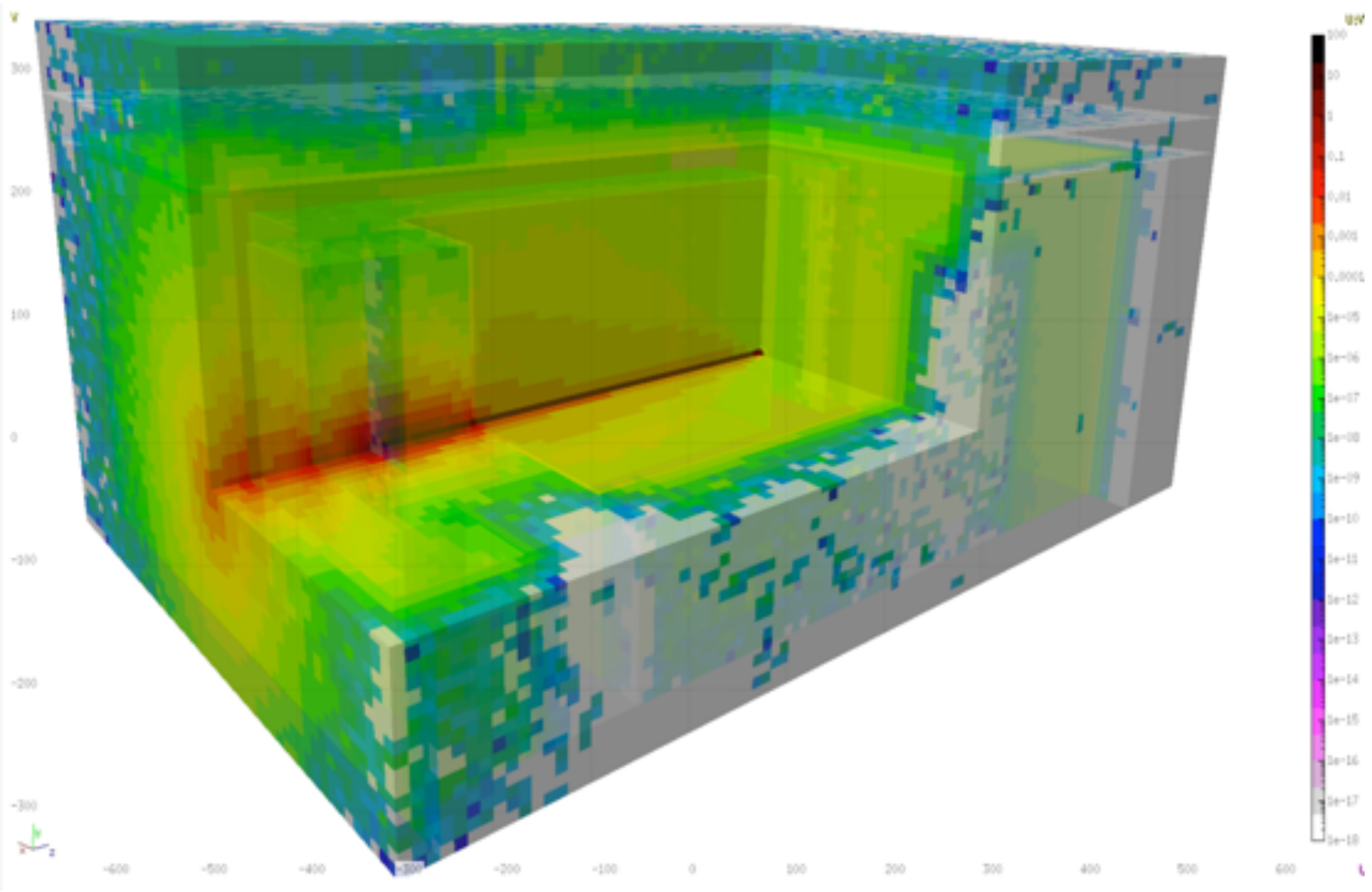

**Figure 3.28.c:** 3D dose rate map (pSv/primary) for the new BTF experimental hall, simulated in FLUKA, electron beam, 1 GeV energy.

Dose rates have been estimated behind (1 cm distance) the front shielding wall (Fig. 3.29) and behind the lateral wall (Fig. 3.30), for the same three primary energies, showing that even with the higher energy and in the very forward direction, a dose rate at the level of 0.05 pSv/primary is expected, corresponding to <5 nSv/h/primary for 24 Hz operation.

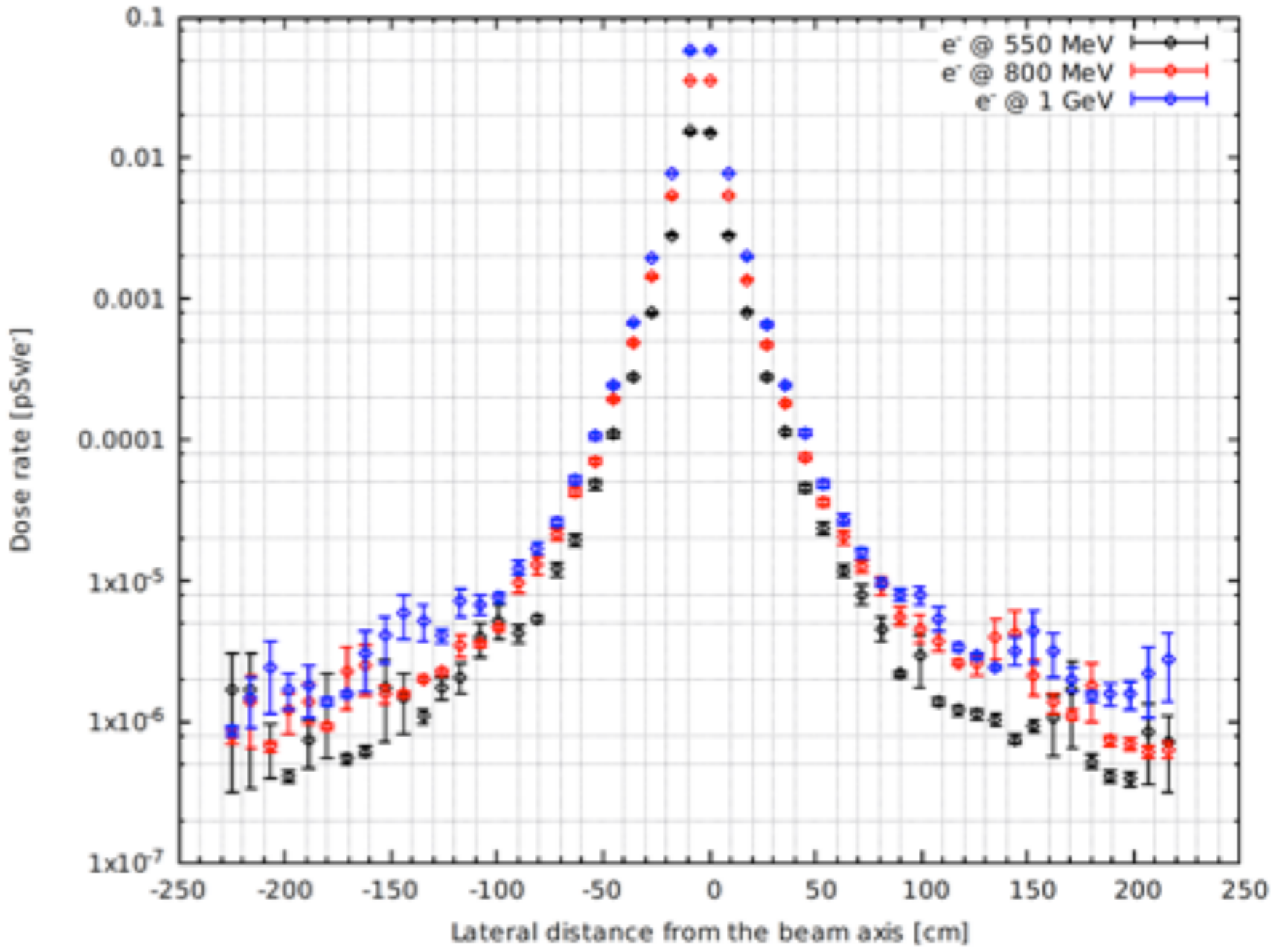

**Figure 3.29:** Dose rate (pSv/primary) for the new BTF experimental hall, simulated in FLUKA, electron beam, 550 MeV, 800 MeV and 1 GeV energy, as a function of the lateral distance along the front wall (1 cm distance, behind the shielding).

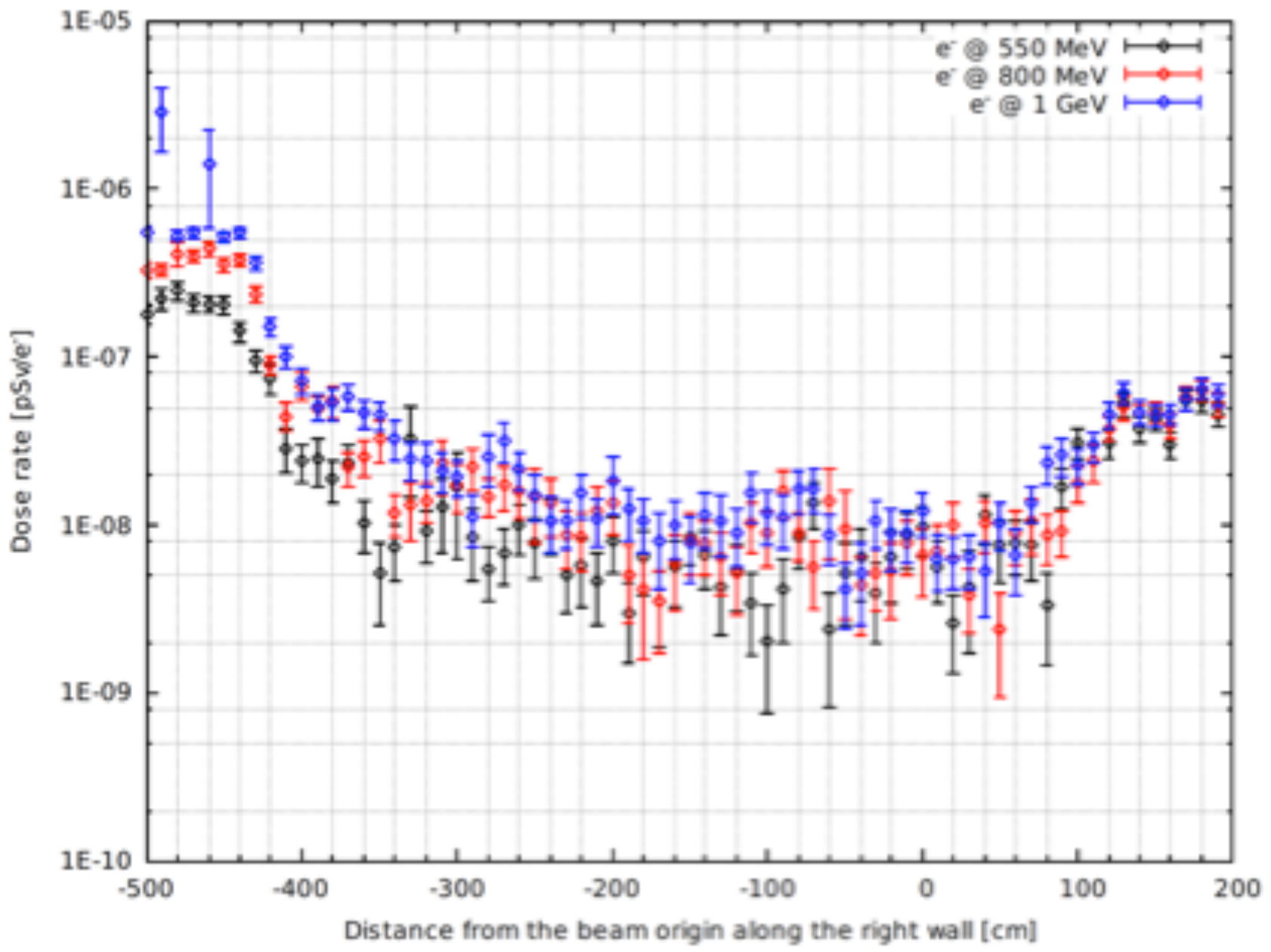

**Figure 3.30:** Dose rate (pSv/primary) for the new BTF experimental hall, simulated in FLUKA, electron beam, 550 MeV, 800 MeV and 1 GeV energy, as a function of the distance from the beam origin, along the lateral wall (1 cm distance, behind the shielding).

A better longitudinal containment seems to be desirable, in order to have a less peaked dose rate map, increasing the forward shielding, also including a dedicated beam dump before the concrete wall.

### 3.3.2 Cost evaluation

Costs for the new elements, as well as a first estimate of the resources for adapting the former control room to an useful experimental hall, are summarized in Tab. 3.5.

| Item | N. | Unit cost (k€) | Cost (k€) |
|---|---|---|---|
| 45° dipoles and power supplies | 4 | 100 | 400 |
| Quadrupoles (new line) | 6 | 70 | 420 |
| Correctors (new line) | 3 | 10 | 30 |
| New shielding, civil engineering | | | 160 |
| New collimators, supports, pipes | | | 40 |
| Electrical systems upgrade | | | 100 |
| Cooling systems upgrade | | | 260 |
| **Total WP3** | | | **1410** |

**Table 3.5:** BTF lines splitting costs (WP3).

Concerning the estimate for the cooling system, it should be underlined that the quoted cost includes the refurbishing of the two BTF experimental halls, the complete redesign of the temperature control of the modulator hall, the upgrade of the secondary cooling circuit of the LINAC and the extension of the primary cooling system for the additional four accelerating sections in the LINAC tunnel.

# 4 COSTS AND PLANNING

## 4.1 SUMMARY OF COSTS

Adding the costs for the three work packages listed in Tabb. 3.2, 3.3 and 3.4, the overall financial resources for the project sum up to about 4 millions, as shown in Tab. 4.1.

| Item | N. | Unit cost (k€) | Cost (k€) |
|---|---|---|---|
| New supplies for PFN charging | 3 | 40 | 120 |
| New supplies for PFN charging (existing modulators) | 8 | 40 | 320 |
| New readout system for LINAC signals | 1 | 50 | 50 |
| New modulator control system | 1 | 100 | 100 |
| New readout and DAQ for BPMs | 1 | 50 | 50 |
| New gun pulser 1 ns - 4.5 us | 1 | 70 | 70 |
| Secondary cooling upgrade for RF structure and SLED | 1 | 100 | 100 |
| Modulator | 1 | 600 | 600 |
| Klystron filament and core bias supplies | 10 | 6 | 60 |
| **Total WP1** | | | **1470** |
| | | | |
| **Item** | **N.** | **Unit cost (k€)** | **Cost (k€)** |
| SLED | 1 | 200 | 200 |
| Klystron | 1 | 200 | 200 |
| Accelerating sections | 4 | 100 | 400 |
| Waveguide network | 1 | 150 | 150 |
| Quadrupoles (LINAC) | 4 | 75 | 300 |
| Correctors (LINAC) | 8 | 10 | 80 |
| **Total WP2** | | | **1330** |
| | | | |
| **Item** | **N.** | **Unit cost (k€)** | **Cost (k€)** |
| 45° dipoles and power supplies | 4 | 100 | 400 |
| Quadrupoles (new line) | 6 | 70 | 420 |
| Correctors (new line) | 3 | 10 | 30 |
| New shielding, civil engineering | | | 160 |
| New collimators, supports, pipes | | | 40 |
| Electrical systems upgrade | | | 100 |
| Cooling systems upgrade | | | 260 |
| **Total WP3** | | | **1410** |
| | | | |
| **Grand total** | | | **4210** |

**Table 4.1:** Overall costs for the project (WP1+WP2+WP3).

## 4.2 Planning

In order to complete the consolidation of the modulators, the following work package (WP1) should be realized:

- Replace the charge circuit of the modulators with new generation HV supplies
- Upgrade the power supplies for the filament and core bias
- Completely replace the analogic system for the interlocks of the modulator with a re-designed, more compact and simpler one, based on digital electronics components and FPGA, adapting it to the existing modulators
- Renew the readout system of the modulator signals and the control with digital I/O, now based on CAMAC-bus
- Upgrade the control system of LINAC, reflecting the hardware upgrades in the modulators
- Upgrade of the diagnostics, in particular of the digitizer and DAQ of the LINAC BPMs
- Upgrade of the pulser of the gun, up to 4.5 μs
- Revise and upgrade the secondary cooling system.

Since all the new components should be not only installed and tested, but also integrated in the existing RF stations, prior to the final installation, an extensive system test for the validation of all new hardware and software parts should be performed.

In order to reduce the down-time of the LINAC, it is fundamental to implement the new systems on a fifth RF station. Only after having validated the new solutions on a complete RF station, it will be possible to proceed to the replacement of power supplies and control components on the existing four modulators.

The overall planning is shown in Figg. 4.1.a,b,c: in order to minimize the impact on the operation of the LINAC, and thus of the entire DAFNE complex, the installation phases inside the modulators hall and on the RF systems are foreseen in correspondence with the usual Summer and Christmas shutdown of the machine.

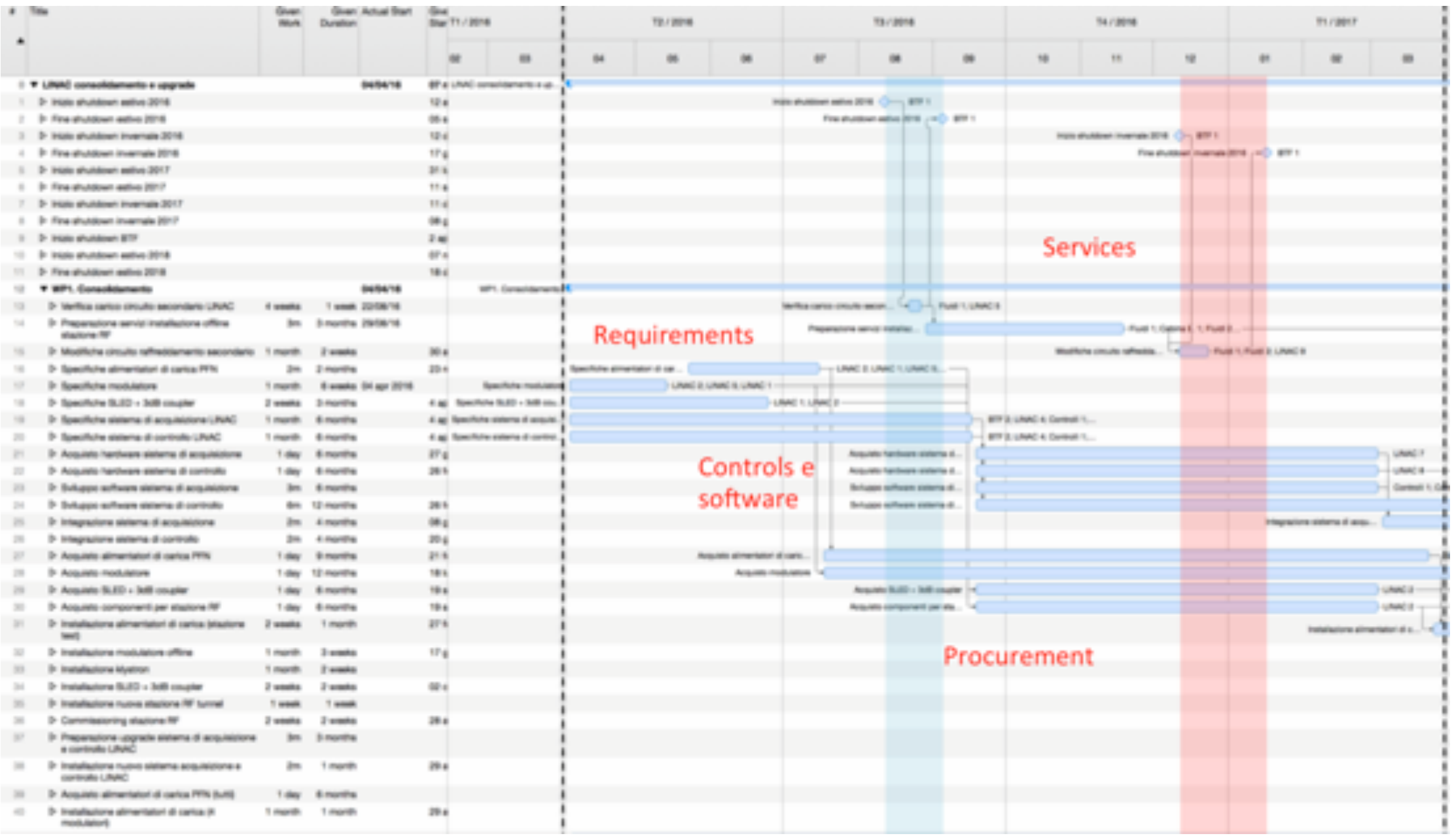

**Figure 4.1.a:** Planning of LINAC modulators consolidation and interlock upgrade (WP1), year 1.

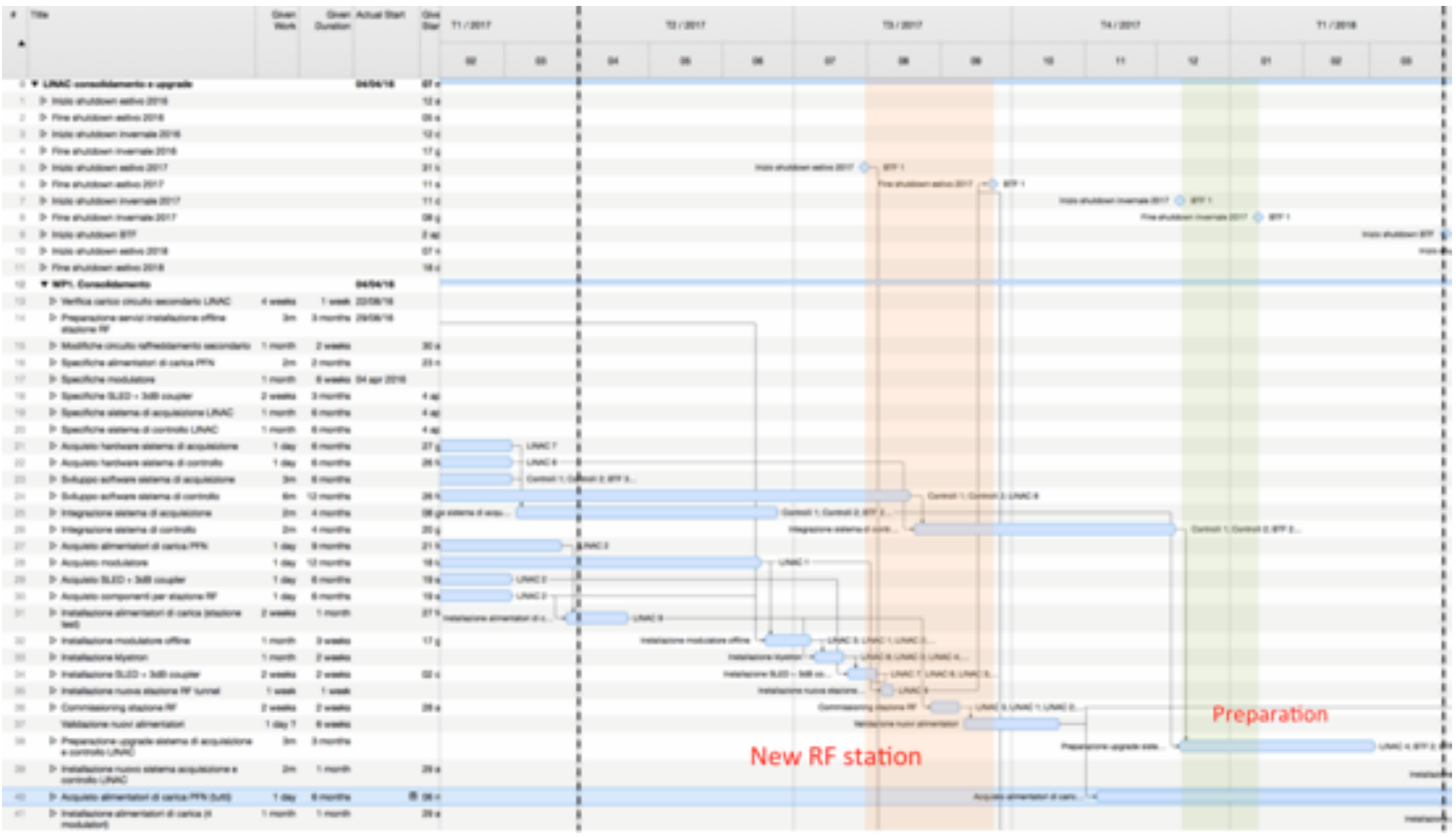

**Figure 4.1.b:** Planning of LINAC modulators consolidation and interlock upgrade (WP1), year 2.

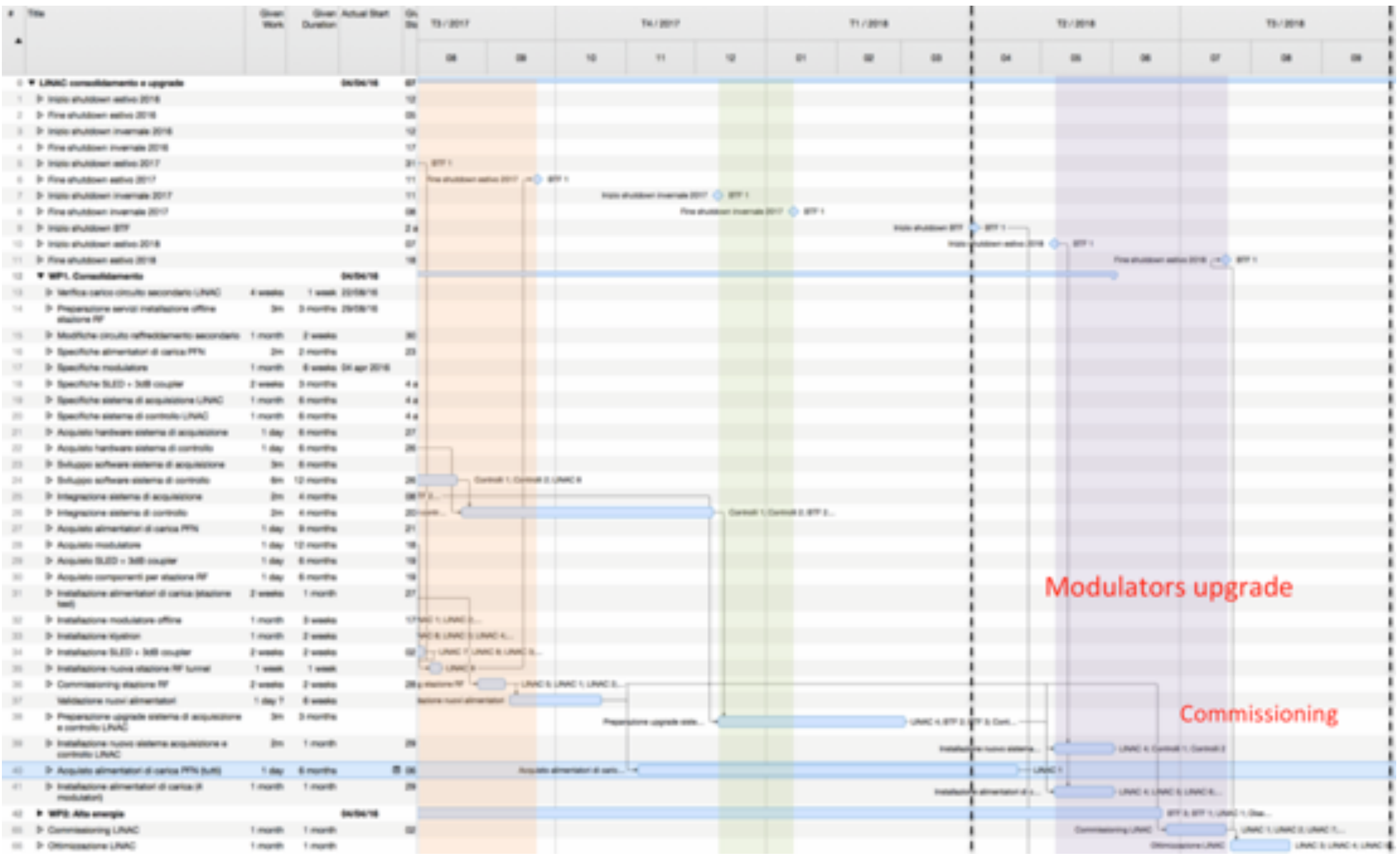

**Figure 4.1.c:** Planning of LINAC modulators consolidation and interlock upgrade, (WP1) year 3.

Sticking to the baseline solution, four additional accelerating sections, identical to the existing 16 ones should be built and installed in the LINAC tunnel. A fifth RF station, including a full modulator, the klystron, the tank with pulse transformer and thyratron, the SLED and all other components, has to be realized and installed in the modulators hall. The new RF station should then be connected through a new waveguide network to the four new accelerating sections. Vacuum, cooling and control and other ancillary systems should of course be extended accordingly for servicing the new portion of the LINAC.

The work package (WP2) is summarized as follows:

- In a preliminary phase, verify the transport optics with four additional sections, with two QW type quadrupoles (see Fig. 3.13)
- Verify the overall layout, duplicating the RF station from one of the two in the "high energy" of the LINAC, in particular the power splitting to the four sections
- Design of supports and ancillary systems, preparation of services (vacuum, cooling, controls in particular)
- Requirements and procurement of:
  - Accelerating sections and loads
  - Waveguide network
  - Quadrupoles and power supplies
  - Diagnostics (BPM, flag)
  - Vacuum pumps
- Dismounting of the 15 m drift, installation of the four accelerating sections
- Connection to the waveguide network
- Procurement, installation and commissioning of the fifth RF station, i.e. execution of WP1
- Commissioning and optimization.

The overall planning of the WP2 spans three years (as shown in Figg. 4.2.a,b,c), concentrating the installation and commissioning of the new sections, waveguides and RF in the shutdowns of the DAΦNE complex, usually foreseen before the August and Christmas stops.

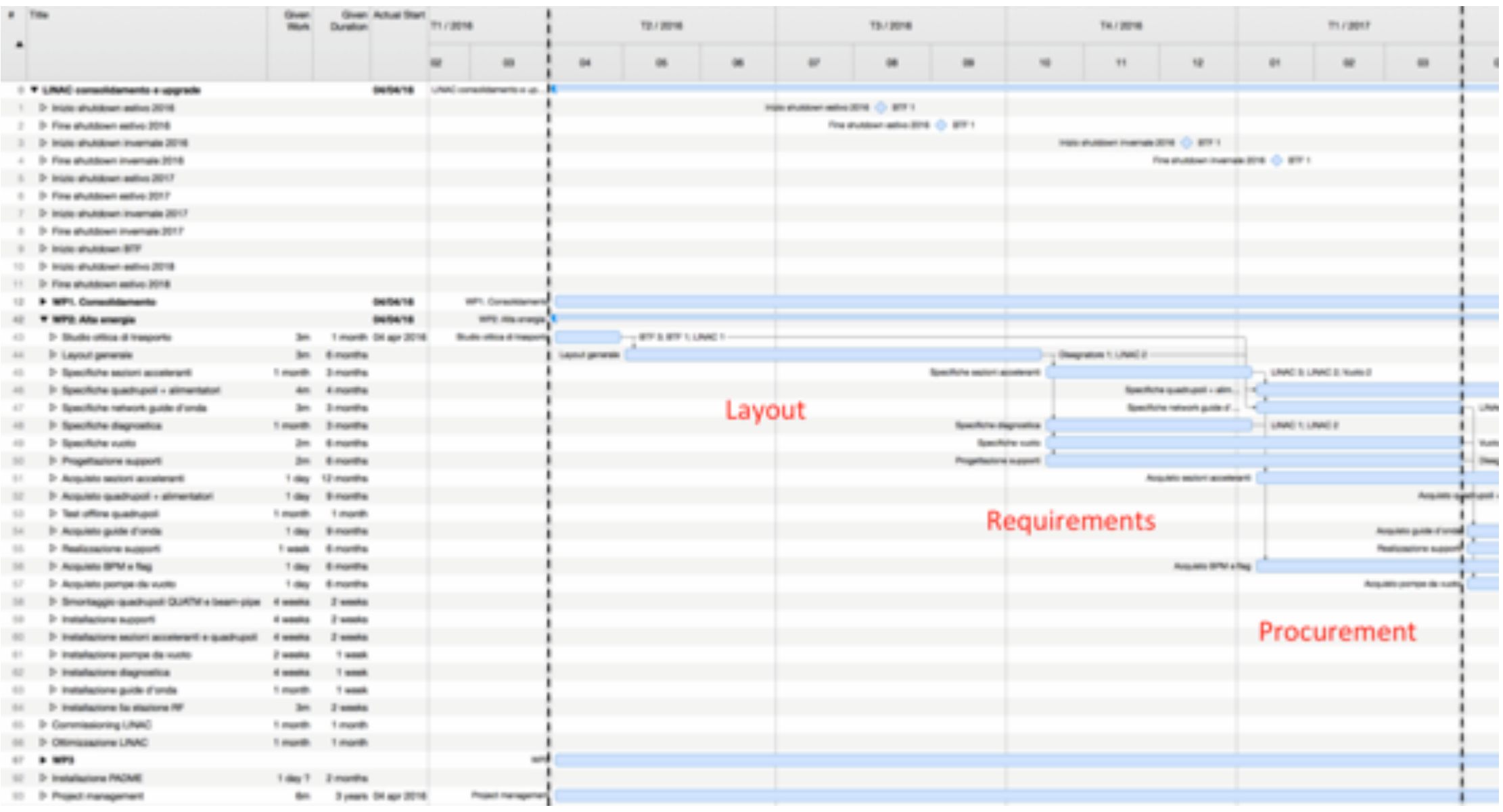

**Figure 4.2.a:** Planning of LINAC energy upgrade, (WP2) year 1.

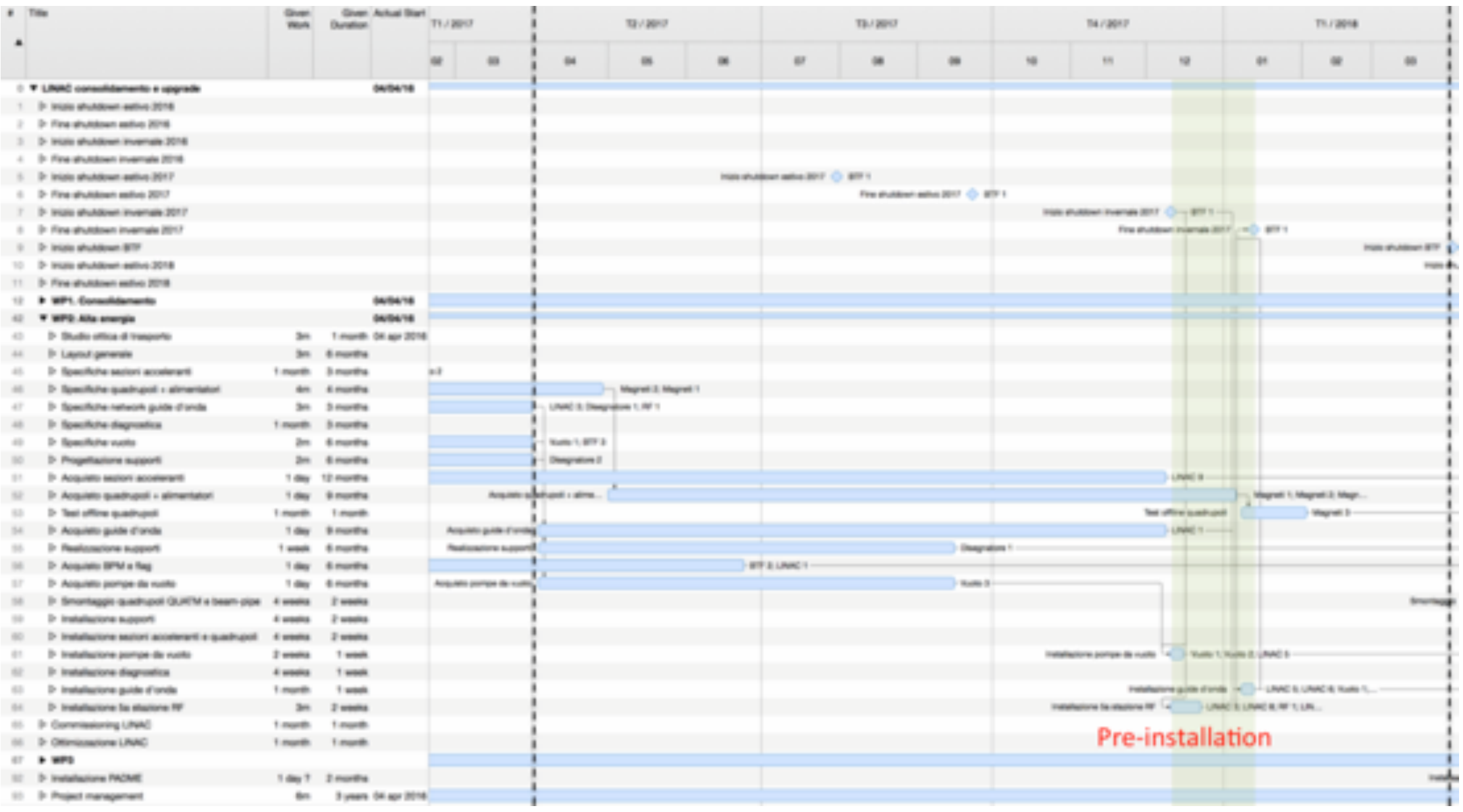

**Figure 4.2.b:** Planning of LINAC energy upgrade, (WP2) year 2.

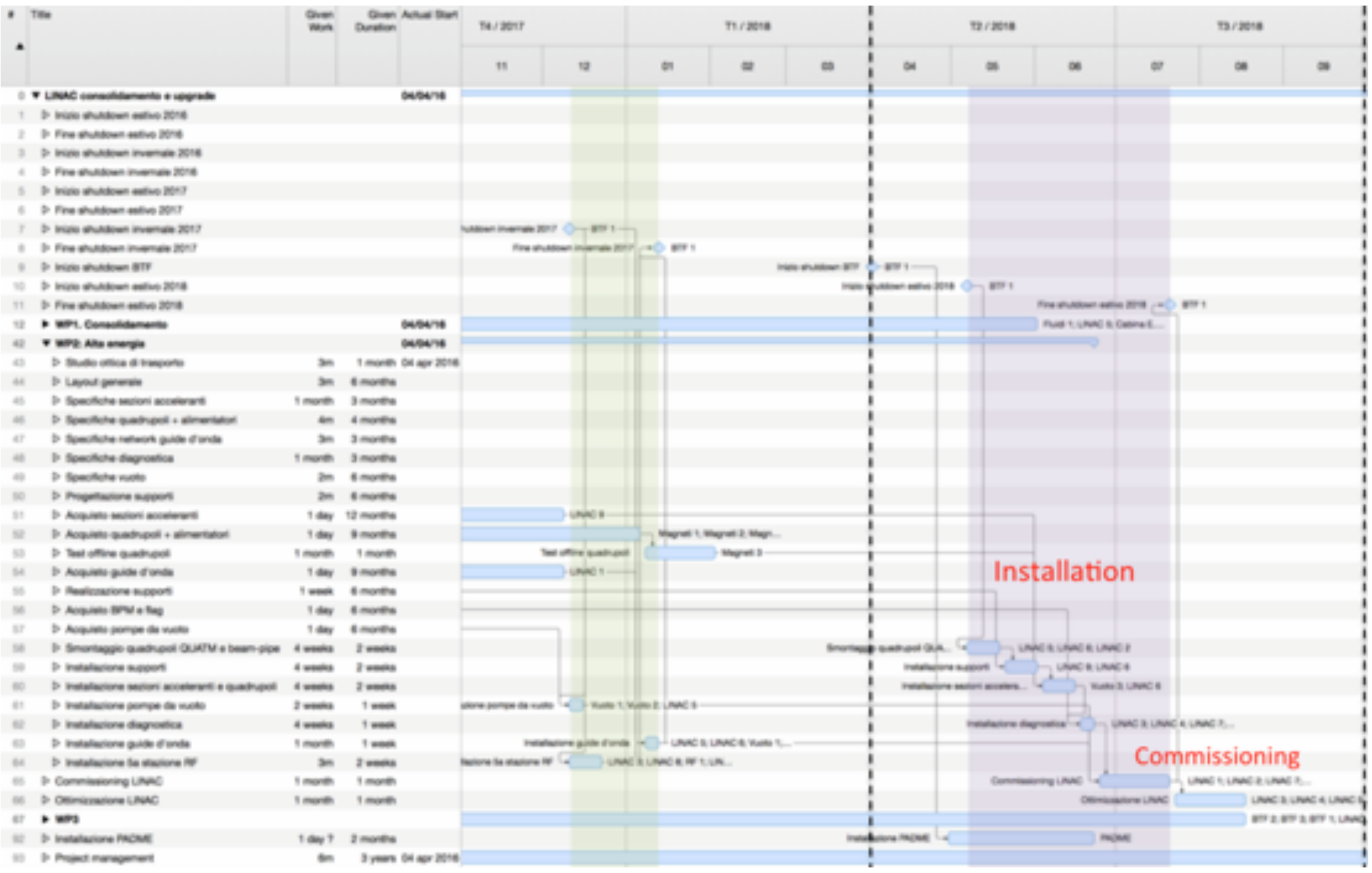

**Figure 4.2.c:** Planning of LINAC energy upgrade, (WP2) year 3.

After having completed the definition of the new lines layout, the requirements of all new elements (in particular the new magnets and relative supplies) have to be prepared, then the procurement can start (year 1, see Fig. 4.3.a). After the purchase of all components, WP3 requires to completely dismount the existing BTF transfer line, from the last collimator (SLTTB04), just downstream of the selecting dipole (DHSTB01) to the end of the beam-line in the BTF hall.

In order to place the first (splitting) dipole immediately at the entrance of the experimental hall, the two FODO doublets are replaced by a quadrupole triplet. This portion of the line is inside the LINAC tunnel (year 2, see Fig. 4.3.b), thus requiring a relatively short stop of the LINAC (and DAΦNE) activities, prior to the final installation and commissioning of the new lines in the BTF area (year 3, see Fig. 4.3.c). For this final phase, an early stop of the BTF activities has to be planned, as well as commissioning phase of the new lines at the end of the installation of the new elements.

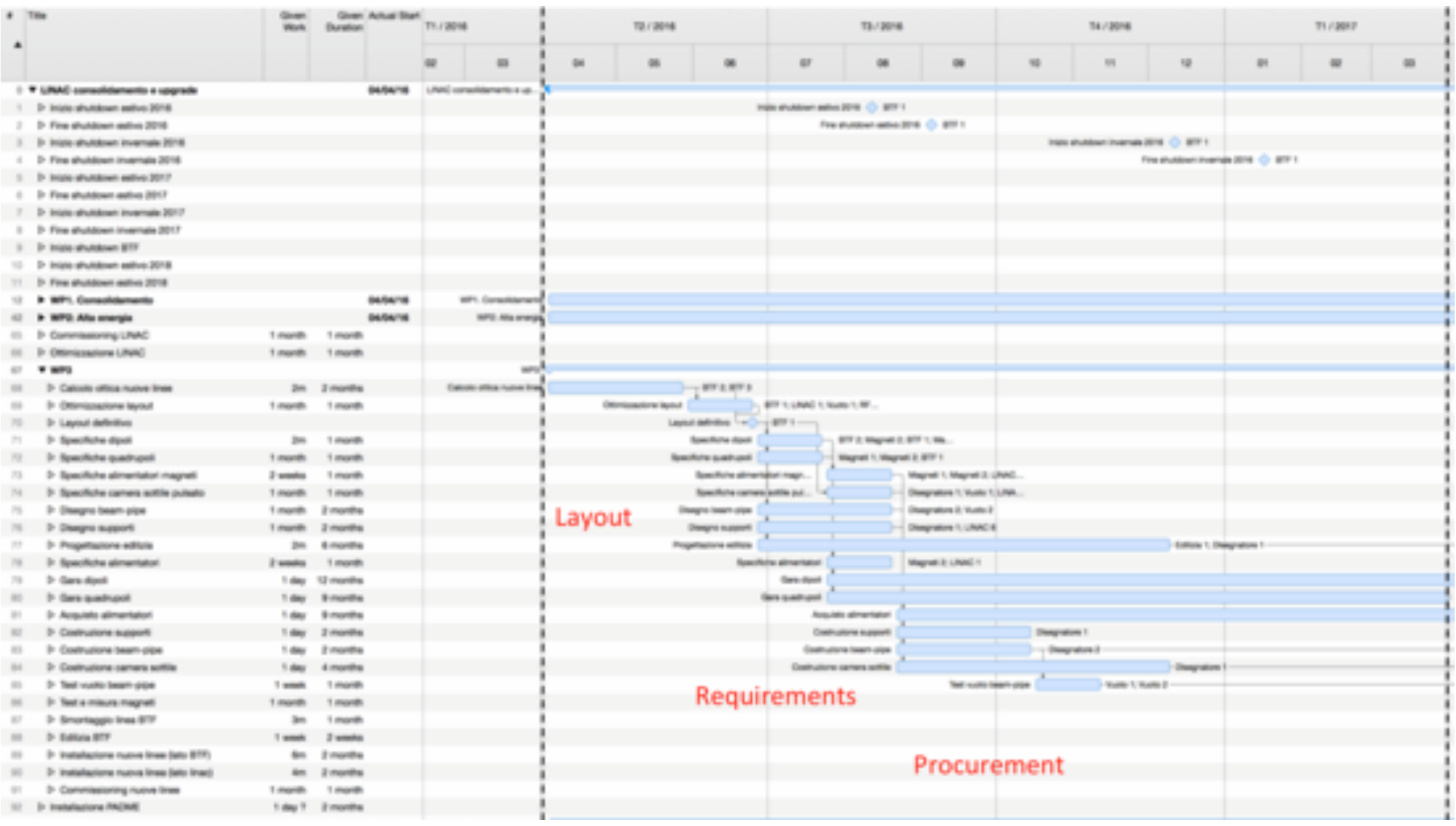

**Figure 4.3.a:** Planning of BTF lines splitting, (WP3) year 1.

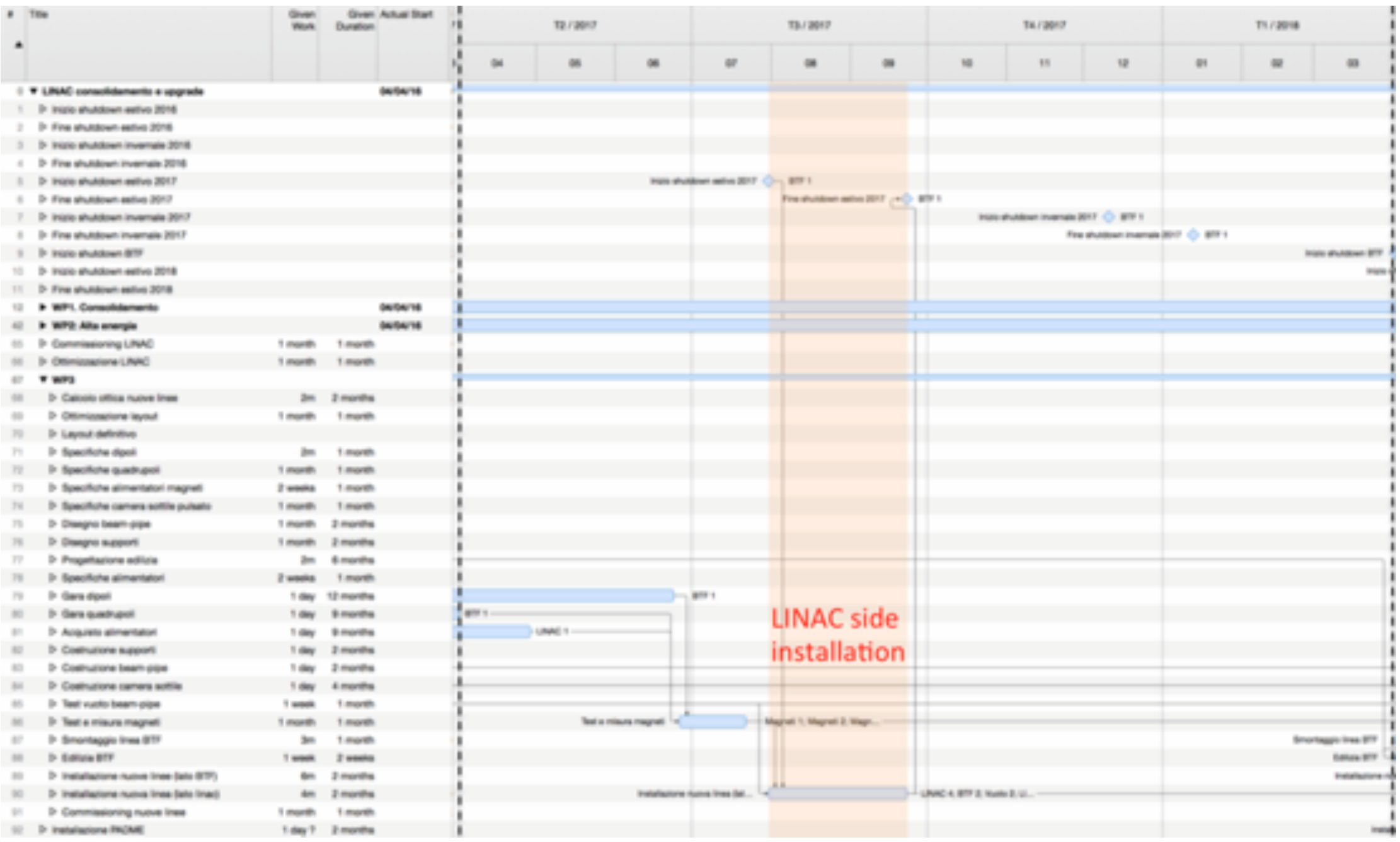

**Figure 4.3.b:** Planning of BTF lines splitting, (WP3) year 2.

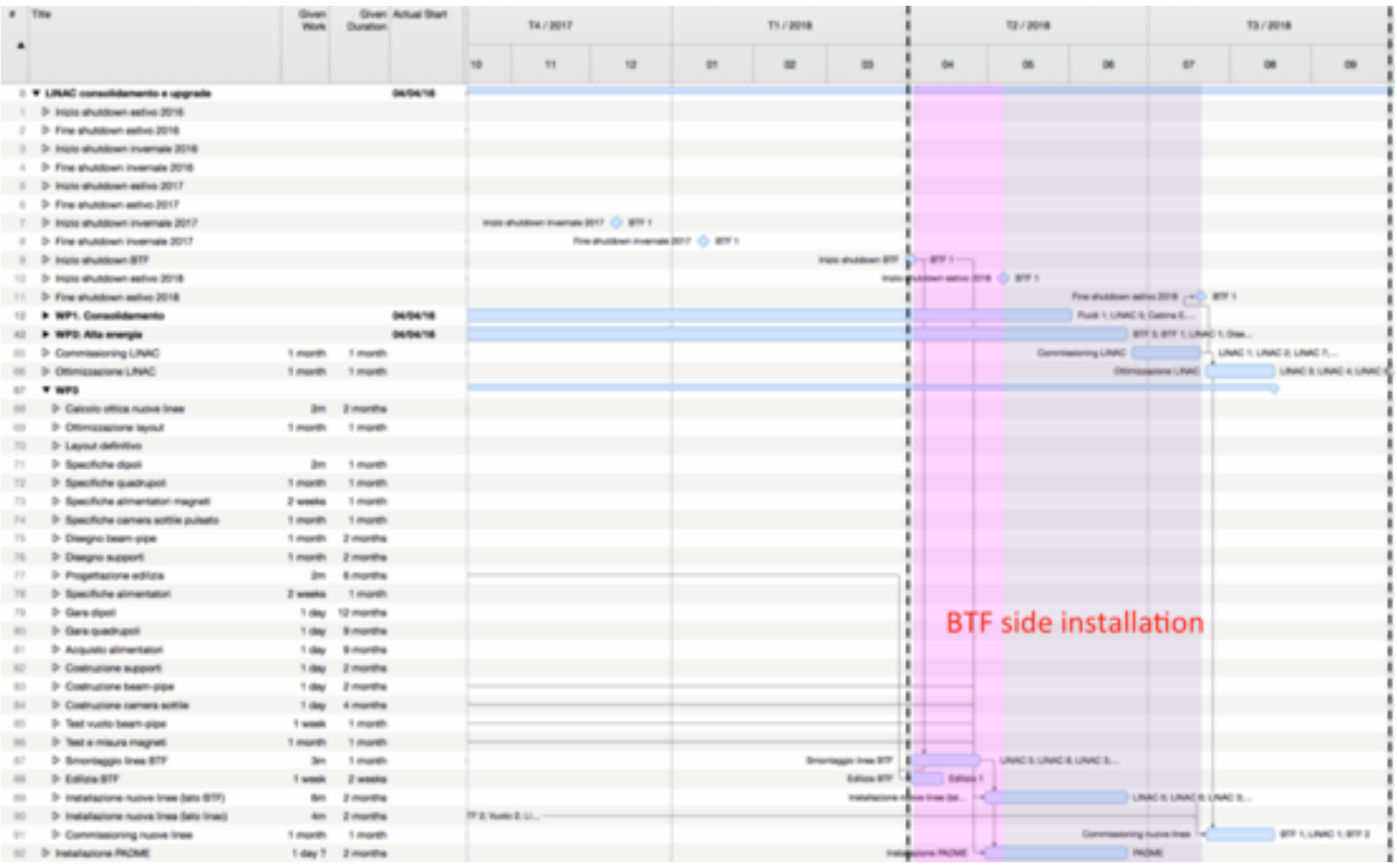

**Figure 4.3.c:** Planning of BTF lines splitting, (WP3) year 3.

## 4.3 HUMAN RESOURCES

The human resources needed to complete all three work packages can be estimated in about 100 man months, covering all the competences in the Accelerator and Technical Divisions. In Fig. 4.4 the summary of the needed resources, tagging each person with a label referring to the expertise area (more or less coinciding with the technical service), is reported.

As it can be expected, the load is maximum on the LINAC and BTF staff, involved in all of the three WPs, but since the activities are diluted in a three years period, the overall commitment is not above 30% of the full time equivalent.

| Title | Actual Start | Actual End | % completed | Actual Work | Remaining Work | Actu |
|---|---|---|---|---|---|---|
| BTF 2 | | | | | 5,83 months | |
| LINAC 1 | | | | | 5,61 months | |
| BTF 4 | | | | | 5,58 months | |
| LINAC 2 | 04/01/16 | | 0,00% | 0 days | 5,43 months | |
| BTF 1 | 04/01/16 | | 0,00% | 0 days | 5,42 months | |
| Magneti 2 | | | | | 5,17 months | |
| LINAC 3 | | | | | 4,68 months | |
| Disegnatore 1 | | | | | 4,35 months | |
| LINAC 4 | | | | | 4,26 months | |
| BTF 3 | | | | | 4,09 months | |
| LINAC 8 | | | | | 3,95 months | |
| LINAC 6 | | | | | 3,61 months | |
| Cabina E. 1 | 09/01/16 | | 0,00% | 0 days | 3,43 months | |
| LINAC 5 | | | | | 3,42 months | |
| Vuoto 2 | | | | | 3,39 months | |
| Disegnatore 2 | | | | | 3,3 months | |
| Controlli 1 | | | | | 3,28 months | |
| Controlli 2 | | | | | 3,28 months | |
| LINAC 9 | | | | | 2,76 months | |
| RF 1 | | | | | 1,95 months | |
| LINAC 7 | | | | | 1,85 months | |
| Magneti 3 | | | | | 1,75 months | |
| Magneti 1 | 30/01/16 | | 0,00% | 0 days | 1,62 months | |
| Vuoto 1 | | | | | 1,27 months | |
| Edilizia 1 | | | | | 1 month | |
| Fluidi 1 | 04/01/16 | | 0,00% | 0 days | 3,53 weeks | |
| Fluidi 2 | 09/01/16 | | 0,00% | 0 days | 3,05 weeks | |
| Vuoto 3 | | | | | 2,1 weeks | |

**Figure 4.4**: Human resources assignment.